\documentclass[letterpaper,11pt]{article}

\usepackage{import}
\usepackage{preamble}

\title{Bayesian model averaging of risk set probabilities using a geometric representation of multivariate extremes}
\author{Elizabeth S. Lawler and Benjamin A. Shaby}
\date{\today}

\begin{document}
\maketitle{}
	
\begin{abstract}
Modeling multivariate extremes using a geometric perspective leverages the shape of the multivariate point cloud to make inference on joint tail probabilities. While the original statistical framework for geometric extremes was fully parametric, relying on a gauge function that uniquely defines the shape for a given density, newer methods have introduced semi- and non-parametric alternatives to increase flexibility. We propose a modeling approach that retains the simplicity of the parametric framework but adds flexibility by using Bayesian model averaging (BMA) to improve prediction of tail risk probabilities. In contrast to previous works that use a truncated radial likelihood, we propose using a censored likelihood, which we find has consistently better performance in terms of predictive mean squared error, particularly in small-sample settings, although at the expense of increased bias. To generate predictions, we use a simple importance sampling scheme that seems to match the accuracy of more bespoke methods. Finally, we apply our approach to two fire weather indices, which are designed to capture somewhat orthogonal aspects of fire risk, to illustrate our method's practical utility in environmental applications.
\end{abstract}

\section{Introduction}\label{sec:intro}
Predicting multivariate risk set probabilities requires understanding the shape and behavior of the joint tail.  Our primary goal in this work is to accurately estimate joint tail probabilities by combining simple parametric models using Bayesian model averaging (BMA). Classical multivariate extreme value theory relies on frameworks like multivariate regular variation (MRV) and max-stable models, both of which are limited to use in asymptotically dependent (AD) models \citep{coles1999}, as they do not apply in nontrivial asymptotically independent (AI) cases. In environmental applications, as we consider here, statisticians have increasingly argued against the use of max-stable models \citep{huser2025}, in favor of more flexible models.

More recent multivariate extreme value modeling has emphasized working in light-tailed margins---as opposed to heavy-tailed in the MRV framework---allowing for more nuance in modeling the joint tail behavior and the ability to capture characteristics of both AD and AI classes. Most of this work is grounded in the limit set representation of multivariate extremes \citep{davis1988, kinoshita1991, balkema2010, balkema2010a, balkema2012}, exploiting a radial-angular decomposition of the data. Other approaches, which are not based in limit set theory, also exploit this decomposition in light-tailed margins to develop semi-parametric angular-radial models \citep{mackay2024, murphy-barltrop2024b}. We restrict our attention to the limit set representation.

Building on earlier work of \cite{nolde2014} and \cite{simpson2021}, \cite{nolde2022} leveraged the limiting shape of the point cloud to provide theoretical connections to key concepts in multivariate extremes, including tail dependence measures, MRV, and conditional extremes. \cite{wadsworth2024} introduced a parametric framework for statistical inference and risk set probability estimation using limit sets, proposing a Gamma approximation of the radial distribution in polar coordinates. Semi-parametric methods followed, with some exploiting the aforementioned Gamma approximation \citep{majumder-2025a, campbell2025} and others leveraging the generalized Pareto distribution \citep{simpson2024a, papastathopoulos2025}. 

\cite{majumder-2025a} and \cite{simpson2024a} focus solely on estimating the shape of the limiting point cloud, while others pursue predictive inference approaches that rely on model selection procedures, introducing potential bias and model uncertainty \citep{wadsworth2024,campbell2025,papastathopoulos2025}. The latter three propose methods of estimating risk set probabilities, either through a bespoke importance sampling scheme \citep{wadsworth2024, campbell2025} or through other numerical and Monte Carlo integration \citep{papastathopoulos2025}. Notably, all of the aforementioned approaches discard observations below some suitably high threshold and make inference using truncated likelihoods. With small datasets, as is often the case in applied settings, incorporating data below the threshold may improve model fitting and subsequent prediction and inference. 

In this work, we also use a radial-angular decomposition of the multivariate point cloud, necessitating a model for the angles and a model for the radii given the angles. We develop a method that maintains the simplicity of the parametric framework of \cite{wadsworth2024}, while adding flexibility through Bayesian model averaging (BMA).

The radial-angular decomposition yields a representation of the joint density in terms of two independent components: an angular component and a radial component given the angles. \cite{wadsworth2024} show that the distribution of radial exceedances, conditioned on the angle, can be approximated by a truncated Gamma distribution. The rate parameter of the Gamma is connected to the shape of the limit set, which we discuss further in Section~\ref{sec:stat_inf}. To avoid introducing potential bias by including non-extremal data points, the authors fit large radii using a likelihood consisting of the right tail of a Gamma distribution, left truncated at a high quantile. However, as angles corresponding to data below the threshold may be useful in informing the shape of the multivariate point cloud, we introduce an alternative likelihood which uses all of the angles and assumes the corresponding radii come from a Gamma distribution which is left censored, rather than truncated, at a high threshold. We conjecture that this censored likelihood \citep[Ch.~9~of][]{beirlant_statistics_2004} makes more efficient use of the data, which is particularly important with smaller sample sizes.

We apply the Gamma approximation  for the radii conditional on the angles to fit a suite of parametric models of limit set shapes. We then use BMA to combine predictions from the entire suite of models, rather than selecting a single model deemed best, as theory suggests that BMA provides an optimal predictor in terms of predictive log scores \citep{madigan1994}. BMA circumvents many limitations that arise when using standard model selection procedures. With standard modeling procedures, one picks a single model from several candidates according to a metric like AIC. This has a couple of potential pitfalls. First the selection criterion could choose the ``wrong'' model, even if the true data-generating model were among the candidates. Second, choosing a single model ignores variability inherent in model selection, resulting in overconfident predictions \citep{madigan1994}. Instead, BMA optimizes over the posterior likelihoods of all models under consideration, and appropriately weights each model to maximize the combined likelihood. The resulting weights can then be used similarly to the weights of a mixture model, but without the potential model-fitting or identifiability issues that may arise in standard mixture models.  Thus, BMA may be preferable to choosing a single model, particularly in cases like multivariate extreme value analysis where paucity of data makes it difficult to check the adequacy of any individual model, and any existing parametric model is unlikely to capture all key tail features of the data-generating process. 

The use of BMA in multivariate extremes remains limited; \cite{apputhurai2011} combine various AD and AI models through a weighted version of the summary measure $\psi$, \cite{sabourin2013} average over the spectral measure of two AD models, and \cite{vettori2020} average over various depths of the nested logistic model \citep{tawn1990}. Our implementation differs from these previous works in that we apply BMA to limit set modeling to improve the prediction of extremely small risk set probabilities. 

The remainder of this paper is arranged as follows. Section~\ref{sec:BMA-limit-sets} provides a brief overview of limit sets and relevant background on Bayesian model averaging.
Section~\ref{sec:stat_inf} describes the construction of the likelihood necessary for model fitting under a given parametric model, in which we decompose the likelihood into radial and angular components. Section~\ref{sec:predictions} details our importance sampling scheme under a given parametric model, which we use to compute our predictions of risk set probabilities. We demonstrate the performance of our approach through a comprehensive simulation study in Section~\ref{sec:sim_study}. We apply our method to fire weather indices and discuss various diagnostics in Section~\ref{sec:data}. And finally, we discuss the utility of our methods and proposed avenues of future research in Section~\ref{sec:discussion}.

\section{Averaging Across Parametric Models for Limit Sets}\label{sec:BMA-limit-sets}

\subsection{Limiting Shape of the Point Cloud}\label{sec:limit_shape}
Let  $\bX_1, \dots, \bX_n$ be $n$ independent random vectors in $\mathbb{R}^d$ with exponential margins. If we scale these random vectors by $\log(n)$ and consider the scaled point cloud $N_n = \{\bX_1/\log(n), \dots, \bX_n/\log(n)\}$, $N_n$ converges onto the limit set $G$ as $n \rightarrow \infty$ \citep{davis1988,kinoshita1991}. $G$ has the following characteristics:
\begin{enumerate}
    \item $G$ is star-shaped: $\mathbf{x} \in G \implies t\mathbf{x}\in G$ for $t \in (0,1)$. 
    \item $G$ can be characterized by a continuous \emph{gauge function}, $g$: $G = \{\mathbf{x} \in \mathbb{R}^d_+:g(\mathbf{x}) \leq 1\},$ meaning the bouldary $\partial G$ is the unit level set of $g$ (i.e. $\partial G = \{\mathbf{x} \in \mathbb{R}^d_+:g(\mathbf{x}) = 1\})$.
    \item $g$ is 1-homogeneous: $g(c\bx) = cg(\bx), c >0$.
\end{enumerate}

\begin{figure}
    \centering
    \includegraphics[width=0.85\linewidth]{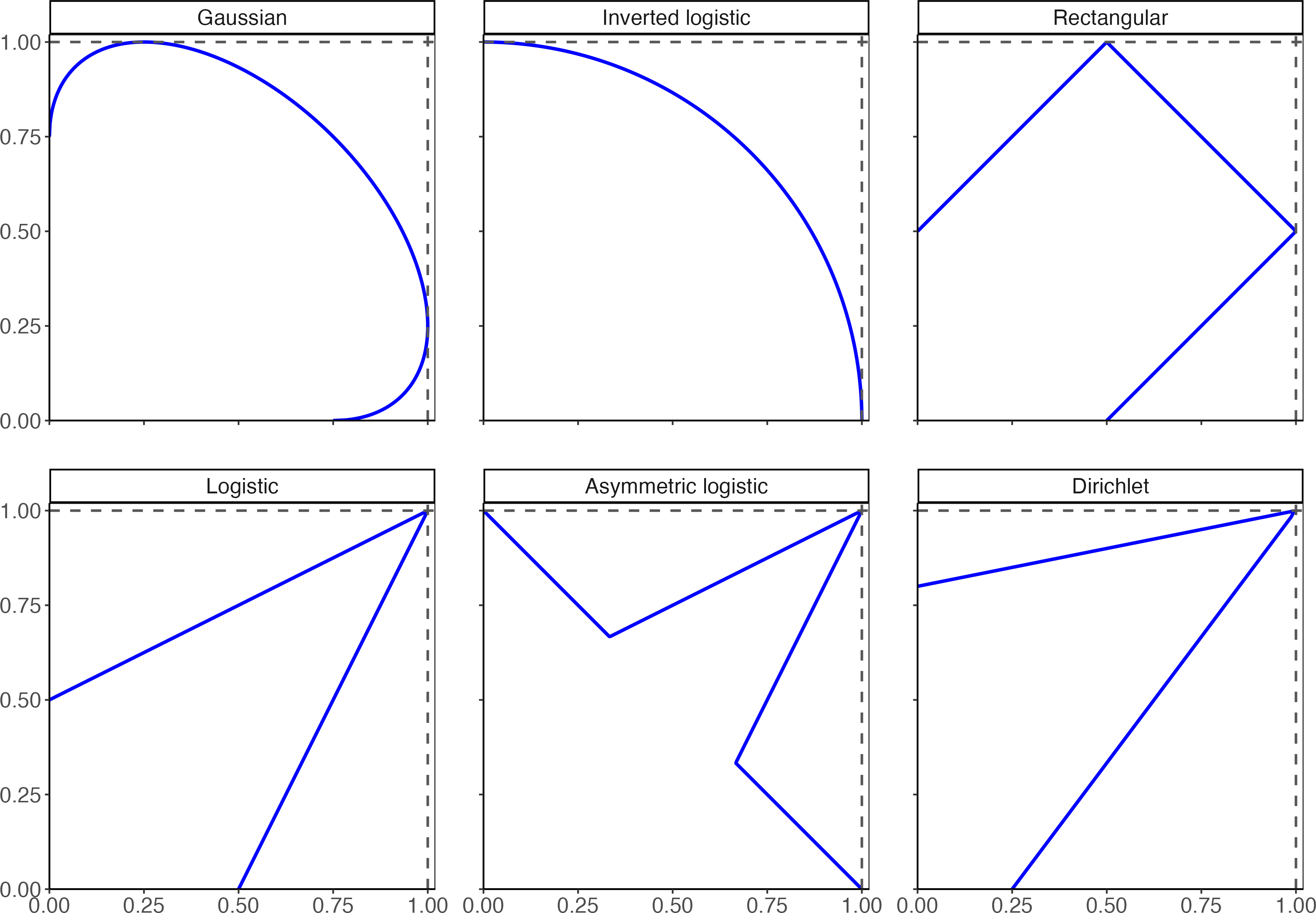}
    \caption{Unit level sets of six parametric gauge functions in blue. The boundary of the unit box is in gray. Top row corresponds to examples of AI gauge functions and bottom to AD.}
    \label{fig:level_sets}
\end{figure}

The shape of $G$ is directly related to the asymptotic dependence classes mentioned in Section~\ref{sec:intro}. In the bivariate case, AD and AI are commonly understood through the measure $\chi = \lim\limits_{u \to 1}\prob(F_1(X_1) > u \given F_2(X_2) > u)$, or the probability of one variable exceeding a suitably high threshold $u$ given that the other exceeds $u$. AI arises when $\chi = 0$, and AD arises when $0 < \chi \leq 1$. Visually, one can easily distinguish between the limit sets of these classes (Figure~\ref{fig:level_sets}): AI (top row) yields a \textit{blunt} shape, meaning it does not reach the upper right corner of the unit box (apart from a few edge cases), while AD shapes touch the point $(1,1)$ in two dimensions \citep{balkema2010a,balkema2012}.

A sufficient condition for convergence of $N_n$ onto $G$ conveniently relates the joint Lebesgue density of $\bX$, if it exists, to the gauge function \citep{nolde2014,nolde2022}:
\begin{equation}\label{eq:lebesgue_gauge}
    \lim\limits_{t \to \infty} \frac{-\log f_{\bX}(t\bx)}{t} = g(\bx), \ \ \bx \in [0, \infty)^d.
\end{equation} Assuming a tractable joint density and continuous gauge functions, one can determine the appropriate form of $g$ and the associated $G$ from \eqref{eq:lebesgue_gauge} \citep{nolde2022}. Notably, each joint density has a unique gauge function, but the converse is not true.

For illustrative purposes, we provide the Gaussian and logistic gauge functions below and direct the reader to \cite{wadsworth2024} and \cite{nolde2022} for associated derivations and other parametric gauge functions. In exponential margins, the gauge function corresponding to data with a Gaussian dependence structure and correlation parameter $\rho$ takes the general formula \[g(\bx \given \rho)= \frac{x_1 + x_2 - 2\rho\sqrt{x_1x_2}}{1 - \rho^2}\] in two dimensions. For data with a logistic dependence structure and dependence parameter $\lambda$, \[g(\bx \given \lambda)=\frac{1}{\lambda}(x_1 + x_2) + \text{min}(x_1,x_2)(1 - \frac{2}{\lambda}).\] In two dimensions, all of the gauge functions in Figure~\ref{fig:level_sets} have one dependence parameter except for Dirichlet, which has two. Throughout this work, we use $\btheta$ to refer to dependence parameters generically and use $\rho$ or $\lambda$ in the respective special cases of the Gaussian correlation parameter and the logistic or H\"usler-Reiss dependence parameter.

In this work, we consider a suite of six parametric families of gauge functions: Gaussian, inverted logistic, and rectangular in the AI class, and logistic, asymmetric logistic, and Dirichlet in the AD class.  Unit level sets for these gauge functions are shown in Figure \ref{fig:level_sets}.

\subsection{Bayesian model averaging}\label{sec:BMA}
Given a library of candidate parametric models, the question remains of how (if at all) to pick the ``best'' model for a given dataset. With standard modeling procedures, we would select a model based on a metric like AIC. Model selection procedures introduce uncertainty and bias through the choice of comparison metric and potentially excluding (unknowingly) the data-generating model from consideration. We use BMA to overcome the limitations inherent in model selection.

In this work, the prediction task is to estimate the probability of a region $B$ in the joint tail, $\prob(\bX \in B)$. Here, we describe BMA this context. Suppose we have access to models $\mathcal{M}= \{M_1,\dots,M_K\}$, each corresponding to a parametric gauge function, and data $\by := (r, \bw, r_\tau(\bw))$ (Section~\ref{sec:decomp_model}). The average posterior prediction under each model, averaged across the models, is then 
\[
\prob(\bX \in B \given \by) = \sum\limits_{k=1}^K \prob(\bX \in B \given M_k, \by)\prob(M_k \given \by).
\]
Predictions are weighted by the posterior probability of their corresponding models. This weighted average has superior predictive performance than any single model $M_k$ on its own \citep{madigan1994} and is the optimal predictor \citep[in terms of log predictive score;][]{hoeting1999}. 


The posterior probability of model $M_k$ is $$\prob(M_k\given \by) = \frac{f(\by | M_k)\prob(M_k)}{\sum\limits_{k=1}^K f(\by \given M_k)\prob(M_k)},$$ and the marginal likelihood of $M_k$ is 
\[
f(\by \given M_k) = \int f(\by \given \bbeta_k, M_k)\pi(\bbeta_k \given M_k)d\bbeta_k,
\]
where $\bbeta_k$ are the parameters for $M_k$. This integral is often intractable, so several approximations have been proposed \citep{hoeting1999, yao2018}, including Pseudo-BMA and Pseudo-BMA+. Pseudo-BMA is an AIC-type weighting that uses the product of Bayesian leave-one-out cross-validated predictive densities in place of the marginal likelihoods. Pseudo-BMA+ stabilizes these estimates using Bayesian bootstrapping.

\cite{yao2018} also propose stacking weights, a predictive distribution analog of \textit{stacking of means}:
$$\stackrel{\text{max}}{\bs \in [0,1]^K} \biggl[ \frac{1}{n} \sum\limits_{i=1}^n \log \sum\limits_{k=1}^K s_k f(y_i|\by_{-i}, M_k)\biggr]; \ \ \sum\limits_{k=1}^K s_k = 1.$$ Stacking weights use the corresponding leave-one-out predictive distribution as an empirical approximation of the full predictive distribution. These weights are asymptotically optimal (with respect to the predictive log score) compared to generic BMA weights, even when the data-generating model is not in $\mathcal{M}= \{M_1,\dots,M_K\}$.  We calculate all three variations of BMA weights for each model from the posterior samples of log-likelihood values using the \texttt{loo} package.

For each parametric model $M_k$ in our library $\mathcal{M}$, then, we need to obtain a posterior sample of its parameters, and given those parameters, a posterior sample of predictions.  This requires both a definition for the joint likelihood of the observations $\bX_1, \ldots, \bX_n$ under each model $M_k \in \mathcal{M}$, along with a method of producing predictions from a posterior sample under each model.

\section{Defining the Likelihood for Each Parametric Model}\label{sec:stat_inf}

\subsection{Radial-angular Decomposition}\label{sec:decomp_model}
When working with star-shaped sets, it is often convenient to transform the data from Euclidean to polar coordinates.  Under the $L_1$-norm, this gives $R=\sum\limits_{j=1}^d X_j$ and $\bW = \bX / R$. In two dimensions, $\bW_2 = 1 - \bW_1$. The joint density of $\bX$ can then be re-expressed in terms of independent radial and angular components:
\begin{equation}\label{multi_vector_density}
    f_{\bX}(\bx) \rightarrow f_{\bW,R}(\bw, r) = f_{R \given \bW}(r \given \bw)f_{\bW}(\bw),
\end{equation} necessitating models for both the radii and the angles.

\subsubsection{Radial Likelihood}\label{sec:radial}
First, we describe our modeling approach for the radial component, conditional on angle, of Equation~\eqref{multi_vector_density}. \cite{wadsworth2024} show the conditional density of $R \given \bW = \bw$ satisfies
\begin{equation}\label{eq:asym_gamma_approx}
    f_{R \given \bW}(r \given \bw) \propto r^{d-1}\text{exp}\{-rg(\bw)[1 + o(1)]\}, \ \ r \rightarrow \infty.    
\end{equation}
They further show that for important parametric cases, the $[1 + o(1)]$ term can be moved out of the exponent and is, in most cases, negligible. Thus, for suitably large thresholds $r_\tau(\bw)$, where $\tau \in (0,1)$ is some quantile level, a truncated Gamma distribution can approximate \eqref{eq:asym_gamma_approx}.

To establish the asymptotic behavior of $f_{R \given \bW}(r \given \bw)\text{ as } r\rightarrow \infty$ for various parametric dependence structures on exponential margins, \citep{wadsworth2024} derive a suite of parametric gauge functions, of which we will use six---three corresponding to AI and three to AD dependence structures (Figure~\ref{fig:level_sets}). Since several known parametric AD models share the same shape of their limit set, the other known AD gauge functions are reparametrizations of the three we use and are therefore not included in our models. 

Let $f(\cdot)$ and $F(\cdot)$ be the density and cumulative distribution function (cdf) of a Gamma distribution, respectively. The truncated Gamma model defines the following likelihood for a radius $r_i, \ \ i=\{1,\dots,n\}$, where the unknown dependence parameter $\btheta_1$ is written as a vector to account for the Dirichlet parametrization:
\begin{equation}\label{trunc_lhood}
    r_i \given \bw_i, r_i > r_\tau(\bw_i), \btheta_1 \sim \frac{f(r_i \given \alpha, g(\bw_i;\btheta_1))}{1 - F(r_\tau(\bw_i) \given \alpha, g(\bw_i;\btheta_1))},
\end{equation}
where $\alpha$ is an unknown shape parameter.

We wanted to investigate whether using a censored likelihood approach could improve model fit relative to the truncated likelihood. In contrast to typical censored likelihood approaches in multivariate extremes \citep[Ch.~9~of][]{beirlant_statistics_2004}, the censored likelihood in this radial-angular decomposition uses all of the angular information below the threshold, while censoring radii that are below the threshold.  The intuition is that including more angular information than the truncated approach may lead to reduced variance.  A concern may be that the reduced variance may come at the expense of additional bias. We address these concerns later in this section.

Using similar notation as above, the censored model defines the following likelihood for a radius $r_i$, as
\begin{equation}\label{cens_lhood}
    r_i \given \bw_i, r_\tau(\bw_i), \btheta_1 \sim 
    \begin{cases}
        f(r_i \given \alpha, g(\bw_i;\btheta_1)) & \text{if } r_i \geq r_\tau(\bw_i) \\
        F(r_\tau(\bw_i) \given \alpha, g(\bw_i;\btheta_1)) & \text{if } r_i < r_\tau(\bw_i). \\
    \end{cases}
\end{equation}

In both the truncated and censored cases, we rely on the asymptotic Gamma approximation for our likelihood.  One might wonder how well this approximation performs in finite-sample settings---specifically in recovering the unknown dependence parameter $\btheta_1$. To investigate this, we simulated data from Gaussian and logistic dependence structures at varying dependence levels, as described in Section~\ref{sec:data_gen}, and fit the two likelihoods with the data-generating gauge function. We then compared the posterior coverage rate of $\theta_1$ to the nominal coverage level by calculating credible intervals at 19 levels, from 0.05 to 0.95. We also examined boxplots of the posterior parameter estimates (i.e. posterior means) in comparison to the true value. We present results for two scenarios below (Figures~\ref{fig:gauss_coverage} and~\ref{fig:logistic_coverage}), with additional results in Supplementary Material~\ref{append:asym_approx_results}.

\begin{figure}[!htbp]
    \centering
    \includegraphics[width=0.75\linewidth]{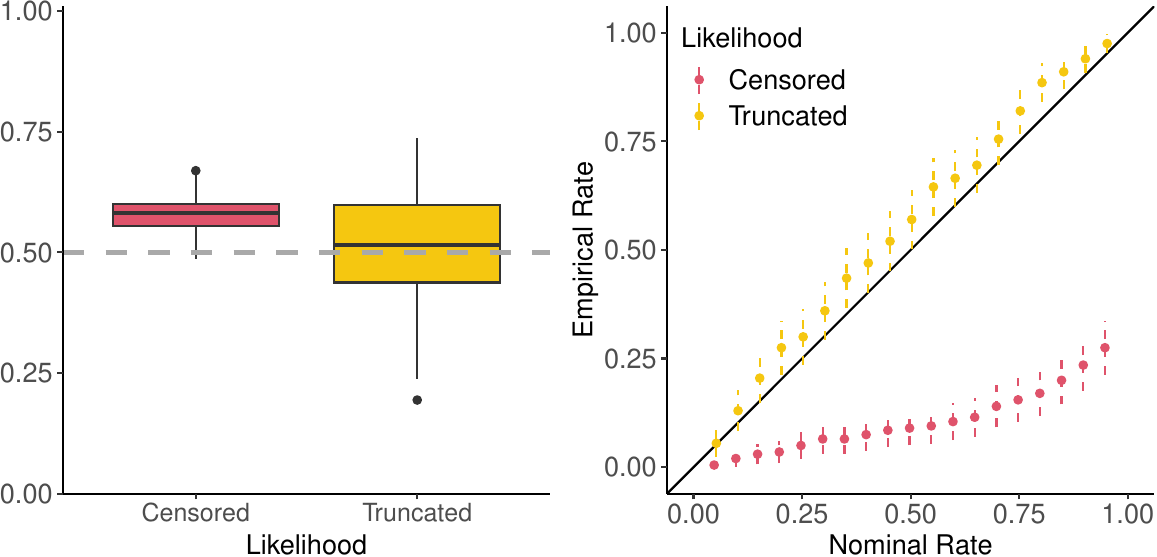}
    \caption{Posterior estimates (left) of the dependence parameter for 200 datasets generated from a Gaussian dependence structure with $\rho=0.5$. Posterior coverage of the true $\rho$ value is on the right.}
    \label{fig:gauss_coverage}
\end{figure}

\begin{figure}[!htbp]
    \centering
    \includegraphics[width=0.75\linewidth]{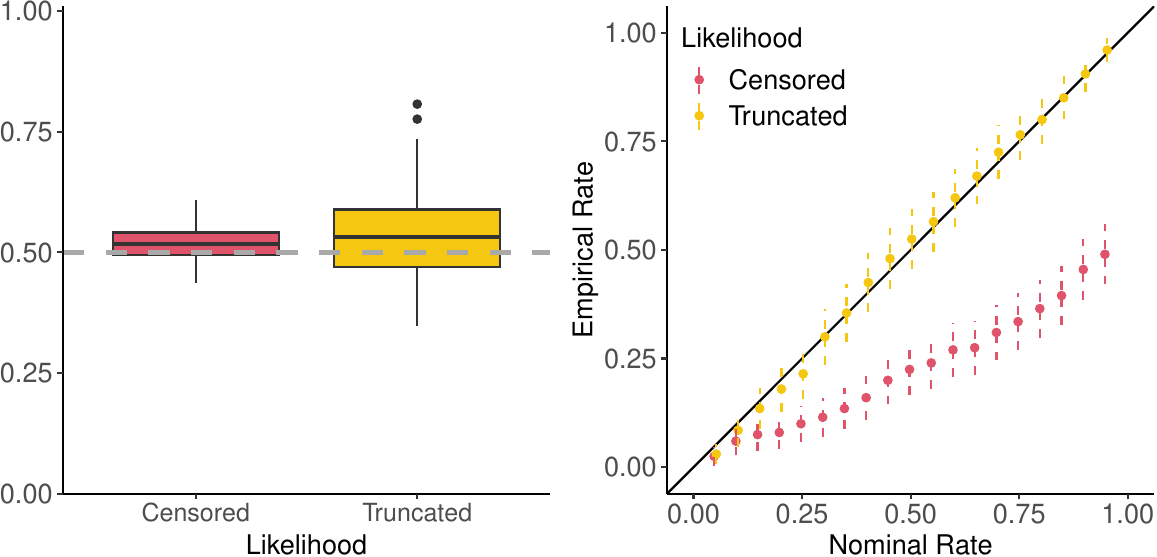}
    \caption{Posterior estimates (left) of the dependence parameter for 200 datasets generated from a logistic dependence structure with $\lambda=0.5$. Posterior coverage of the true $\lambda$ value is on the right.}
    \label{fig:logistic_coverage}
\end{figure}

Overall, the truncated likelihood provides substantially better calibration of the true parameter compared to the censored likelihood, which severely under-covers. This is a manifestation of a classic bias-variance tradeoff---while the censored likelihood tends to exhibit bias, it yields less variable estimates and may still be advantageous in small-sample settings.

\begin{figure}[!htbp]
    \centering
    \includegraphics[width=0.95\linewidth]{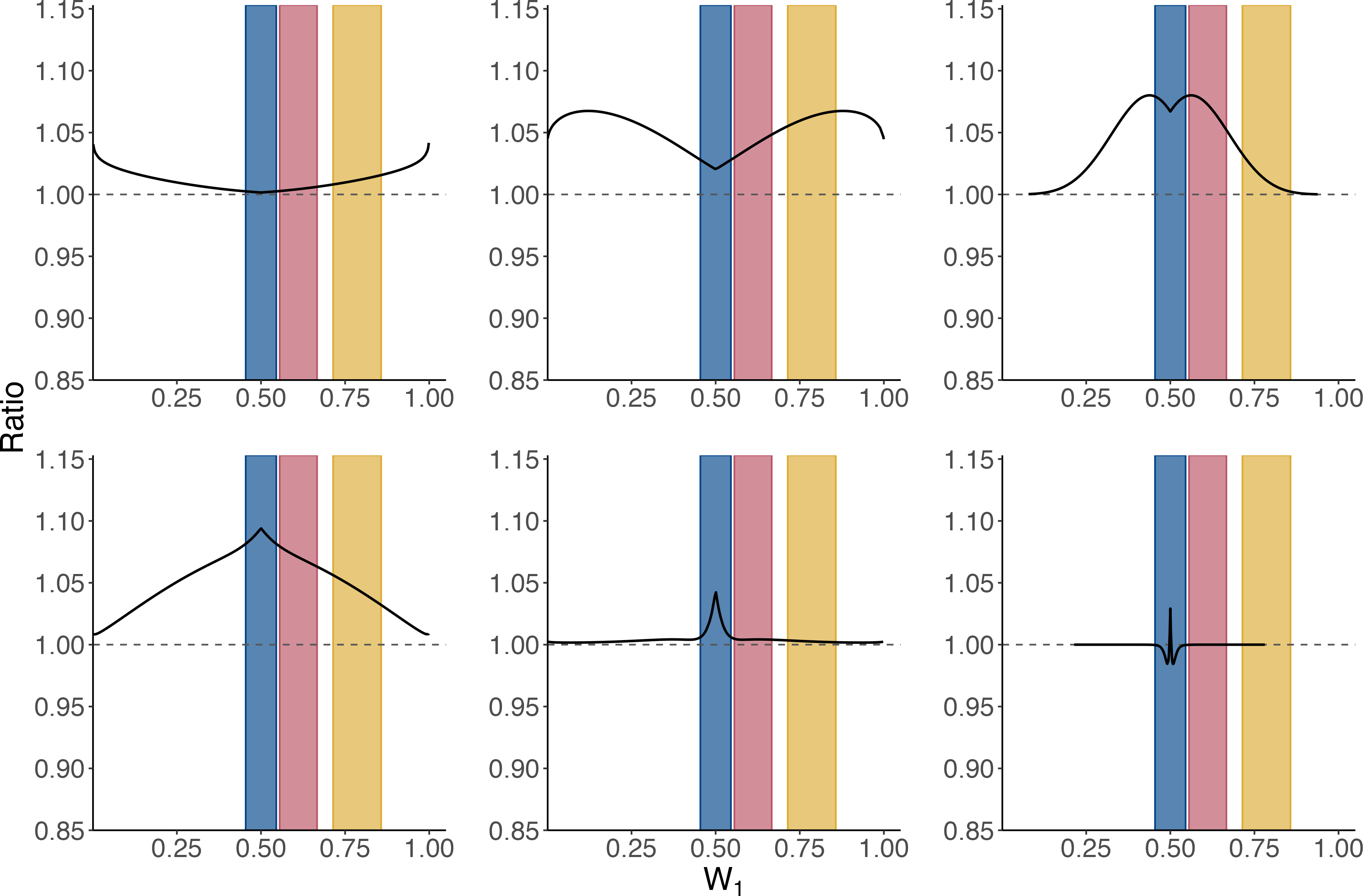}
    \caption{Ratios of the true cdf to the approximate Gamma cdf, under Gaussian (top row) and logistic (bottom row) dependence structures. Columns show low, mid, and high-dependence (low: $\rho = 0.1, \ \lambda=0.9$, mid:$\rho = \lambda = 0.5$, and high: $\rho = 0.9, \ \lambda=0.1$, for Gaussian correlation and logistic dependence parameters, respectively). Colored regions correspond to risk sets $B$ from Section~\ref{sec:sim_study}. Gray dotted line indicates ratio of one, i.e., no difference between the two probabilities.}
    \label{fig:exceed_ratio}
\end{figure}

Another potential drawback of the censored approach is that it assumes that the $(\tau \times 100)^\text{th}$ conditional quantile of the radii is equal to the $(\tau \times 100)^\text{th}$ quantile of the Gamma distribution.  The Gamma approximation holds as $r \rightarrow \infty$, so this assumption may be inaccurate for some models and some angles.  The truncated version ignores observations below the threshold, so is potentially less sensitive to deviations from the Gamma approximation. To explore this potential issue, we visualize the ratio of the true cdf arising from the conditional density in Equation~\eqref{eq:asym_gamma_approx} (calculated using numerical integration) to the cdf under the Gamma approximation, i.e., Gamma$(\text{shape} = 2, \text{rate} = g(\bw; \btheta_1))$, where $\btheta_1$ is the true data-generating parameter. The cdf is the second case in Equation~\eqref{cens_lhood}. Figure~\ref{fig:exceed_ratio} illustrates how this ratio varies across angles under Gaussian and logistic dependence at three levels of dependence.


If the Gamma approximation were perfect, the ratio would be one. Ratios less than one mean the Gamma model \textit{overestimates} the cdf; ratios greater than one mean it \textit{underestimates}. Across all dependence scenarios, the ratio is close to one, with the largest deviations seen under the mid-dependence Gaussian setting at angles close to $0$ and $1$, and under the high-dependence Gaussian and low-dependence logistic settings at angles close to $0.5$. These deviations could lead to bias in estimating the third term of Equation~\eqref{eqn:prob-derivation}.


Despite these potential concerns, we proceed with both likelihoods in our modeling approach. In particular, we include the censored likelihood to assess whether incorporating data below the threshold can yield more stable probability estimates, which we revisit in Section~\ref{sec:data}. 


Finally, to complete the model, we specify the following priors for the truncated and censored likelihoods:
\begin{align}\label{eq:priors}
    \alpha &\sim \text{Gamma}(4, \text{rate}=2) \nonumber \\
    \btheta_1 &\sim \begin{cases}
        \text{Half-Student-t}(\nu=4,\text{scale}=4\text{\ or \ }2) &\quad \theta_{1,1}, \theta_{1,2},\text{ Dirichlet} \\
        \text{Uniform}(0,1) &\quad \theta_{1,1} \text{ otherwise}
    \end{cases}
\end{align}

\subsubsection{Angular Likelihood}\label{sec:angular}
Next, we detail our modeling approach for the angular component of Equation~\eqref{multi_vector_density}. \cite{papastathopoulos2025} and \cite{campbell2025} propose an angular density which depends on the volume of some star-shaped set $\mathcal{L}$ and its associated gauge function (Equation~\eqref{eqn:star}) \citep{balkema2010a}. Since  $\bw = [w_1 \quad w_2]$ is fully specified in the two-dimensional case by the first angular component $w_1 \in (0,1)$, we also propose a mixture of Beta densities for $w_1$, with weights $\bp$ (Equation~\eqref{eqn:mixture}). We investigate both forms of the angular density in our approach, with
\begin{numcases}{\bw_i \given \btheta_2 \sim \,}
   \frac{g_\mathcal{L}(\bw_i \given \btheta_2)^{-2}}{2|\mathcal{L}|} \quad &\text{Star-shaped}\label{eqn:star}
   \\
   \sum_{j=1}^J p_j \,\text{Beta}(w_{1 ,i} \given \alpha_j, \beta_j), \quad &\text{Mixture} \label{eqn:mixture}
\end{numcases}
where in \eqref{eqn:star} $|\mathcal{L}|$ denotes the volume of the set $\mathcal{L}$ and $\btheta_2$ denotes the parameters of the gauge function $g_\mathcal{L}$, and in \eqref{eqn:mixture} $\btheta_2$ contains the  mixing weight $p_j$ and the Beta parameters $\alpha_j$ and $\beta_j$, for $j=1, \ldots, J$.

The star-shaped angular density relationship \eqref{eqn:star} holds exactly for \textit{some} star-shaped set $\mathcal{L}$ and its associated gauge function $g_\mathcal{L}$, but $g_\mathcal{L}$ is  not necessarily the same gauge function $g(\bx)$ that defines the shape of the multivariate point cloud. Specifically, when the joint density of $\bx$ is not homothetic, as is often the case when transforming from original to exponential margins, the gauge function $g_\mathcal{L} \neq g(\bx)$, and plugging $g(\bx)$ into \eqref{eqn:star}  fails to capture the full density of angles \citep[see Section~2.6 of][for a more comprehensive explanation]{papastathopoulos2025}.

\begin{figure}[!htbp]
    \centering
    \includegraphics[width=0.95\linewidth]{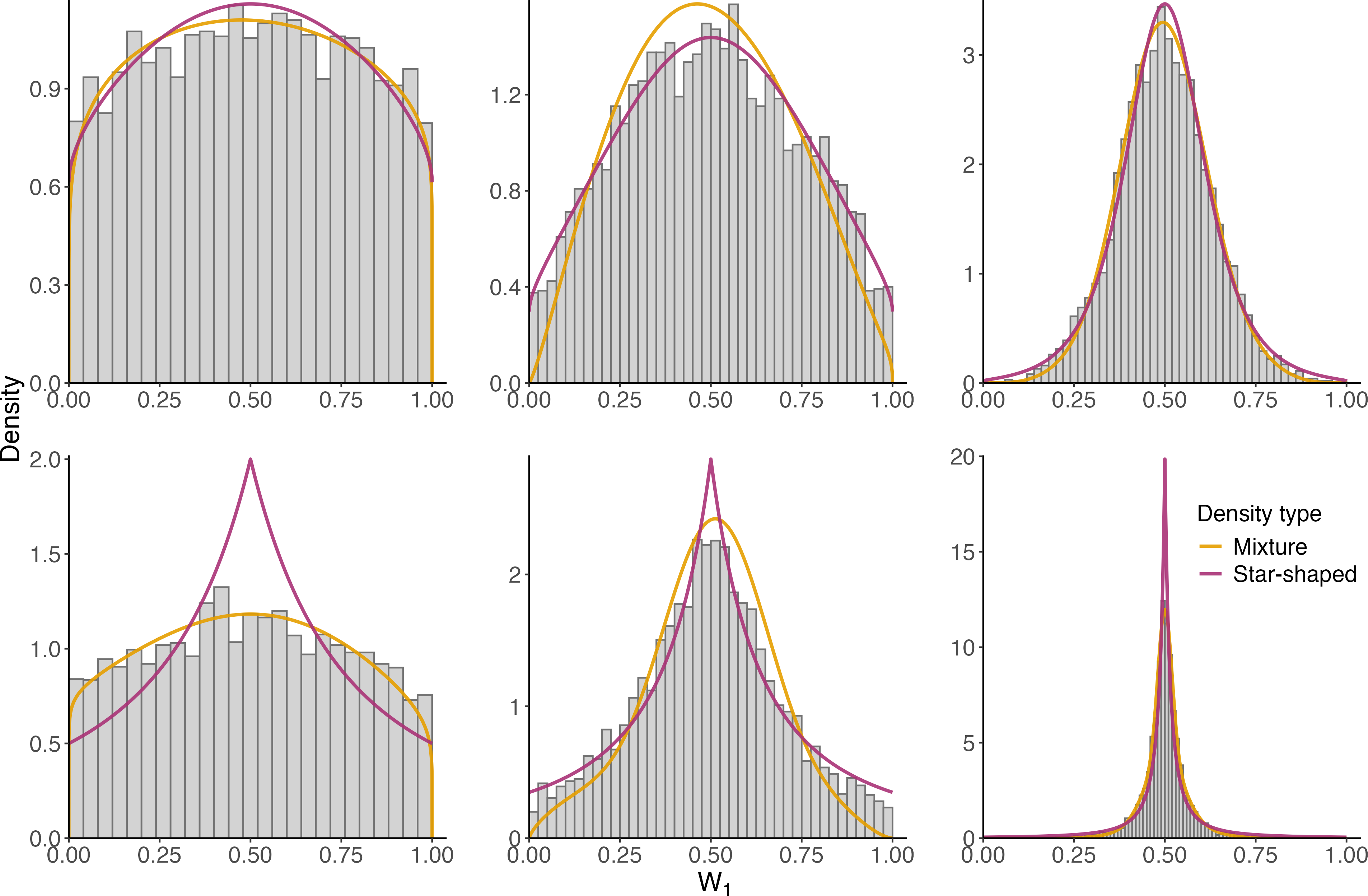}
    \caption{Posterior angular densities, as defined by Equation~\eqref{eqn:mixture} and sampled according to \ref{sec:angular}, for Gaussian (top) and logistic (bottom) dependence structures. Rows and columns presented as in Figure~\ref{fig:exceed_ratio}.}
    \label{fig:star_and_mix_dens}
\end{figure}

The prior on $\btheta_2$ is analogous to that in \eqref{eq:priors}. Again, the above holds exactly for some star-shaped set $\mathcal{L}$ with potentially unknown gauge function $g_\mathcal{L}$. In our implementation, we use the same family for this $g_\mathcal{L}$ as in the radial density, but allow the dependence parameters to differ. The main computational difficulty lies calculating the volume $|\mathcal{L}|$. \cite{papastathopoulos2025} estimate the volume of $\mathcal{L}$ through a latent parameter using Gaussian random fields, while \cite{campbell2025} leverage the piecewise-linear nature of their gauge function estimation and, hence, calculate the volume in closed form. However, we have found that the volume of $\mathcal{L}$ can be estimated quite accurately and quickly using numerical integration methods, as we do here: generate a $(0,1)\times(0,1)$ grid of $n=10,000$ equally spaced points, $\by_i$, and evaluate the gauge function at those points. Then, $|\mathcal{L}| \approx \frac{1}{n}\sum_{i=1}^{n}\one\{g(\by_i) \leq 1\}$.

  We include the Beta mixture density as a flexible alternative in case the mis-specification of $g_\mathcal{L}$ leads to poor performance. This is straightforward in two dimensions, however the higher-dimensional generalization of the Beta mixture is a mixture of Dirichlet densities, which could introduce additional computational difficulties due to the larger parameter space. 
In practice, imposing the convexity restriction on the mixing weights $\bp$, $\sum_{j=1}^J p_j = 1$, poses mixing and identifiability challenges. To effectively implement this restriction, we place a stick-breaking prior on the weights and incorporate a latent cluster assignment $z_i \in j=\{1,\dots,J\}$, as proposed by \cite{ishwaran2003}. We also reparametrize the Beta density using location and variance-like parameters, $\mu$ and $\tau$, to allow for efficient mixing \citep{kottas2006}. Now, the likelihood for $w_{1,i}$ becomes:
\begin{equation*}
    w_{1,i} \given z_i, \bmu,\btau \sim \text{Beta}(w_{1,i} \given \mu_{z_i}\tau_{z_i}, (1-\mu_{z_i})\tau_{z_i}),
\end{equation*}
with priors specified as:
\begin{align*}
    \mu_j &\sim \text{Uniform}(0,1) \\
    \tau_j &\sim \text{Inv-Gamma}(2, \beta)\\
    z_i &\sim \text{Categorical}(\bp) \\
    p_1&=v_1, \quad p_j=v_j \prod_{k=1}^{j-1}(1-v_k), \quad p_J=\prod_{k=1}^{J -1}(1-v_k) \\
    v_k &\sim \text{Beta}(1, \alpha), \quad k \in \{1,\dots, J-1\} \\
    \alpha &\sim \text{Gamma}(2, 2)\\
    \beta &\sim \text{Exponential}(1/8).
\end{align*}
Empirically, we found that $J=10$ provides sufficient flexibility.

To illustrate the potential differences between these two angular density models, Figure~\ref{fig:star_and_mix_dens} shows the empirical distribution of angles at varying dependence levels for 5000 points randomly generated from Gaussian and logistic dependence structures. Overlayed on each histogram is the posterior star-shaped density, estimated using the family as the data-generating gauge function (Gaussian and logistic, respectively), along with the posterior Beta mixture density. In the Gaussian setting, the densities are fairly similar, and both can capture the shape of the distribution quite well at varying levels of dependence. On the other hand, the star-shaped density struggles to capture the angular distribution in the logistic setting, while the Beta mixture does not.

In contrast to the case of estimating the radial density, where we either censor or truncate the data below a threshold, for estimating the angular densities we use all of the angles. It seems reasonable to conjecture that using all angles, rather than only angles associated with radii above the threshold, would allow for a less variable fitting posterior angular density because of the increased sample size. We discuss additional considerations surrounding this in Section~\ref{sec:IS scheme}, along with potential drawbacks in \ref{sec:diag}.

\subsection{Threshold Selection}
The gamma approximation in \eqref{eq:asym_gamma_approx} applies to large values, so we must choose a high threshold, $r_\tau(\bw)$, above which we model $R > r_\tau(\bw)$ according to \eqref{trunc_lhood} and \eqref{cens_lhood}. \cite{wadsworth2024} use additive quantile regression to specify $r_\tau(\bw)$, with $\bW$ treated as a covariate. \cite{campbell2025} use a kernel density estimation approach, as a higher-dimensional analogue of the additive quantile regression approach. \cite{papastathopoulos2025} use Gamma quantile regression in a Bayesian framework to choose sets of exceedances. 

Similar to \cite{wadsworth2024} and \cite{campbell2025}, we let $\tau=0.95$ and select a threshold $r_\tau(\bw)$ that yields the largest $5\%$ of points overall ($\approx 250$ when $n=5000$). However, we instead use a simple \textit{marginal} threshold---with roughly balanced exceedances across both margins---as opposed to a quantile regression threshold. There are two key advantages to this. In contrast to the quantile regression approaches, implementing a marginal threshold is trivial, even in higher dimensions. Secondly, unlike more complicated thresholds, the marginal threshold does not jointly depend on both variables.  As a result, transformation from the original to exponential margins could potentially be implemented into a hierarchical Bayesian model, along with the radial and angular components, thereby fully accounting for the compounded variability inherent in the marginal transformation.  This variability is ignored when performing the two-steep inference procedure of first performing the marginal transformation and subsequently estimating gauge function and angular model parameters. 

\begin{figure}
    \centering
    \includegraphics[width=0.95\linewidth]{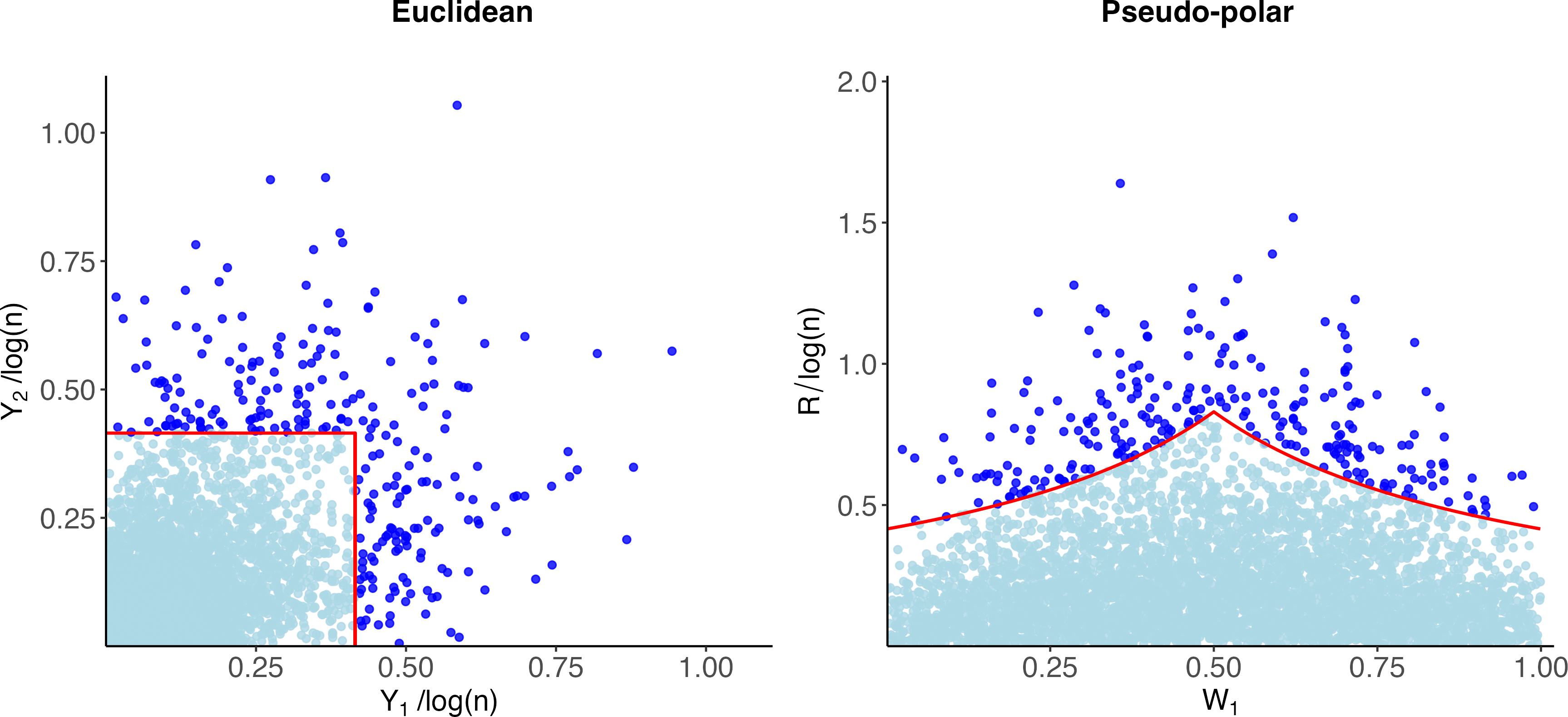}
    \caption{Marginal threshold, indicated by the red line in both plots, shown in both Euclidean and pseudo-polar coordinates. Truncated likelihood only uses the points above the threshold (dark blue points) in evaluating Equation~\ref{trunc_lhood}, while censored likelihood uses the dark blue points in evaluating the first case of Equation~\ref{cens_lhood} and the associated point on the threshold line for points below the threshold in evaluating the second.}
    \label{fig:marginal_threshold}
\end{figure}

\cite{majumder-2025a} provide a comprehensive comparison of inference using a marginal threshold versus a quantile regression threshold, and they find that the marginal threshold is comparable to the quantile threshold for estimating the residual tail coefficient. While not included in this paper, we performed a similar comparison in early simulation studies and found no perceptible advantage to using the quantile threshold over the marginal threshold for model fitting.  This near-total insensitivity to thresholding scheme his may or may not generalize to other contexts, but it appears to hold here. For quantile $\tau$, the marginal threshold is simply
\begin{equation*}
    r_\tau(\bw) = 
    \begin{cases}
        \frac{\tau}{w_1} & \text{if } w_1 >0.5 \\
        \frac{\tau}{1 - w_1} & \text{if } w_1 < 0.5.
    \end{cases}
\end{equation*}
Figure~\ref{fig:marginal_threshold} illustrates this threshold in both Euclidean and pseudo-polar coordinates.

\subsection{Combining Radial and Angular Models}\label{sec:joint_model}

The combination of two radial likelihoods with two angular likelihoods yields four joint likelihoods for each of the $K=6$ parametric gauge functions from Section~\ref{sec:decomp_model}.  In total, then, 24 fits are applied to each dataset. 

Since $R \given \bW$ and $\bW$ are independent, we model the radial and angular likelihoods separately and join them together after sampling. 
For the censored and truncated radial likelihoods, along with the star-shaped angular likelihood, we fit $K=6$ variations of the respective likelihoods, corresponding to each of the six gauge functions, to each dataset. We fit the mixture angular likelihood only once, as its form does not depend on the limit set. We run these models in parallel, using \texttt{nimble} for their stick-breaking implementation of the Beta mixture likelihood, and \texttt{Rcpp} for the radial and the star-shaped angular models, implemented via an adaptive Metropolis-Hastings algorithm \citep{shaby2010}.


\section{Prediction for Each Parametric Model}\label{sec:predictions}

\subsection{Importance Sampling Scheme}\label{sec:IS scheme}
\begin{figure}[!htbp]
    \centering
    \includegraphics[width=0.95\linewidth]{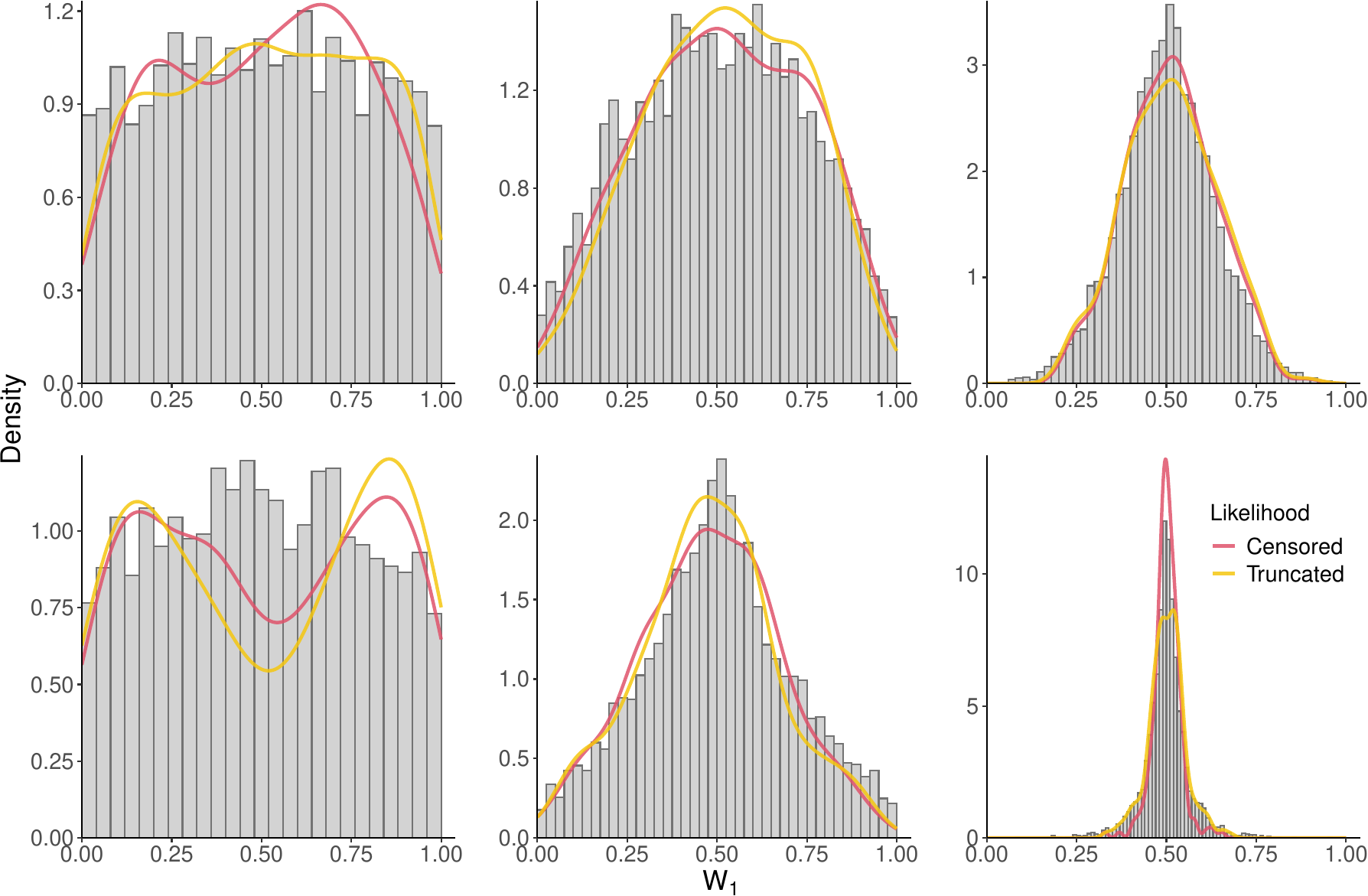}
    \caption{Histogram of empirical angles for one realization of the 200 simulated datasets generated for each dependence type and level. Top and bottom rows follow the same arrangement as Figure~\ref{fig:exceed_ratio}. The two lines overlayed on each histogram are the kernel density estimates of angles from points exceeding the threshold, using the posterior estimates of $\alpha$ and $\btheta_1$ to find $r_q(\bw)$ as the $95$th quantile of a Gamma($\alpha_\text{post}$, $\btheta_{1,\text{post}}$).  The two colors correspond to the two likelihoods used to obtain the posterior estimates of $\alpha$ and $\btheta_1$.  See Figure~\ref{fig:post_quantiles} for visualization of these thresholds.}
    \label{fig:angles_above_threshold}
\end{figure}

\begin{figure}[!htbp]
    \centering
    \includegraphics[width=0.95\linewidth]{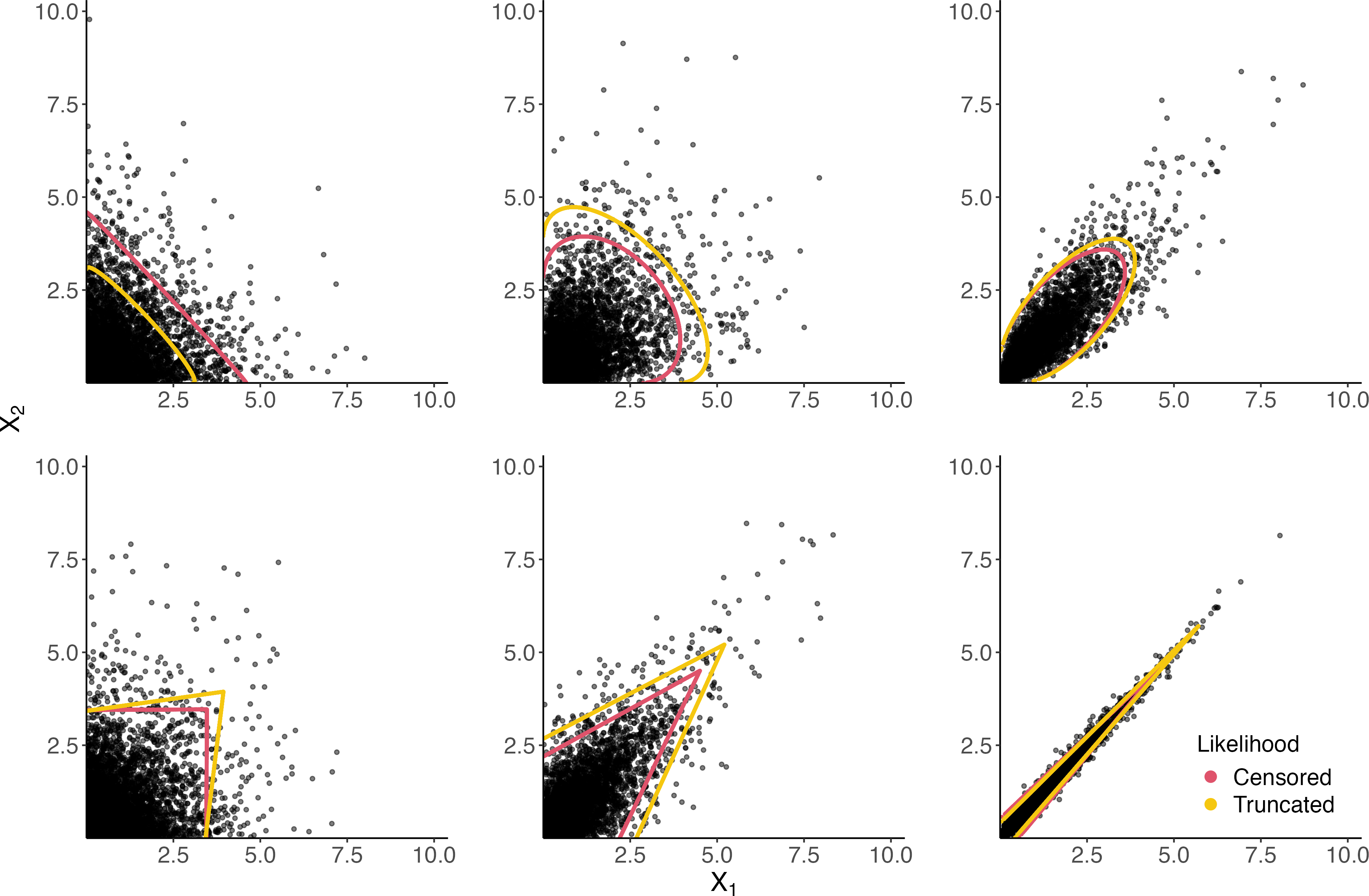}
    \caption{Posterior $r_q(\bw)$ determined as the $95$th quantile of a Gamma distribution, using the posterior estimates of$\alpha$ and $\btheta_1$. Top and bottom rows follow the same arrangement as Figure~\ref{fig:exceed_ratio}.}
    \label{fig:post_quantiles}
\end{figure}

The prediction target of interest here is the risk set probability for some region on the far joint tail.  Letting $l \in \{\text{cens}, \text{trunc}\}$ denote the radial likelihood type (censored or truncated) and $a \in \{\text{mix}, \text{star}\}$ the angular density (mixture or star-shaped), we define the prediction target $\prob_{l,a}(\bX \in B)$ as the probability that $\bX \in B$, under the combination of posterior parameters under radial likelihood $l$ and angular density $a$.  To estimate these probabilities, we use the following key assumption:
\begin{assumption}\label{assump:w_above_quant_thres}
Consider all angles $\bW$. For any $q\in(0,1),\, \prob(R \geq r_q(\bW)) = 1 - q$, where $r_q(\bw)$ is the $q$ conditional quantile of the radii at angle $\bw$. Then, as $q \rightarrow 0$, $\bW \given R \geq r_q(\bW) \overset{d}{=} \bW$ (Prop.~5, \cite{papastathopoulos2025}).
\end{assumption}
Assumption~\ref{assump:w_above_quant_thres} holds exactly when $r_q(\bw)$ is the conditional quantile of the true data-generating distribution. While our model fitting procedure does not rely on approximations to conditional quantiles in specifying the throld $r_q(\bw)$, our importance sampling schema below \textit{does}. Figure~\ref{fig:angles_above_threshold} visualizes the density of angles above this quantile threshold (with the threshold shown in Figure~\ref{fig:post_quantiles}), calculated from a Gamma distribution quantile parametrized with the posterior estimates of the radial parameters. Figure ~\ref{fig:angles_above_threshold} suggests that Assumption \eqref{assump:w_above_quant_thres} seems to hold in most scenarios except for the logistic dependence structure with low dependence (lower left panel).  In this case,  the assumption breaks down when $w_1$ is close to $0.5$, aligning with the dip in the cdf ratios seen in Figure~\ref{fig:exceed_ratio}.

To estimate the probability $\prob_{l,a}(\bX \in B)$ in regions beyond the range of our data, we use a simple importance sampling procedure. Let $S_B = \{\bw \in \mathcal{S}^{d-1}: r\bw \in B, r \in (0, \infty]\}, \ I_B(\bW) = \{r > 0: r\bw \in B, \bw \in S_B\}$. This implies that if $R\bW \in B$, then $\bW \in S_B$ and $R \in I_B(\bW)$. With this decomposition, we write: 
\begin{align}{\label{eqn:prob-derivation}}
    \prob_{l,a}(\bX \in B) &= \prob_{l,a}(\bX \in B \given R \geq r_q(\bw))\prob_l(R \geq r_q(\bw)) \nonumber \\
    &= \prob_{l,a}(R\bW \in B \given R \geq r_q(\bw))\prob_l(R \geq r_q(\bw)) \nonumber \\
    &= \prob_l(R \in I_B(\bW) \given \bW \in S_B, R \geq r_q(\bw))\prob_a(\bW \in S_B \given R \geq r_q(\bw))\prob_l(R \geq r_q(\bw)) \nonumber
    \intertext{Using Assumption~\ref{assump:w_above_quant_thres},}
    \prob_{l,a}(\bX \in B) &= \prob_l(R \in I_B(\bW) \given \bW \in S_B, R \geq r_q(\bw))\prob_a(\bW \in S_B)\prob_l(R \geq r_q(\bw))
\end{align}

In two dimensions, we choose a bivariate normal distribution as our importance or proposal distribution, $H$, but this can easily be extended to higher-dimensional settings with the multivariate normal analogue.  We sample $\bX^*$ from $H:= \text{MVN}(\bmu = \text{midpoint of }B, \ \bSigma = \mathbf{c}_\text{dist}\times\bI_2)$, where $\mathbf{c}_\text{dist}$ is a 2-dimensional vector; the first element corresponds to the width of the box in the $x$-direction and the second to the $y$-direction. In our simulation study, we set $\mathbf{c}_\text{dist} = (2,\ 2)$, but allow it to vary in our data analyses.

Define $r_q(\bw^*)$ as the $q$ quantile of a Gamma distribution with shape $\hat{\alpha}_{\text{post}}$ and rate $g(\bw^*; \hat{\btheta}_{1, \text{post}})$, where $\bw^*$ are the corresponding angles of $\bX^*$ (and $r^*$ the corresponding radii). Let $h_{R, \bW}$ denote the importance density transformed to the $L_1-$norm. Then, suppressing the $l,a$ for readability, 
\begin{align*}
    \begin{split}
    \eqref{eqn:prob-derivation} &= \frac{1}{n} \sum_{i=1}^{n} \one\{r_i^* \in I_B(\bw_i^*) \given \bw_i^* \in S_B, r_i^* \geq r_q(\bw_i^*)\} \ \times \one\{\bw_i^* \in S_B\} \\
    &\qquad\qquad\times \frac{f_{R\given \bW}(r_i^* \given \bw_i^*, r_q(\bw_i^*), \alpha_\text{post}, \btheta_{1,\text{post}})f_{\bW}(\bw_i^*\given \btheta_{2,\text{post}})}{h_{R,\bW}(r_i^*, \bw_i^*)} \ \times (1 - q)
    \end{split}
    \\
    \begin{split}
      &= \frac{1}{n} \sum_{i=1}^{n} \one\{\bx_i^* \in B \given r_i^* \geq r_q(\bw_i^*)\} \\
        &\qquad\qquad\times \frac{f_{R\given \bW}(r_i^* \given \bw_i^*, r_q(\bw_i^*), \alpha_\text{post}, \btheta_{1,\text{post}})f_{\bW}(\bw_i^*\given \btheta_{2,\text{post}})}{h_{\bX^*}(\bx_i^*) \ r_i^*} \ \times (1-q)
    \end{split}
\end{align*}

As mentioned in Section~\ref{sec:radial}, the Gamma approximation is not perfect, introducing the potential for bias in our probability prediction estimates. The most obvious place for this bias to arise is in the $(1-q)$ term of Equation~\eqref{eqn:prob-derivation}; when the ratio of true cdf to the approximated cdf (Figure~\ref{fig:exceed_ratio}) is less than one, our estimate of $(1-q)$ may be underbiased, while overbias may occur in the opposite situation. We tried to account for the variation seen in this ratio across angles by using self-normalized importance sampling \citep{owen2013}. However, we did not see any improvement in predictive performance, and in some cases, the estimates were worse.

\section{Simulation Study}\label{sec:sim_study}
\begin{figure}[!htbp]
    \centering
    \includegraphics[width=0.95\linewidth]{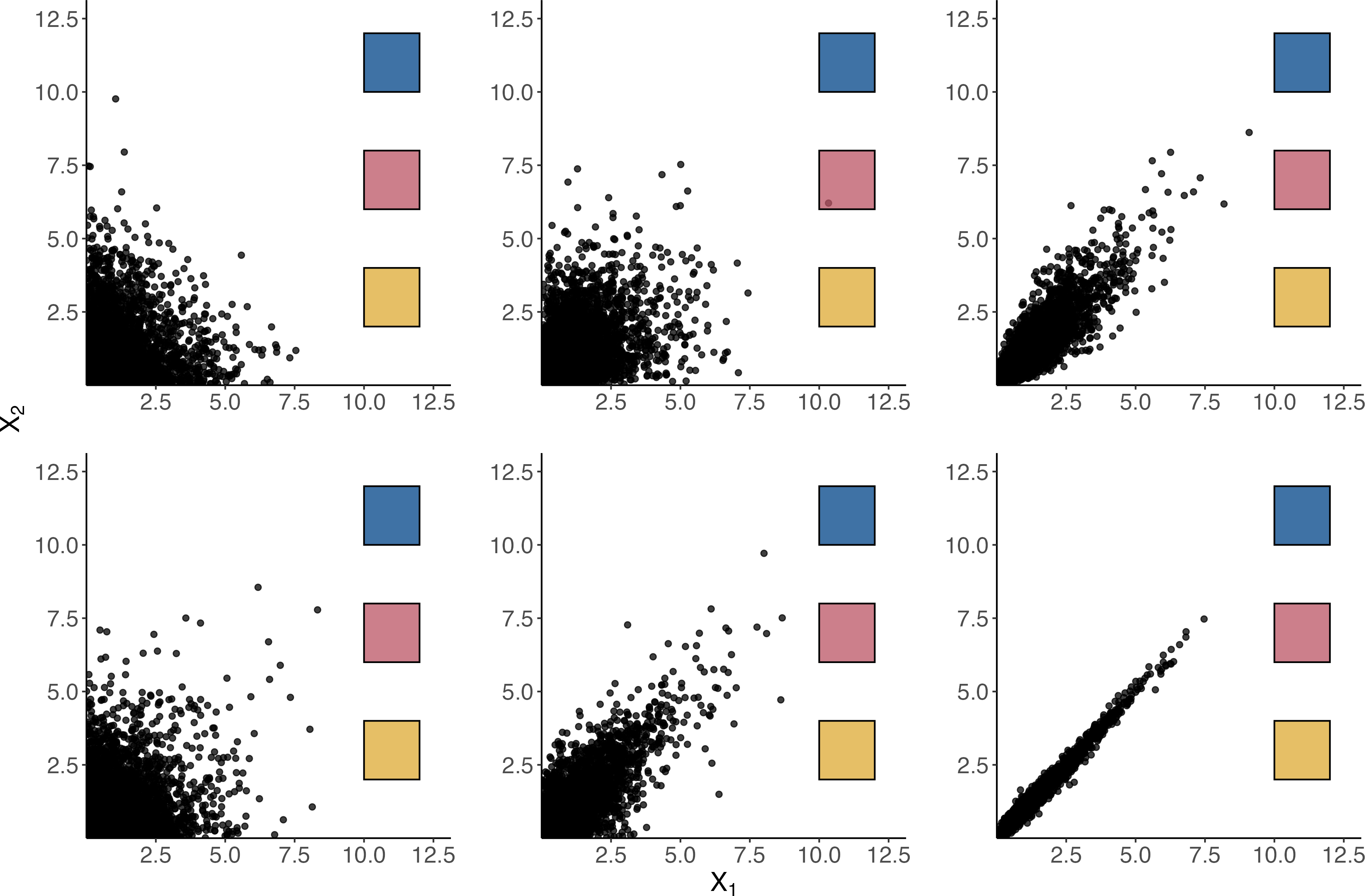}
    \caption{Prediction regions under the simulation study: $B_1=(10,12)\times(10,12)$, $B_2=(10,12)\times(6,8)$, and $B_3=(10,12)\times(2,4)$, represented by the blue, red, and yellow boxes, respectively. Panels follow the arrangement as Figure~\ref{fig:exceed_ratio}. Data are generated according to \ref{sec:data_gen}.}
    \label{fig:sim_pred_task}
\end{figure}

In this section, we evaluate different combinations of radial likelihoods, angular models, and BMA weighting against two existing approaches from the literature: BezELS \citep{majumder-2025a} and C-W \citep{campbell2025}. \cite{majumder-2025a}  use their approach to semi-parametrically estimate the gauge function, but do not calculate risk set probability predictions.  To compare like with like, we adopt their existing mechanics for the truncated radial likelihood and couple that with our Beta mixture angular density, subsequently generating predictions using our importance sampling procedure detailed in Section~\ref{sec:IS scheme}. \cite{campbell2025} propose multiple model variants, so, for the sake of comparison, we implement their best-fitting model (SS4) and use their proposed importance sampling scheme for prediction. We compare all methods based on their ability to capture the true probability of lying in three tail regions (Figure~\ref{fig:sim_pred_task}): $B_1=(10,12)\times(10,12), \ B_2=(10,12)\times(6,8), \text{ and }B_3=(10,12)\times(2,4)$.

\subsection{Data-generating Procedure}\label{sec:data_gen}
To provide a comprehensive comparison, we examine nine different combinations of dependence structures and dependence levels:
\begin{itemize}
    \item Gaussian: $\rho=0.1 \ (\text{low}), \ \rho=0.5 \ (\text{mid}), \text{ and }\rho=0.9 \ (\text{high})$
    \item Logistic: $\lambda=0.9 \ (\text{low}), \ \lambda=0.5 \ (\text{mid}), \text{ and }\lambda=0.1 \ (\text{high})$
    \item H\"usler-Reiss: $\lambda=0.1 \ (\text{low}), \ \lambda=1 \ (\text{mid}), \text{ and }\lambda=3 \ (\text{high})$ (Figure~\ref{fig:hr_pred_task})
\end{itemize}
Under each dependence structure and level, we generate $200$ datasets of $5,000$ points each, picking the top $250 \,(\approx 5\% \text{ or } 1-\tau=0.05)$ points as the marginal threshold. 

\subsection{Results}\label{sec:sim_results}
With three boxes of interest for each dependence type and level, there are a total of 27 sets of comparison scenarios. As an initial overview of the results, we first compare the various methods using RMSE of the estimated tail probabilities. We calculate RMSE for each of our four likelihood combinations (censored vs. truncated $\times$ star-shaped vs. mixture) under the three BMA weighting methods, along with the two previously proposed methods, using the 200 sets of predictions generated within each approach. Since the true and predicted probabilities are extremely small, the resulting RMSEs are all very small and difficult to interpret. We therefore standardize each RMSE value by its corresponding true estimand, to yield a unitless metric that may be interpreted as percent difference from the truth. Tables~\ref{tab:gauss_mse},\ref{tab:logistic_mse}, and \ref{tab:husler_reiss_mse} provide these comparisons, with a corresponding key in Table~\ref{tab:rmse_key}.

There is no uniform winner among the different likelihood/model/weighting combinations, but some trends are discernible. Within our censored models, the best-performing angular density and BMA weight method varies by dependence structure. In the Gaussian setting, most scenarios favored the star-shaped angular density, followed closely by the mixture; Pseudo-BMA and Pseudo-BMA+ outperform stacking weights. In the logistic setting, the mixture angular density is heavily preferred, and stacking performs best in the low-dependence setting. The mixture density has a slight advantage over the star-shaped in the H\"usler-Reiss scenarios, and stacking performs best under high dependence.

Our censored model outperforms BezELS and C-W except in all but three of the 27 scenarios. C-W slightly outperforms our censored model with the angular mixture density using stacking weights in the logistic low-dependence setting for $B_3$ (RMSE: 0.810 vs 0.983). C-W's RMSE is about half that of our best model (truncated, angular mixture, Pseudo-BMA weights) under the H\"usler-Reiss mid-dependence case for $B_3$, and is about $25\%$ better than ours under the H\"usler-Reiss, high-dependence scenario in $B_2$. Notably, this is also the only scenario in which our truncated likelihood outperforms our censored likelihood.

\begin{table}[!htbp]
\centering
\begin{tabular}{lllp{3cm}}
\toprule
\textbf{BMA Method} & \textbf{Likelihood} & \textbf{Ang. Dens.} & \shortstack{\textbf{Other}\\\textbf{Methods}}\\
\midrule
\cellcolor[HTML]{DF526B}$\text{Pseudo-BMA}$ & \textit{Censored}   & Mixture: $\circledcirc$ & \cellcolor[HTML]{FFFFFF}C-W \\
\cellcolor[HTML]{26BCE5}$\text{Pseudo-BMA+}$ & \textbf{Truncated} & Star-shaped: $\bigstar$ & \cellcolor[HTML]{FFFFFF}$\circledcirc$ BezELS \\
\cellcolor[HTML]{F5C70E}$\text{Stacking}$ &  &  & \\
\bottomrule
\end{tabular}
\caption{Key for the formatting in the RMSE comparison tables among our models, BezELS, and C-W. All models developed in this work are represented by a combination of one aspect from each of the first three columns. The semi-parametric methods are depicted as shown in the last column.}
\label{tab:rmse_key}
\end{table}

\begin{table}[!htbp]
    \centering 
\begin{tabular}[t]{c|c|c||c|c|c||c|c|c}
\toprule
\multicolumn{3}{c}{\textbf{Low}} & \multicolumn{3}{c}{\textbf{Mid}} & \multicolumn{3}{c}{\textbf{High}} \\
\cmidrule(l{3pt}r{3pt}){1-3} \cmidrule(l{3pt}r{3pt}){4-6} \cmidrule(l{3pt}r{3pt}){7-9}
$B_1$ & $B_2$ & $B_3$ & $B_1$ & $B_2$ & $B_3$ & $B_1$ & $B_2$ & $B_3$\\
\midrule
\cellcolor[HTML]{26BCE5}$\bigstar\,\textit{1.135}$ & \cellcolor[HTML]{26BCE5}$\circledcirc\,\textit{0.557}$ & \cellcolor[HTML]{F5C70E}$\bigstar\,\textit{0.211}$ & \cellcolor[HTML]{DF526B}$\bigstar\,\textit{0.637}$ & \cellcolor[HTML]{F5C70E}$\circledcirc\,\textit{0.318}$ & \cellcolor[HTML]{DF526B}$\bigstar\,\textit{0.160}$ & \cellcolor[HTML]{F5C70E}$\bigstar\,\textit{0.188}$ & \cellcolor[HTML]{26BCE5}$\circledcirc\,\textit{0.184}$ & \cellcolor[HTML]{DF526B}$\bigstar\,\textit{0.670}$\\
\cellcolor[HTML]{DF526B}$\bigstar\,\textit{1.138}$ & \cellcolor[HTML]{DF526B}$\circledcirc\,\textit{0.608}$ & \cellcolor[HTML]{F5C70E}$\circledcirc\,\textit{0.225}$ & \cellcolor[HTML]{26BCE5}$\bigstar\,\textit{0.637}$ & \cellcolor[HTML]{DF526B}$\bigstar\,\textit{0.336}$ & \cellcolor[HTML]{26BCE5}$\bigstar\,\textit{0.160}$ & \cellcolor[HTML]{DF526B}$\bigstar\,\textit{0.191}$ & \cellcolor[HTML]{DF526B}$\circledcirc\,\textit{0.185}$ & \cellcolor[HTML]{26BCE5}$\bigstar\,\textit{0.670}$\\
\cellcolor[HTML]{FFFFFF}$4.830$ & \cellcolor[HTML]{26BCE5}$\bigstar\,\textit{0.656}$ & \cellcolor[HTML]{26BCE5}$\circledcirc\,\textit{0.230}$ & \cellcolor[HTML]{DF526B}$\circledcirc\,\textit{1.090}$ & \cellcolor[HTML]{26BCE5}$\bigstar\,\textit{0.336}$ & \cellcolor[HTML]{DF526B}$\circledcirc\,\textit{0.196}$ & \cellcolor[HTML]{26BCE5}$\bigstar\,\textit{0.191}$ & \cellcolor[HTML]{DF526B}$\bigstar\,\textit{0.203}$ & \cellcolor[HTML]{DF526B}$\circledcirc\,\textit{0.689}$\\
\cellcolor[HTML]{26BCE5}$\bigstar\,\textbf{10.960}$ & \cellcolor[HTML]{DF526B}$\bigstar\,\textit{0.658}$ & \cellcolor[HTML]{26BCE5}$\bigstar\,\textit{0.233}$ & \cellcolor[HTML]{FFFFFF}$1.201$ & \cellcolor[HTML]{26BCE5}$\circledcirc\,\textit{0.356}$ & \cellcolor[HTML]{26BCE5}$\circledcirc\,\textit{0.198}$ & \cellcolor[HTML]{DF526B}$\circledcirc\,\textit{0.199}$ & \cellcolor[HTML]{26BCE5}$\bigstar\,\textit{0.203}$ & \cellcolor[HTML]{26BCE5}$\circledcirc\,\textit{0.825}$\\
\cellcolor[HTML]{DF526B}$\bigstar\,\textbf{11.085}$ & \cellcolor[HTML]{F5C70E}$\circledcirc\,\textit{1.738}$ & \cellcolor[HTML]{DF526B}$\bigstar\,\textit{0.233}$ & \cellcolor[HTML]{26BCE5}$\circledcirc\,\textit{1.333}$ & \cellcolor[HTML]{DF526B}$\circledcirc\,\textit{0.394}$ & \cellcolor[HTML]{F5C70E}$\circledcirc\,\textit{0.199}$ & \cellcolor[HTML]{F5C70E}$\circledcirc\,\textit{0.202}$ & \cellcolor[HTML]{F5C70E}$\circledcirc\,\textit{0.206}$ & \cellcolor[HTML]{F5C70E}$\circledcirc\,\textit{3.233}$\\
\cellcolor[HTML]{DF526B}$\circledcirc\,\textit{13.588}$ & \cellcolor[HTML]{FFFFFF}$2.536$ & \cellcolor[HTML]{DF526B}$\circledcirc\,\textit{0.240}$ & \cellcolor[HTML]{F5C70E}$\circledcirc\,\textit{1.886}$ & \cellcolor[HTML]{F5C70E}$\bigstar\,\textit{0.411}$ & \cellcolor[HTML]{F5C70E}$\bigstar\,\textit{0.225}$ & \cellcolor[HTML]{26BCE5}$\circledcirc\,\textit{0.207}$ & \cellcolor[HTML]{F5C70E}$\bigstar\,\textit{0.224}$ & \cellcolor[HTML]{FFFFFF}$12.863$\\
\cellcolor[HTML]{26BCE5}$\circledcirc\,\textit{17.190}$ & \cellcolor[HTML]{26BCE5}$\bigstar\,\textbf{3.428}$ & \cellcolor[HTML]{FFFFFF}$0.827$ & \cellcolor[HTML]{DF526B}$\bigstar\,\textbf{2.300}$ & \cellcolor[HTML]{FFFFFF}$0.658$ & \cellcolor[HTML]{FFFFFF}$0.466$ & \cellcolor[HTML]{FFFFFF}$0.684$ & \cellcolor[HTML]{FFFFFF}$0.624$ & \cellcolor[HTML]{F5C70E}$\bigstar\,\textit{41.065}$\\
\cellcolor[HTML]{F5C70E}$\circledcirc\,\textit{46.196}$ & \cellcolor[HTML]{DF526B}$\bigstar\,\textbf{3.435}$ & \cellcolor[HTML]{26BCE5}$\bigstar\,\textbf{0.847}$ & \cellcolor[HTML]{26BCE5}$\bigstar\,\textbf{2.322}$ & \cellcolor[HTML]{26BCE5}$\bigstar\,\textbf{1.338}$ & \cellcolor[HTML]{F5C70E}$\bigstar\,\textbf{0.841}$ & \cellcolor[HTML]{FFFFFF}$\circledcirc\,1.463$ & \cellcolor[HTML]{DF526B}$\circledcirc\,\textbf{1.431}$ & \cellcolor[HTML]{FFFFFF}$\circledcirc\,140.260$\\
\cellcolor[HTML]{F5C70E}$\bigstar\,\textbf{76.072}$ & \cellcolor[HTML]{F5C70E}$\bigstar\,\textbf{4.515}$ & \cellcolor[HTML]{DF526B}$\bigstar\,\textbf{0.847}$ & \cellcolor[HTML]{F5C70E}$\bigstar\,\textit{3.676}$ & \cellcolor[HTML]{DF526B}$\bigstar\,\textbf{1.338}$ & \cellcolor[HTML]{26BCE5}$\bigstar\,\textbf{0.878}$ & \cellcolor[HTML]{26BCE5}$\circledcirc\,\textbf{1.893}$ & \cellcolor[HTML]{26BCE5}$\circledcirc\,\textbf{1.484}$ & \cellcolor[HTML]{26BCE5}$\circledcirc\,\textbf{245.925}$\\
\cellcolor[HTML]{F5C70E}$\bigstar\,\textit{335.088}$ & \cellcolor[HTML]{DF526B}$\circledcirc\,\textbf{11.312}$ & \cellcolor[HTML]{F5C70E}$\bigstar\,\textbf{0.850}$ & \cellcolor[HTML]{F5C70E}$\bigstar\,\textbf{4.859}$ & \cellcolor[HTML]{F5C70E}$\bigstar\,\textbf{1.388}$ & \cellcolor[HTML]{DF526B}$\bigstar\,\textbf{0.879}$ & \cellcolor[HTML]{DF526B}$\circledcirc\,\textbf{1.969}$ & \cellcolor[HTML]{DF526B}$\bigstar\,\textbf{1.584}$ & \cellcolor[HTML]{DF526B}$\circledcirc\,\textbf{248.301}$\\
\cellcolor[HTML]{FFFFFF}$\circledcirc\,412.551$ & \cellcolor[HTML]{26BCE5}$\circledcirc\,\textbf{12.729}$ & \cellcolor[HTML]{DF526B}$\circledcirc\,\textbf{0.888}$ & \cellcolor[HTML]{FFFFFF}$\circledcirc\,8.620$ & \cellcolor[HTML]{F5C70E}$\circledcirc\,\textbf{1.549}$ & \cellcolor[HTML]{26BCE5}$\circledcirc\,\textbf{0.958}$ & \cellcolor[HTML]{F5C70E}$\circledcirc\,\textbf{2.064}$ & \cellcolor[HTML]{26BCE5}$\bigstar\,\textbf{1.584}$ & \cellcolor[HTML]{F5C70E}$\circledcirc\,\textbf{292.333}$\\
\cellcolor[HTML]{DF526B}$\circledcirc\,\textbf{514.489}$ & \cellcolor[HTML]{F5C70E}$\bigstar\,\textit{12.974}$ & \cellcolor[HTML]{26BCE5}$\circledcirc\,\textbf{0.892}$ & \cellcolor[HTML]{26BCE5}$\circledcirc\,\textbf{12.112}$ & \cellcolor[HTML]{26BCE5}$\circledcirc\,\textbf{1.576}$ & \cellcolor[HTML]{DF526B}$\circledcirc\,\textbf{0.995}$ & \cellcolor[HTML]{26BCE5}$\bigstar\,\textbf{2.573}$ & \cellcolor[HTML]{FFFFFF}$\circledcirc\,1.585$ & \cellcolor[HTML]{F5C70E}$\bigstar\,\textbf{314.683}$\\
\cellcolor[HTML]{26BCE5}$\circledcirc\,\textbf{540.811}$ & \cellcolor[HTML]{F5C70E}$\circledcirc\,\textbf{14.493}$ & \cellcolor[HTML]{F5C70E}$\circledcirc\,\textbf{1.037}$ & \cellcolor[HTML]{DF526B}$\circledcirc\,\textbf{12.475}$ & \cellcolor[HTML]{DF526B}$\circledcirc\,\textbf{1.581}$ & \cellcolor[HTML]{FFFFFF}$\circledcirc\,1.082$ & \cellcolor[HTML]{DF526B}$\bigstar\,\textbf{2.574}$ & \cellcolor[HTML]{F5C70E}$\bigstar\,\textbf{1.585}$ & \cellcolor[HTML]{26BCE5}$\bigstar\,\textbf{315.351}$\\
\cellcolor[HTML]{F5C70E}$\circledcirc\,\textbf{778.821}$ & \cellcolor[HTML]{FFFFFF}$\circledcirc\,18.575$ & \cellcolor[HTML]{FFFFFF}$\circledcirc\,1.516$ & \cellcolor[HTML]{F5C70E}$\circledcirc\,\textbf{12.655}$ & \cellcolor[HTML]{FFFFFF}$\circledcirc\,1.722$ & \cellcolor[HTML]{F5C70E}$\circledcirc\,\textbf{1.193}$ & \cellcolor[HTML]{F5C70E}$\bigstar\,\textbf{2.581}$ & \cellcolor[HTML]{F5C70E}$\circledcirc\,\textbf{1.649}$ & \cellcolor[HTML]{DF526B}$\bigstar\,\textbf{315.390}$\\
\bottomrule
\end{tabular}
    \caption{Comparisons across low, mid, and high levels of Gaussian dependence structures, in three boxes of interest (Figure~\ref{fig:sim_pred_task}).}
    \label{tab:gauss_mse}
\end{table}

\begin{table}[!htbp]
    \centering
\centering
\begin{tabular}[t]{c|c|c||c|c|c||c|c|c}
\toprule
\multicolumn{3}{c}{\textbf{Low}} & \multicolumn{3}{c}{\textbf{Mid}} & \multicolumn{3}{c}{\textbf{High}} \\
\cmidrule(l{3pt}r{3pt}){1-3} \cmidrule(l{3pt}r{3pt}){4-6} \cmidrule(l{3pt}r{3pt}){7-9}
$B_1$ & $B_2$ & $B_3$ & $B_1$ & $B_2$ & $B_3$ & $B_1$ & $B_2$ & $B_3$\\
\midrule
\cellcolor[HTML]{F5C70E}$\circledcirc\,\textit{0.467}$ & \cellcolor[HTML]{F5C70E}$\bigstar\,\textit{0.399}$ & \cellcolor[HTML]{FFFFFF}$0.810$ & \cellcolor[HTML]{DF526B}$\circledcirc\,\textit{0.194}$ & \cellcolor[HTML]{DF526B}$\circledcirc\,\textit{0.335}$ & \cellcolor[HTML]{DF526B}$\circledcirc\,\textit{0.805}$ & \cellcolor[HTML]{26BCE5}$\circledcirc\,\textit{0.182}$ & \cellcolor[HTML]{26BCE5}$\bigstar\,\textit{40.125}$ & --\\
\cellcolor[HTML]{F5C70E}$\bigstar\,\textit{0.524}$ & \cellcolor[HTML]{F5C70E}$\circledcirc\,\textit{0.485}$ & \cellcolor[HTML]{F5C70E}$\circledcirc\,\textit{0.983}$ & \cellcolor[HTML]{26BCE5}$\circledcirc\,\textit{0.207}$ & \cellcolor[HTML]{26BCE5}$\circledcirc\,\textit{0.418}$ & \cellcolor[HTML]{26BCE5}$\circledcirc\,\textit{0.821}$ & \cellcolor[HTML]{DF526B}$\circledcirc\,\textit{0.191}$ & \cellcolor[HTML]{DF526B}$\bigstar\,\textit{42.381}$ & --\\
\cellcolor[HTML]{FFFFFF}$0.635$ & \cellcolor[HTML]{26BCE5}$\circledcirc\,\textit{0.503}$ & \cellcolor[HTML]{26BCE5}$\circledcirc\,\textit{1.099}$ & \cellcolor[HTML]{F5C70E}$\circledcirc\,\textit{0.287}$ & \cellcolor[HTML]{F5C70E}$\circledcirc\,\textit{1.084}$ & \cellcolor[HTML]{F5C70E}$\circledcirc\,\textit{1.527}$ & \cellcolor[HTML]{F5C70E}$\circledcirc\,\textit{0.198}$ & \cellcolor[HTML]{DF526B}$\circledcirc\,\textit{60.439}$ & --\\
\cellcolor[HTML]{26BCE5}$\circledcirc\,\textit{0.696}$ & \cellcolor[HTML]{DF526B}$\circledcirc\,\textit{0.541}$ & \cellcolor[HTML]{DF526B}$\circledcirc\,\textit{1.211}$ & \cellcolor[HTML]{FFFFFF}$0.376$ & \cellcolor[HTML]{DF526B}$\bigstar\,\textbf{1.889}$ & \cellcolor[HTML]{26BCE5}$\bigstar\,\textit{3.429}$ & \cellcolor[HTML]{F5C70E}$\bigstar\,\textit{0.314}$ & \cellcolor[HTML]{26BCE5}$\circledcirc\,\textit{88.483}$ & --\\
\cellcolor[HTML]{F5C70E}$\bigstar\,\textbf{0.770}$ & \cellcolor[HTML]{FFFFFF}$0.550$ & \cellcolor[HTML]{F5C70E}$\bigstar\,\textit{1.274}$ & \cellcolor[HTML]{F5C70E}$\bigstar\,\textit{0.436}$ & \cellcolor[HTML]{26BCE5}$\bigstar\,\textbf{1.952}$ & \cellcolor[HTML]{DF526B}$\bigstar\,\textit{3.610}$ & \cellcolor[HTML]{26BCE5}$\bigstar\,\textit{0.493}$ & \cellcolor[HTML]{F5C70E}$\bigstar\,\textit{4.412e+04}$ & --\\
\cellcolor[HTML]{DF526B}$\bigstar\,\textbf{0.777}$ & \cellcolor[HTML]{26BCE5}$\bigstar\,\textit{0.708}$ & \cellcolor[HTML]{DF526B}$\circledcirc\,\textbf{1.454}$ & \cellcolor[HTML]{26BCE5}$\bigstar\,\textit{0.560}$ & \cellcolor[HTML]{F5C70E}$\bigstar\,\textit{2.051}$ & \cellcolor[HTML]{FFFFFF}$5.682$ & \cellcolor[HTML]{DF526B}$\bigstar\,\textit{0.522}$ & \cellcolor[HTML]{F5C70E}$\circledcirc\,\textit{1.172e+05}$ & --\\
\cellcolor[HTML]{26BCE5}$\bigstar\,\textbf{0.781}$ & \cellcolor[HTML]{26BCE5}$\bigstar\,\textbf{0.720}$ & \cellcolor[HTML]{26BCE5}$\bigstar\,\textit{1.528}$ & \cellcolor[HTML]{DF526B}$\bigstar\,\textit{0.623}$ & \cellcolor[HTML]{F5C70E}$\bigstar\,\textbf{2.177}$ & \cellcolor[HTML]{F5C70E}$\bigstar\,\textit{10.734}$ & \cellcolor[HTML]{FFFFFF}$0.530$ & \cellcolor[HTML]{DF526B}$\bigstar\,\textbf{1.042e+06}$ & --\\
\cellcolor[HTML]{DF526B}$\circledcirc\,\textit{0.878}$ & \cellcolor[HTML]{DF526B}$\bigstar\,\textbf{0.721}$ & \cellcolor[HTML]{DF526B}$\bigstar\,\textit{1.544}$ & \cellcolor[HTML]{FFFFFF}$\circledcirc\,1.035$ & \cellcolor[HTML]{FFFFFF}$2.302$ & \cellcolor[HTML]{26BCE5}$\bigstar\,\textbf{16.002}$ & \cellcolor[HTML]{FFFFFF}$\circledcirc\,0.660$ & \cellcolor[HTML]{26BCE5}$\bigstar\,\textbf{1.598e+06}$ & --\\
\cellcolor[HTML]{26BCE5}$\bigstar\,\textit{0.961}$ & \cellcolor[HTML]{DF526B}$\bigstar\,\textit{0.723}$ & \cellcolor[HTML]{26BCE5}$\circledcirc\,\textbf{1.582}$ & \cellcolor[HTML]{26BCE5}$\circledcirc\,\textbf{1.423}$ & \cellcolor[HTML]{26BCE5}$\bigstar\,\textit{2.898}$ & \cellcolor[HTML]{DF526B}$\bigstar\,\textbf{16.287}$ & \cellcolor[HTML]{26BCE5}$\circledcirc\,\textbf{0.752}$ & \cellcolor[HTML]{F5C70E}$\bigstar\,\textbf{2.354e+06}$ & --\\
\cellcolor[HTML]{DF526B}$\bigstar\,\textit{0.966}$ & \cellcolor[HTML]{F5C70E}$\bigstar\,\textbf{0.732}$ & \cellcolor[HTML]{F5C70E}$\circledcirc\,\textbf{1.618}$ & \cellcolor[HTML]{F5C70E}$\circledcirc\,\textbf{1.479}$ & \cellcolor[HTML]{DF526B}$\circledcirc\,\textbf{3.189}$ & \cellcolor[HTML]{F5C70E}$\bigstar\,\textbf{16.571}$ & \cellcolor[HTML]{DF526B}$\circledcirc\,\textbf{0.781}$ & \cellcolor[HTML]{F5C70E}$\circledcirc\,\textbf{3.214e+06}$ & --\\
\cellcolor[HTML]{FFFFFF}$\circledcirc\,2.132$ & \cellcolor[HTML]{26BCE5}$\circledcirc\,\textbf{1.261}$ & \cellcolor[HTML]{F5C70E}$\bigstar\,\textbf{1.740}$ & \cellcolor[HTML]{DF526B}$\circledcirc\,\textbf{1.628}$ & \cellcolor[HTML]{DF526B}$\bigstar\,\textit{3.275}$ & \cellcolor[HTML]{DF526B}$\circledcirc\,\textbf{26.386}$ & \cellcolor[HTML]{F5C70E}$\circledcirc\,\textbf{0.808}$ & \cellcolor[HTML]{DF526B}$\circledcirc\,\textbf{5.305e+06}$ & --\\
\cellcolor[HTML]{26BCE5}$\circledcirc\,\textbf{2.235}$ & \cellcolor[HTML]{DF526B}$\circledcirc\,\textbf{1.328}$ & \cellcolor[HTML]{DF526B}$\bigstar\,\textbf{1.823}$ & \cellcolor[HTML]{F5C70E}$\bigstar\,\textbf{2.112}$ & \cellcolor[HTML]{26BCE5}$\circledcirc\,\textbf{3.540}$ & \cellcolor[HTML]{26BCE5}$\circledcirc\,\textbf{27.720}$ & \cellcolor[HTML]{F5C70E}$\bigstar\,\textbf{0.958}$ & \cellcolor[HTML]{26BCE5}$\circledcirc\,\textbf{7.779e+06}$ & --\\
\cellcolor[HTML]{DF526B}$\circledcirc\,\textbf{2.436}$ & \cellcolor[HTML]{F5C70E}$\circledcirc\,\textbf{1.332}$ & \cellcolor[HTML]{26BCE5}$\bigstar\,\textbf{1.876}$ & \cellcolor[HTML]{26BCE5}$\bigstar\,\textbf{2.170}$ & \cellcolor[HTML]{F5C70E}$\circledcirc\,\textbf{3.754}$ & \cellcolor[HTML]{F5C70E}$\circledcirc\,\textbf{28.014}$ & \cellcolor[HTML]{26BCE5}$\bigstar\,\textbf{0.990}$ & \cellcolor[HTML]{FFFFFF}$1.922e+07$ & --\\
\cellcolor[HTML]{F5C70E}$\circledcirc\,\textbf{2.655}$ & \cellcolor[HTML]{FFFFFF}$\circledcirc\,2.009$ & \cellcolor[HTML]{FFFFFF}$\circledcirc\,1.920$ & \cellcolor[HTML]{DF526B}$\bigstar\,\textbf{2.216}$ & \cellcolor[HTML]{FFFFFF}$\circledcirc\,4.070$ & \cellcolor[HTML]{FFFFFF}$\circledcirc\,29.302$ & \cellcolor[HTML]{DF526B}$\bigstar\,\textbf{1.002}$ & \cellcolor[HTML]{FFFFFF}$\circledcirc\,2.013e+07$ & --\\
\bottomrule
\end{tabular}
    \caption{Comparisons across low, mid, and high levels of logistic dependence structures, in three boxes of interest (Figure~\ref{fig:sim_pred_task}). The probability for $B_3$ under the high scenario is smaller than \texttt{R}'s machine precision, yielding zero as the true probability; as a result, the RMSE values are not informative and are excluded from this table.}
    \label{tab:logistic_mse}
\end{table}

\begin{table}[!htbp]
    \centering
\begin{tabular}[t]{c|c|c||c|c|c||c|c|c}
\toprule
\multicolumn{3}{c}{\textbf{Low}} & \multicolumn{3}{c}{\textbf{Mid}} & \multicolumn{3}{c}{\textbf{High}} \\
\cmidrule(l{3pt}r{3pt}){1-3} \cmidrule(l{3pt}r{3pt}){4-6} \cmidrule(l{3pt}r{3pt}){7-9}
$B_1$ & $B_2$ & $B_3$ & $B_1$ & $B_2$ & $B_3$ & $B_1$ & $B_2$ & $B_3$\\
\midrule
\cellcolor[HTML]{26BCE5}$\bigstar\,\textit{1.265}$ & \cellcolor[HTML]{26BCE5}$\circledcirc\,\textit{0.555}$ & \cellcolor[HTML]{26BCE5}$\circledcirc\,\textit{0.240}$ & \cellcolor[HTML]{26BCE5}$\circledcirc\,\textit{0.441}$ & \cellcolor[HTML]{DF526B}$\bigstar\,\textit{0.167}$ & \cellcolor[HTML]{FFFFFF}$5.644$ & \cellcolor[HTML]{F5C70E}$\bigstar\,\textit{0.219}$ & \cellcolor[HTML]{FFFFFF}$22.498$ & --\\
\cellcolor[HTML]{DF526B}$\bigstar\,\textit{1.267}$ & \cellcolor[HTML]{DF526B}$\circledcirc\,\textit{0.578}$ & \cellcolor[HTML]{DF526B}$\circledcirc\,\textit{0.245}$ & \cellcolor[HTML]{F5C70E}$\circledcirc\,\textit{0.448}$ & \cellcolor[HTML]{26BCE5}$\bigstar\,\textit{0.170}$ & \cellcolor[HTML]{DF526B}$\circledcirc\,\textbf{10.007}$ & \cellcolor[HTML]{FFFFFF}$0.221$ & \cellcolor[HTML]{F5C70E}$\circledcirc\,\textbf{30.273}$ & --\\
\cellcolor[HTML]{FFFFFF}$4.508$ & \cellcolor[HTML]{DF526B}$\bigstar\,\textit{0.737}$ & \cellcolor[HTML]{F5C70E}$\circledcirc\,\textit{0.262}$ & \cellcolor[HTML]{FFFFFF}$0.449$ & \cellcolor[HTML]{F5C70E}$\bigstar\,\textit{0.223}$ & \cellcolor[HTML]{DF526B}$\bigstar\,\textbf{10.276}$ & \cellcolor[HTML]{26BCE5}$\circledcirc\,\textit{0.236}$ & \cellcolor[HTML]{DF526B}$\circledcirc\,\textbf{31.948}$ & --\\
\cellcolor[HTML]{26BCE5}$\bigstar\,\textbf{20.243}$ & \cellcolor[HTML]{26BCE5}$\bigstar\,\textit{0.737}$ & \cellcolor[HTML]{26BCE5}$\bigstar\,\textit{0.275}$ & \cellcolor[HTML]{F5C70E}$\bigstar\,\textit{0.450}$ & \cellcolor[HTML]{26BCE5}$\circledcirc\,\textit{0.260}$ & \cellcolor[HTML]{F5C70E}$\circledcirc\,\textit{10.297}$ & \cellcolor[HTML]{DF526B}$\circledcirc\,\textit{0.241}$ & \cellcolor[HTML]{26BCE5}$\bigstar\,\textbf{33.644}$ & --\\
\cellcolor[HTML]{DF526B}$\bigstar\,\textbf{20.285}$ & \cellcolor[HTML]{FFFFFF}$1.909$ & \cellcolor[HTML]{DF526B}$\bigstar\,\textit{0.275}$ & \cellcolor[HTML]{DF526B}$\circledcirc\,\textit{0.478}$ & \cellcolor[HTML]{F5C70E}$\circledcirc\,\textit{0.270}$ & \cellcolor[HTML]{26BCE5}$\circledcirc\,\textbf{10.356}$ & \cellcolor[HTML]{F5C70E}$\circledcirc\,\textit{0.242}$ & \cellcolor[HTML]{DF526B}$\bigstar\,\textbf{34.231}$ & --\\
\cellcolor[HTML]{DF526B}$\circledcirc\,\textit{57.390}$ & \cellcolor[HTML]{F5C70E}$\circledcirc\,\textit{2.441}$ & \cellcolor[HTML]{F5C70E}$\bigstar\,\textit{0.506}$ & \cellcolor[HTML]{26BCE5}$\bigstar\,\textit{0.761}$ & \cellcolor[HTML]{DF526B}$\circledcirc\,\textit{0.299}$ & \cellcolor[HTML]{F5C70E}$\circledcirc\,\textbf{10.364}$ & \cellcolor[HTML]{DF526B}$\bigstar\,\textit{0.333}$ & \cellcolor[HTML]{F5C70E}$\bigstar\,\textbf{35.076}$ & --\\
\cellcolor[HTML]{26BCE5}$\circledcirc\,\textit{62.626}$ & \cellcolor[HTML]{F5C70E}$\bigstar\,\textbf{5.934}$ & \cellcolor[HTML]{FFFFFF}$1.014$ & \cellcolor[HTML]{DF526B}$\bigstar\,\textit{0.799}$ & \cellcolor[HTML]{FFFFFF}$0.398$ & \cellcolor[HTML]{26BCE5}$\bigstar\,\textbf{10.602}$ & \cellcolor[HTML]{26BCE5}$\bigstar\,\textit{0.333}$ & \cellcolor[HTML]{F5C70E}$\bigstar\,\textit{36.018}$ & --\\
\cellcolor[HTML]{F5C70E}$\bigstar\,\textbf{72.494}$ & \cellcolor[HTML]{26BCE5}$\bigstar\,\textbf{6.234}$ & \cellcolor[HTML]{DF526B}$\circledcirc\,\textbf{1.151}$ & \cellcolor[HTML]{FFFFFF}$\circledcirc\,2.471$ & \cellcolor[HTML]{26BCE5}$\bigstar\,\textbf{1.620}$ & \cellcolor[HTML]{F5C70E}$\bigstar\,\textbf{10.839}$ & \cellcolor[HTML]{FFFFFF}$\circledcirc\,0.877$ & \cellcolor[HTML]{F5C70E}$\circledcirc\,\textit{36.157}$ & --\\
\cellcolor[HTML]{F5C70E}$\circledcirc\,\textit{94.396}$ & \cellcolor[HTML]{DF526B}$\bigstar\,\textbf{6.242}$ & \cellcolor[HTML]{26BCE5}$\circledcirc\,\textbf{1.174}$ & \cellcolor[HTML]{26BCE5}$\circledcirc\,\textbf{3.368}$ & \cellcolor[HTML]{F5C70E}$\bigstar\,\textbf{1.633}$ & \cellcolor[HTML]{26BCE5}$\circledcirc\,\textit{11.006}$ & \cellcolor[HTML]{26BCE5}$\circledcirc\,\textbf{1.076}$ & \cellcolor[HTML]{26BCE5}$\circledcirc\,\textbf{36.805}$ & --\\
\cellcolor[HTML]{F5C70E}$\bigstar\,\textit{1.470e+03}$ & \cellcolor[HTML]{DF526B}$\circledcirc\,\textbf{17.149}$ & \cellcolor[HTML]{F5C70E}$\circledcirc\,\textbf{1.181}$ & \cellcolor[HTML]{F5C70E}$\circledcirc\,\textbf{3.537}$ & \cellcolor[HTML]{DF526B}$\bigstar\,\textbf{1.654}$ & \cellcolor[HTML]{DF526B}$\circledcirc\,\textit{11.191}$ & \cellcolor[HTML]{F5C70E}$\circledcirc\,\textbf{1.162}$ & \cellcolor[HTML]{26BCE5}$\circledcirc\,\textit{39.026}$ & --\\
\cellcolor[HTML]{DF526B}$\circledcirc\,\textbf{1.612e+03}$ & \cellcolor[HTML]{26BCE5}$\circledcirc\,\textbf{22.004}$ & \cellcolor[HTML]{F5C70E}$\bigstar\,\textbf{1.202}$ & \cellcolor[HTML]{DF526B}$\circledcirc\,\textbf{3.574}$ & \cellcolor[HTML]{26BCE5}$\circledcirc\,\textbf{2.535}$ & \cellcolor[HTML]{F5C70E}$\bigstar\,\textit{12.932}$ & \cellcolor[HTML]{DF526B}$\circledcirc\,\textbf{1.187}$ & \cellcolor[HTML]{DF526B}$\circledcirc\,\textit{42.181}$ & --\\
\cellcolor[HTML]{26BCE5}$\circledcirc\,\textbf{1.802e+03}$ & \cellcolor[HTML]{F5C70E}$\circledcirc\,\textbf{22.167}$ & \cellcolor[HTML]{26BCE5}$\bigstar\,\textbf{1.205}$ & \cellcolor[HTML]{F5C70E}$\bigstar\,\textbf{6.155}$ & \cellcolor[HTML]{DF526B}$\circledcirc\,\textbf{2.669}$ & \cellcolor[HTML]{FFFFFF}$\circledcirc\,13.189$ & \cellcolor[HTML]{F5C70E}$\bigstar\,\textbf{1.807}$ & \cellcolor[HTML]{DF526B}$\bigstar\,\textit{46.009}$ & --\\
\cellcolor[HTML]{FFFFFF}$\circledcirc\,1.909e+03$ & \cellcolor[HTML]{FFFFFF}$\circledcirc\,38.714$ & \cellcolor[HTML]{DF526B}$\bigstar\,\textbf{1.205}$ & \cellcolor[HTML]{26BCE5}$\bigstar\,\textbf{6.608}$ & \cellcolor[HTML]{F5C70E}$\circledcirc\,\textbf{2.743}$ & \cellcolor[HTML]{26BCE5}$\bigstar\,\textit{16.924}$ & \cellcolor[HTML]{26BCE5}$\bigstar\,\textbf{1.954}$ & \cellcolor[HTML]{26BCE5}$\bigstar\,\textit{46.009}$ & --\\
\cellcolor[HTML]{F5C70E}$\circledcirc\,\textbf{2.712e+03}$ & \cellcolor[HTML]{F5C70E}$\bigstar\,\textit{38.782}$ & \cellcolor[HTML]{FFFFFF}$\circledcirc\,1.745$ & \cellcolor[HTML]{DF526B}$\bigstar\,\textbf{7.198}$ & \cellcolor[HTML]{FFFFFF}$\circledcirc\,2.986$ & \cellcolor[HTML]{DF526B}$\bigstar\,\textit{17.359}$ & \cellcolor[HTML]{DF526B}$\bigstar\,\textbf{2.031}$ & \cellcolor[HTML]{FFFFFF}$\circledcirc\,84.377$ & --\\
\bottomrule
\end{tabular}
    \caption{Comparisons across low, mid, and high levels of H\"usler-Reiss dependence structures, in three boxes of interest (Figure~\ref{fig:hr_pred_task}). As in Table~\ref{tab:logistic_mse}, the probability for $B_3$ under the high scenario is smaller than \texttt{R}'s machine precision, yielding zero as the true probability; as a result, the RMSE values are not informative and are excluded from this table.}
    \label{tab:husler_reiss_mse}
\end{table}

The comparison of RMSE allows us to evaluate the bias-variance tradeoff across these methods in a single metric, yet it does not tell the whole story. To investigate the comparisons on a more granular level, we examine boxplots of the results by dependence structure and level. For readability, we present the results from one Gaussian and one logistic setting here; the remaining 25 (including all H\"usler-Reiss settings) are provided in Supplementary Material~\ref{append:sim_study_results}. All boxplots use a $95\%$ winsorization to exclude extreme outliers and provide a better visual comparison.

\begin{figure}[!htbp]
    \centering
    \includegraphics[width=0.75\linewidth]{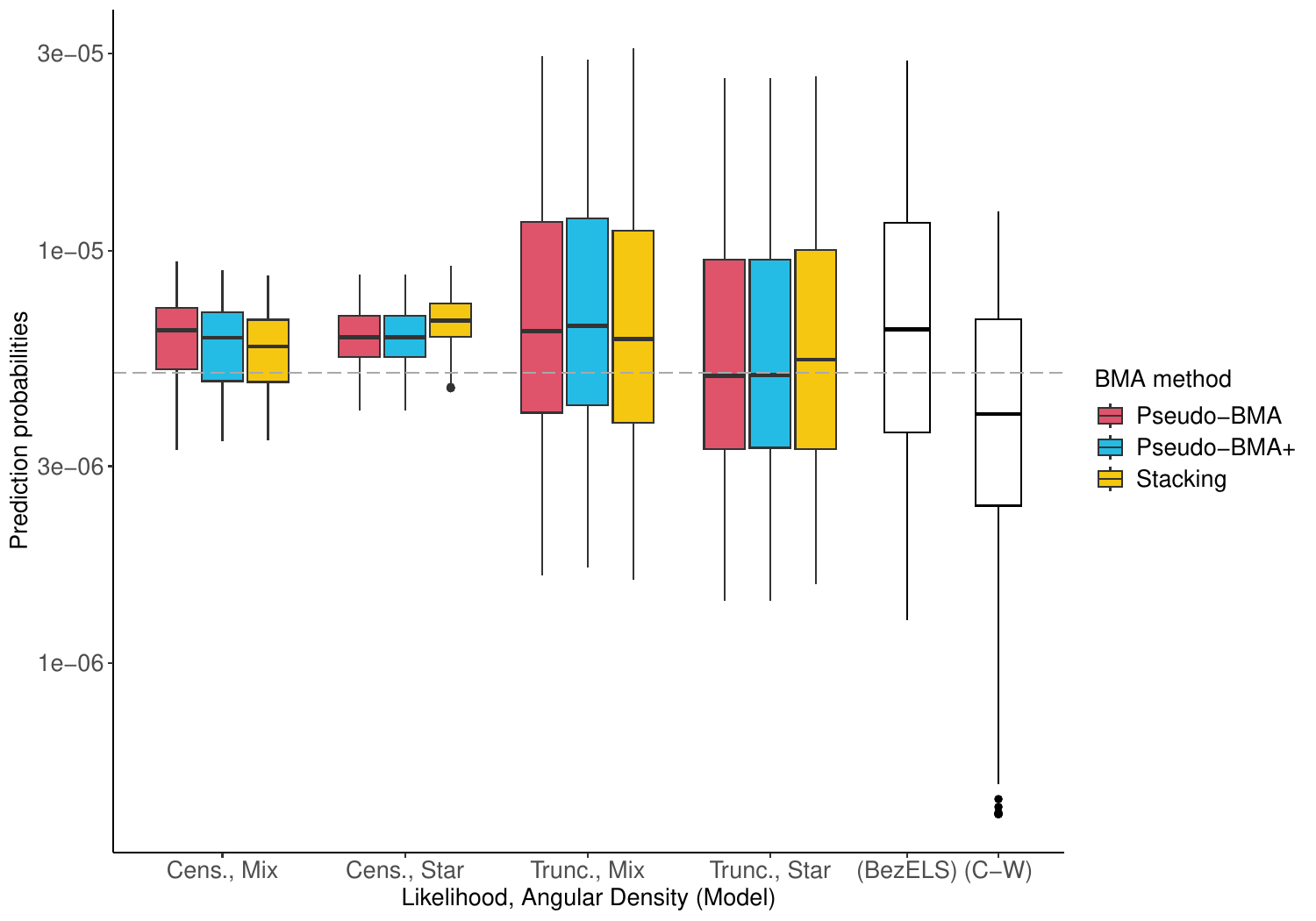}
    \caption{Predicted probabilities of lying in $B_2$, under a Gaussian dependence structure with $\rho=0.5$ (see Figure~\ref{fig:sim_pred_task}), across 200 simulated datasets. The dashed grey line is the true probability. Y-axis is shown on the log scale.}
    \label{fig:gauss_preds_boxplot}
\end{figure}

\begin{figure}[!htbp]
    \centering
    \includegraphics[width=0.75\linewidth]{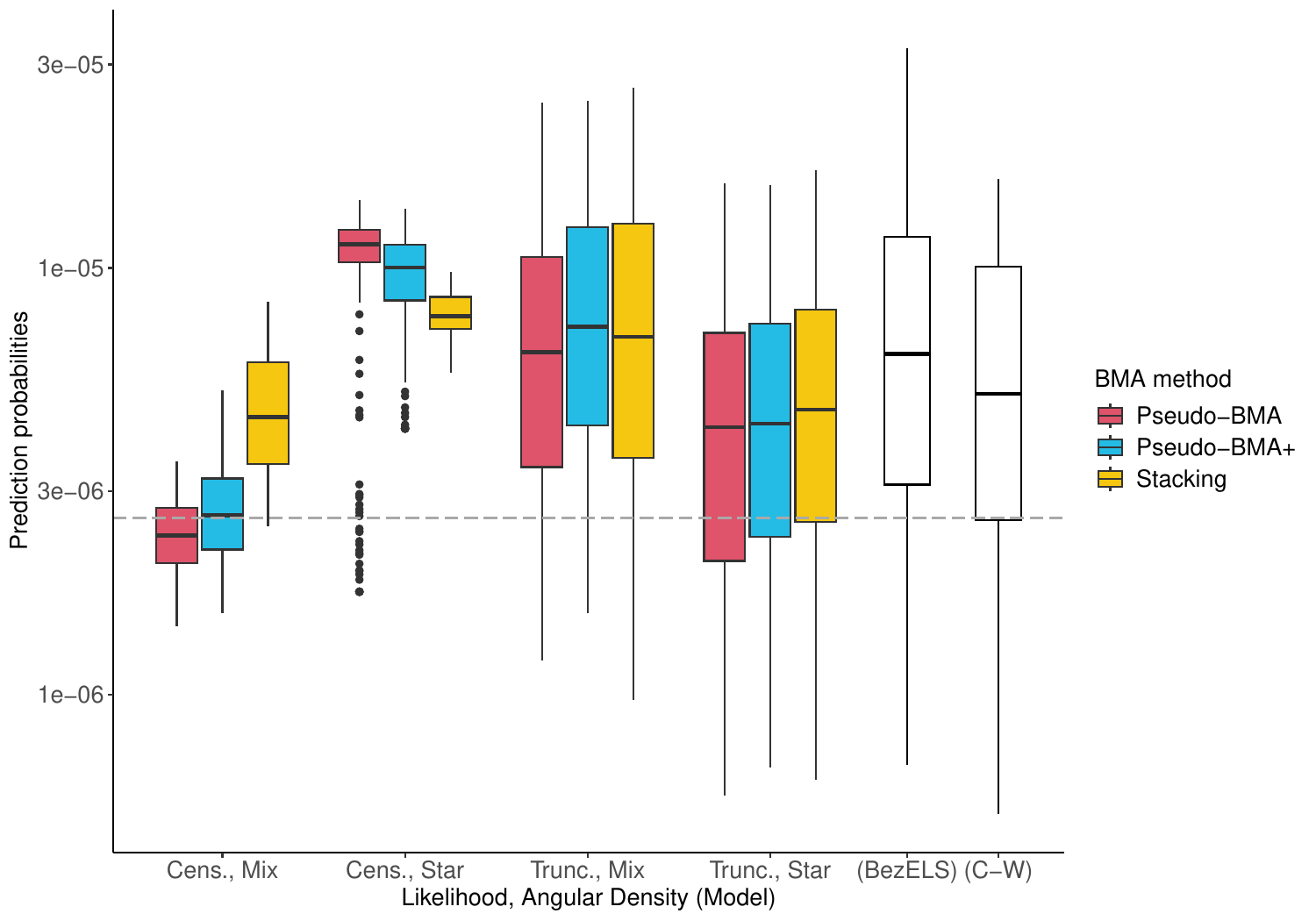}
    \caption{Predicted probabilities of lying in $B_2$, under a logistic dependence structure with $\lambda=0.5$ (see Figure~\ref{fig:sim_pred_task}), across 200 simulated datasets. The dashed grey line is the true probability. Y-axis is shown on the log scale.}
    \label{fig:logistic_preds_boxplot}
\end{figure}

Figures~\ref{fig:gauss_preds_boxplot} and \ref{fig:logistic_preds_boxplot} provide detailed comparisons of the mid-dependence scenarios in $B_2$ for Gaussian and logistic structures, respectively. In both, predictions from the censored models exhibit substantially lower variability than those from the truncated, BezELS, or C-W approaches. In the Gaussian case, all of our censored models (along with the truncated angular mixture and BezELS) are slightly overbiased, while the truncated model with the star-shaped density is essentially unbiased, and C-W is slightly underbiased. In the logistic case, all approaches are heavily overbiased, except for the censored mixture model using Pseudo-BMA+ weights (unbiased) and its Pseudo-BMA counterpart (underbiased).  These boxplots highlight the bias-variance tradeoff that lead to the censored approaches having favorable RMSE.  This nuance is lost when looking at the RMSE comparisons alone, so it is crucial to recognize the bias-variance tradeoff and weigh what is more important for the application setting.

\section{Data Analyses}\label{sec:data}
We apply our modeling and prediction procedures in analyzing the relationship between two fire weather indices: the Energy Release Component \citep[ERC,][]{bradshaw1984} and Canadian Forest Fire Weather Index \citep[FWI,][]{turner1978}. \cite{lawler2024} find regional variations and tradeoffs between the two indices, specifically in predicting the size of wildfire burned areas. Their analysis, based on surface meteorological data at a $4 \text{km} \times 4 \text{km}$ resolution and a single forest fuel model \citep[obtained from GridMET,][]{abatzoglou2013a}, shows that when one index is extreme in a specific region, the other often is not. We use our modeling approach to further investigate this relationship at a more localized scale, using forest fuel models specific to the region and higher-resolution weather data.

\subsection{Fire weather indices}
ERC is a calculated fuel moisture index from the National Fire Danger Rating System, which considers the cumulative drying effect of temperature, precipitation, humidity, and solar radiation as an overall measure of wildfire fuel \citep{abatzoglou2013}. As a relative index, an ERC value of 30 represents twice the potential heat release of a value of 15. FWI captures somewhat orthogonal fire behavior, combining fuel moisture and weather conditions into an index of potential wildfire activity.

We calculate daily ERC and FWI values using FireFamilyPlus \citep[FF+,version 5; ][]{Bradshaw_2000}, based on hourly weather data from the Program for Climate, Ecosystem, and Fire Applications and the Wildland Fires Application portal. After calculating the indices, we restrict these data to the fire season (April - October). We focus on two Remote Automatic Weather Stations (RAWS): Friend Mountain in California and Redstone in Colorado. We selected these sites due to their proximity, respectively, to the August Complex and Cameron Peak fires. These were the largest wildfires in each state’s history, both of which ignited in August 2020. 

Notably, the methodology of FF+ for calculating FWI differs slightly from packages like \texttt{cffdrs}, yielding only integer-valued outputs. While this results in a somewhat discrete appearance in parts of the FWI distribution, the number of observed values is large enough to justify applying a continuous marginal transformation.

\subsection{Marginal Transformation}
\begin{figure}[!htbp]
    \centering
    \includegraphics[width=0.75\linewidth]{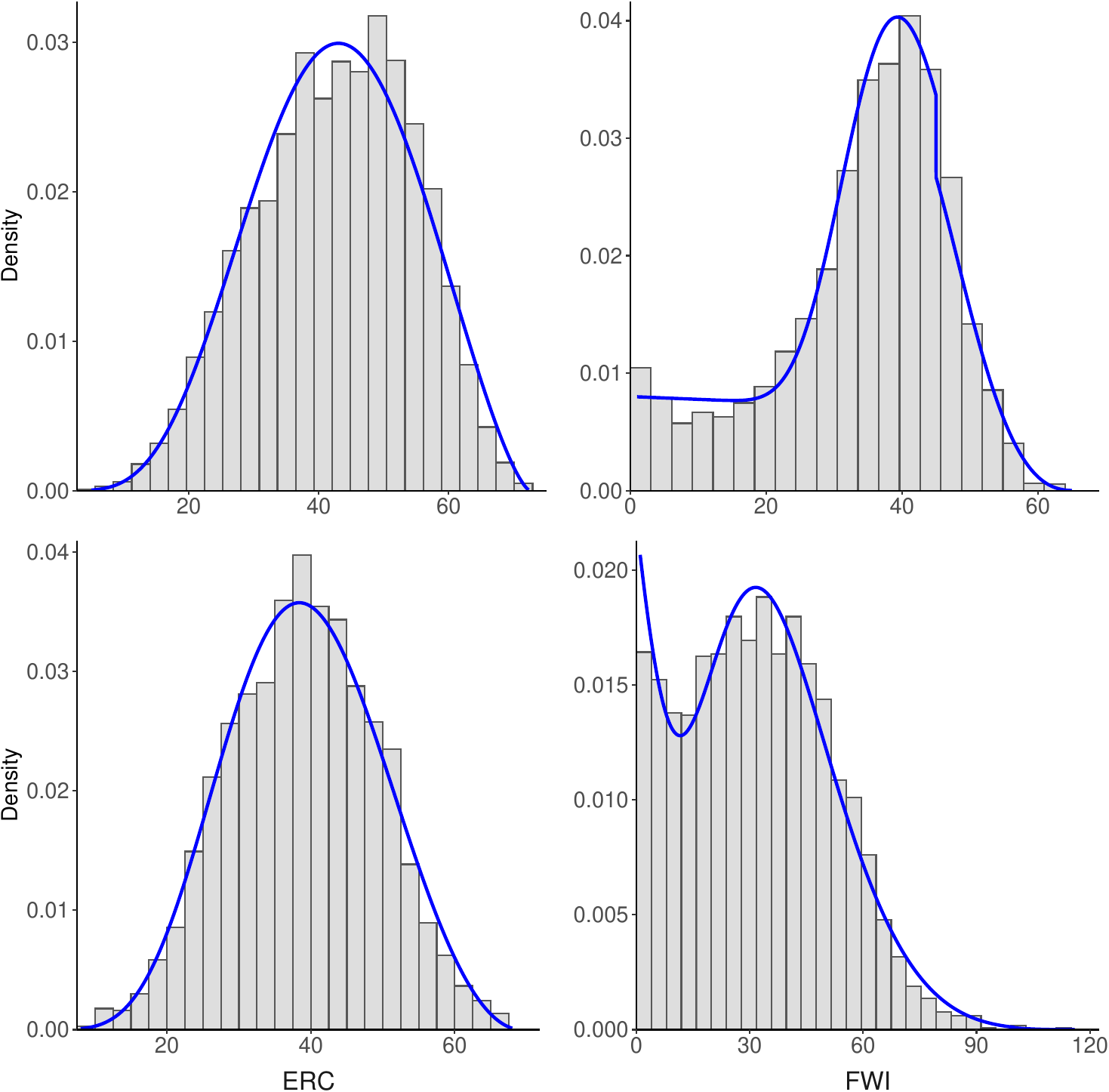}
    \caption{Empirical distribution of the fire indices, with their posterior marginal distributions overlayed in blue. Friend Mountain is the top row, and Redstone the bottom.}
    \label{fig:marg_transform}
\end{figure}

Before applying our model to these fire indices, we must first transform the data to standard exponential margins. We want to capture the tail behavior of each index appropriately while also fitting the bulk of the distribution, so we fit an extended generalized Pareto distribution \citep[EGPD,][]{naveau2016} separately to each fire index. For FWI, we incorporate the EGPD into a mixture model with an upper-truncated exponential distribution (truncated at 45 and 100 for the two RAWS, respectively). We truncate the exponential component so the upper tail behavior is guaranteed to be driven by the EGPD component rather than the exponential component. The usefulness of this mixture is evident in Figure~\ref{fig:marg_transform}, as the mixture marginal distribution estimates the FWI distribution quite well. The marginal EGPD distributions also appear to fit the ERC distributions well, with a potential underestimation around ERC values of 55 and 50, respectively, for each RAWS station.

\begin{figure}[!htbp]
    \centering
    \includegraphics[width=0.75\linewidth]{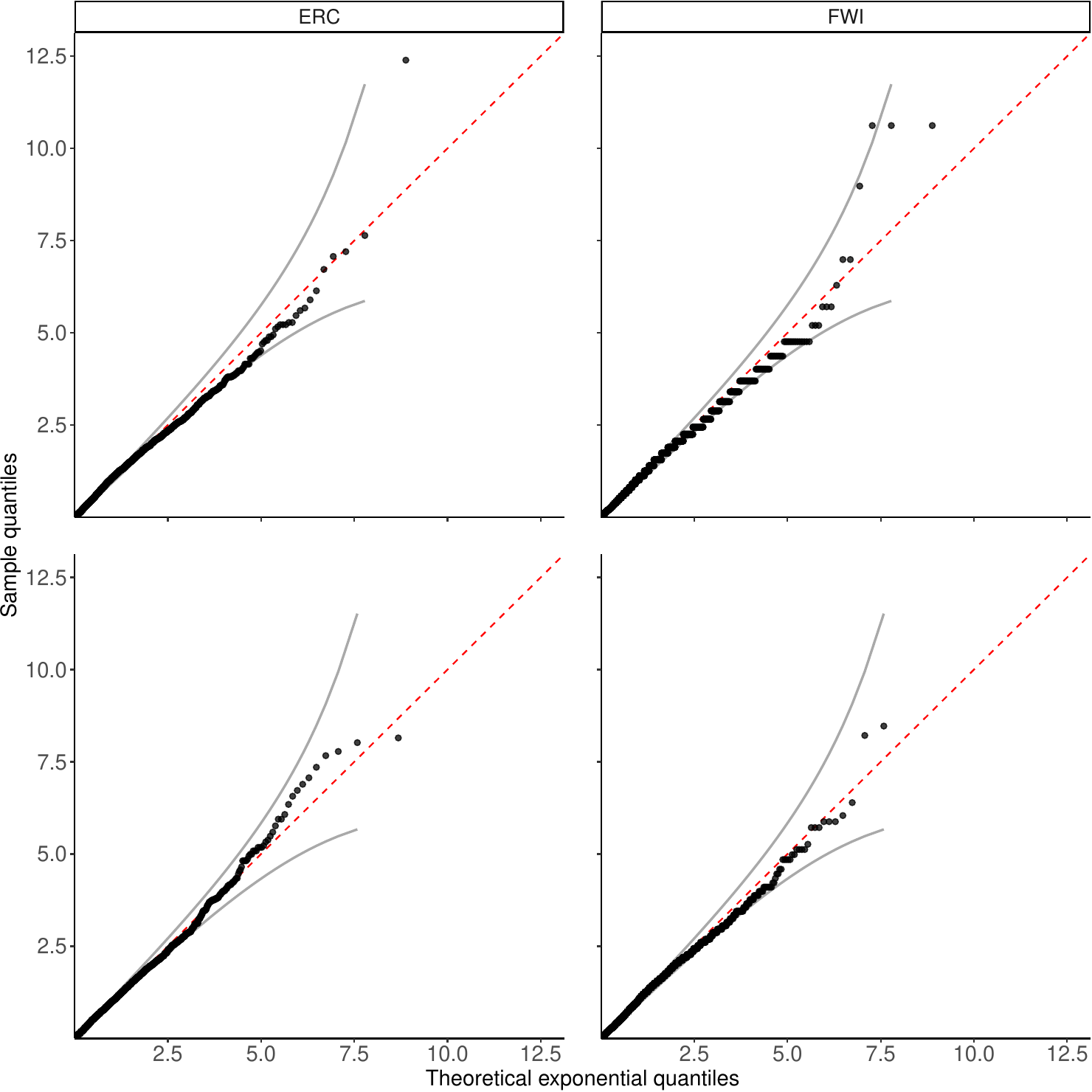}
    \caption{QQ plot comparing the fire indices on exponential margins to the theoretical quantiles of a standard exponential distribution. The identity line is in red and the $95\%$ confidence band in gray. As in Figure~\ref{fig:marg_transform}, top is Friend Mountain and bottom is Redstone.}
    \label{fig:qq_expo_indices}
\end{figure}

Once transformed onto standard exponential margins, the indices appear to follow a standard exponential distribution, apart from a few points in the tails of FWI at both RAWS (Figure~\ref{fig:qq_expo_indices}).

\subsection{Threshold Selection and Model Fitting}
\begin{figure}[!htbp]
    \centering
    \includegraphics[width=0.95\linewidth]{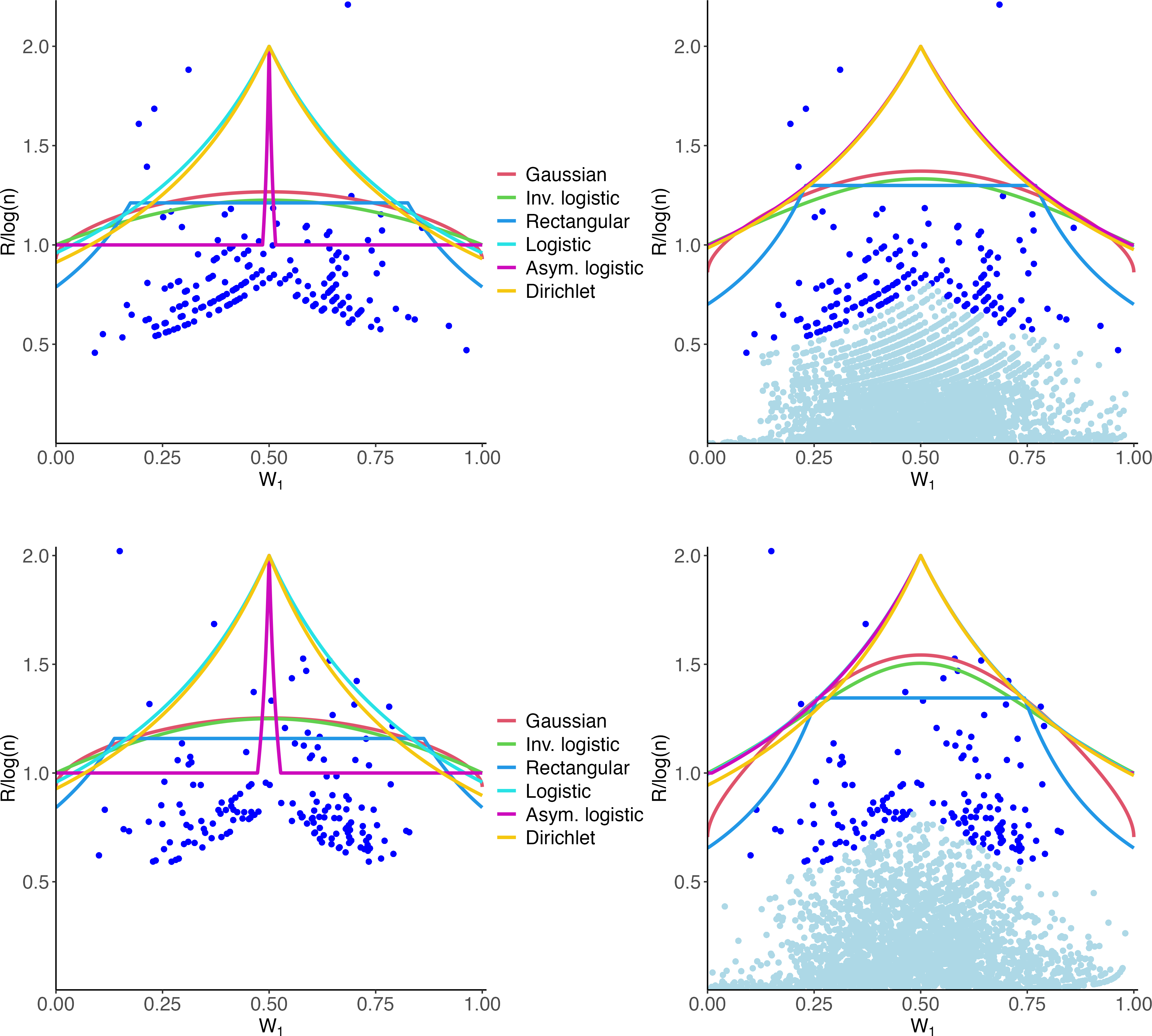}
    \caption{Posterior unit level sets of the six gauge functions fit to Friend Mountain (top) and Redstone (bottom), under the truncated (left) and censored (right) likelihoods.}
    \label{fig:posterior_unit_level_sets}
\end{figure}

These datasets have fewer points than our simulated datasets, with approximately 3600 and 3000 data points, yielding 184 and 155 points in the top $5\%$ of the datasets, respectively. The limitation of the smaller sample size is immediately apparent when comparing the unit level sets of the posterior gauge function fits of the truncated and censored likelihoods (Figure~\ref{fig:posterior_unit_level_sets}), showing a clear distinction between the two approaches.

\subsection{Diagnostics}\label{sec:diag}
\subsubsection{Radial Density}\label{sec:radial_diag}
We assess model fit for the truncated Gamma (tg) models in Equations~\eqref{trunc_lhood} and \eqref{cens_lhood} using probability-probability (PP) and quantile-quantile (QQ) plots, similar to \cite{wadsworth2024} and \cite{majumder-2025a}. We define the fitted distribution function under scenario $\{l,a\}$ as:
\begin{align*}
    F_{\text{tg},l,a}(r \given r_\tau(\bw), \bw, \btheta_{1, \text{post}}, \alpha_\text{post}) & := \prob_{l,a}(R \leq r \given \bW = \bw, \btheta_{1, \text{post}}, \alpha_\text{post}) \\
    &= 1 - \frac{\bar{F}_{l,a}(r; \alpha_\text{post}, g(\bw; \btheta_{1, \text{post}}))}{\bar{F}_{l,a}(r_\tau(\bw); \alpha_\text{post}, g(\bw;\btheta_{1, \text{post}}))}
\end{align*}
For each of the $n_0$ observations with \( R_i \geq r_\tau(\bw_i) \), we compute: $u_i = F_{\text{tg},l,a}(r_i \given r_\tau(\bw_i), \bw_i, \btheta_{1, \text{post}}, \alpha_\text{post})$ and compare it to the empirical estimate $\frac{i}{n_0 + 1}$ to generate the PP plot.

We compute $u_i$ for all four combinations of radial and angular likelihoods, and construct Bayesian model-averaged diagnostics by weighting them accordingly. The resulting BMA-based $u_i$ correspond to a true marginal posterior predictive distribution \textit{only} when the angular density is held fixed, as is the case with the Beta mixture shared across the six parameterizations of the radial density. When angular densities vary across models---as is the case with the star-shaped densities---the BMA-weighted $u_i$ do not correspond to a marginal posterior but can still be interpreted as a model-weighted diagnostic reflecting each model component's contribution to the joint density.

\begin{landscape}
\begin{figure}[!htbp]
     \centering
     \begin{subfigure}{0.65\textwidth}
         \centering
         \includegraphics[width=\textwidth]{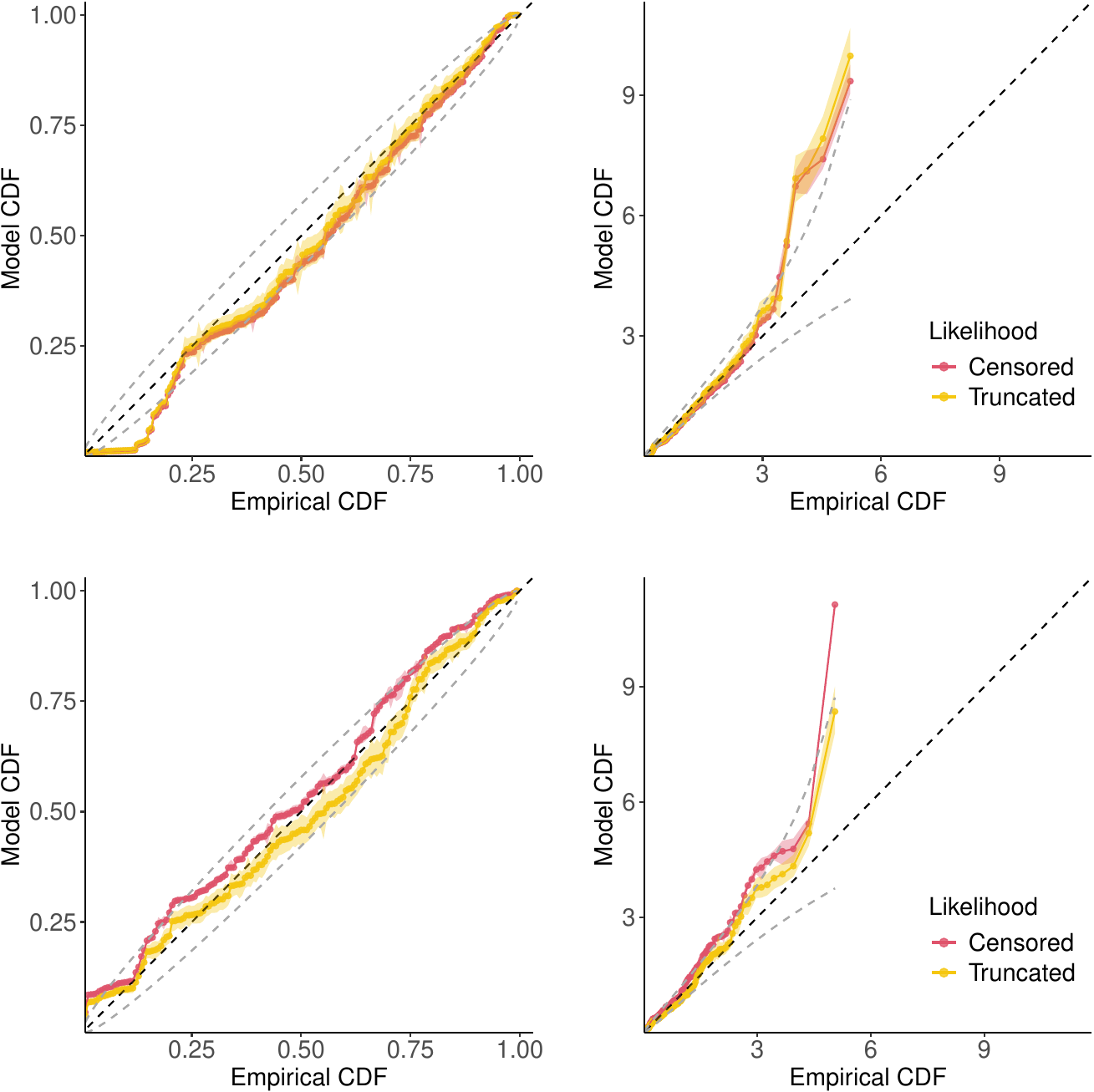}
         \caption{True marginal posterior, constant angular density}
         \label{fig:mix_diag_plots}
     \end{subfigure}
     \hfill
     \begin{subfigure}{0.65\textwidth}
         \centering
         \includegraphics[width=\textwidth]{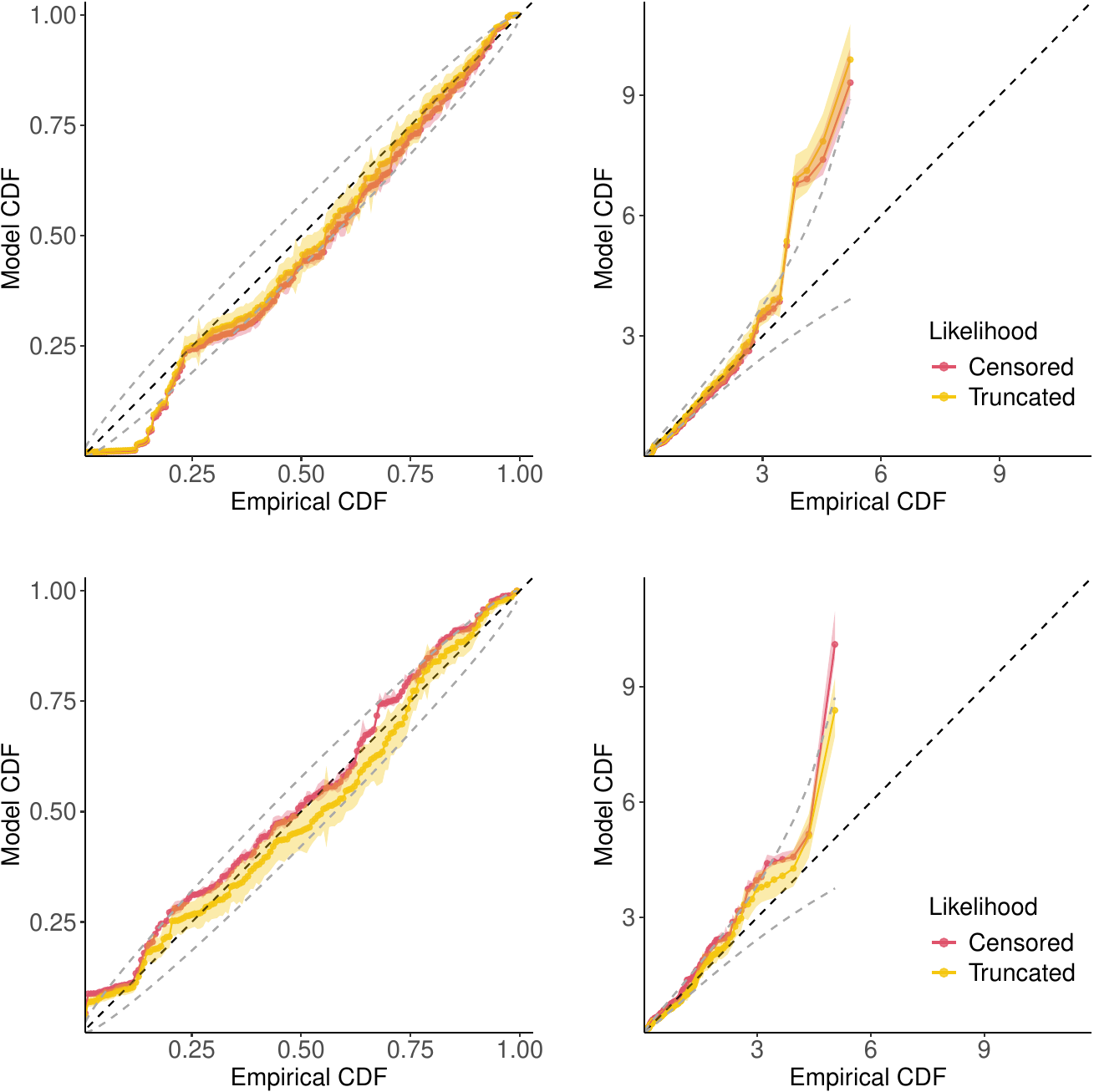}
         \caption{Model-weighted diagnostic, varying star-shaped density}
         \label{fig:star_diag_plots}
     \end{subfigure}
     \caption{PP and QQ plots of Friend Mountain (top) and Redstone (bottom) for the fitted truncated gamma distribution, with uniform margins in the first panel of each subfigure and exponential margins in the second panel. Colored ribbons around estimates represent the $95\%$ credible interval of the model cdf, dotted grey lines represent the $2.5\%$ and $97.5\%$ confidence bands of the empirical cdf, and the black line is the identity.}
     \label{fig:pp_qq_plots}
\end{figure}
\end{landscape}

Figure~\ref{fig:pp_qq_plots} suggests an overall adequate fit across models, with no critical deviations from the diagonal. The only notable signs of lack of fit appear in the upper tail of the Friend Mountain data, where a few observations are heavier-tailed than expected when shown on exponential margins.

Another potential goodness-of-fit diagnostic is based on return-level curves \citep{papastathopoulos2025, murphy-barltrop2023a, simpson2024, campbell2025}, which generalize univariate return levels to the multivariate setting. For a given return period $T$, \cite{campbell2025} derive the return curve in the truncated gamma setting as:
$$\mathcal{R}(T) = \{\bx \in \mathbb{R}^d_+ \given \bx = F_\text{Ga}^{-1}(1 - T^{-1}; d, g(\bw, \btheta)), \bw \in \mathcal{S}_{d-1}\}.$$
Similarly, we compute return curves for 100 values of $T=\{10,20,\dots,1000\}$, and compare the observed proportion of exceedances $\mathcal{R}(T)$ to the expected value of $T^{-1}$. As with the PP and QQ plots, we construct Bayesian model-averaged return curves, with analogous caveats and interpretations with respect to the angular densities.

\begin{figure}[!htbp]
     \centering
     \begin{subfigure}{0.85\textwidth}
         \centering
         \includegraphics[width=\textwidth]{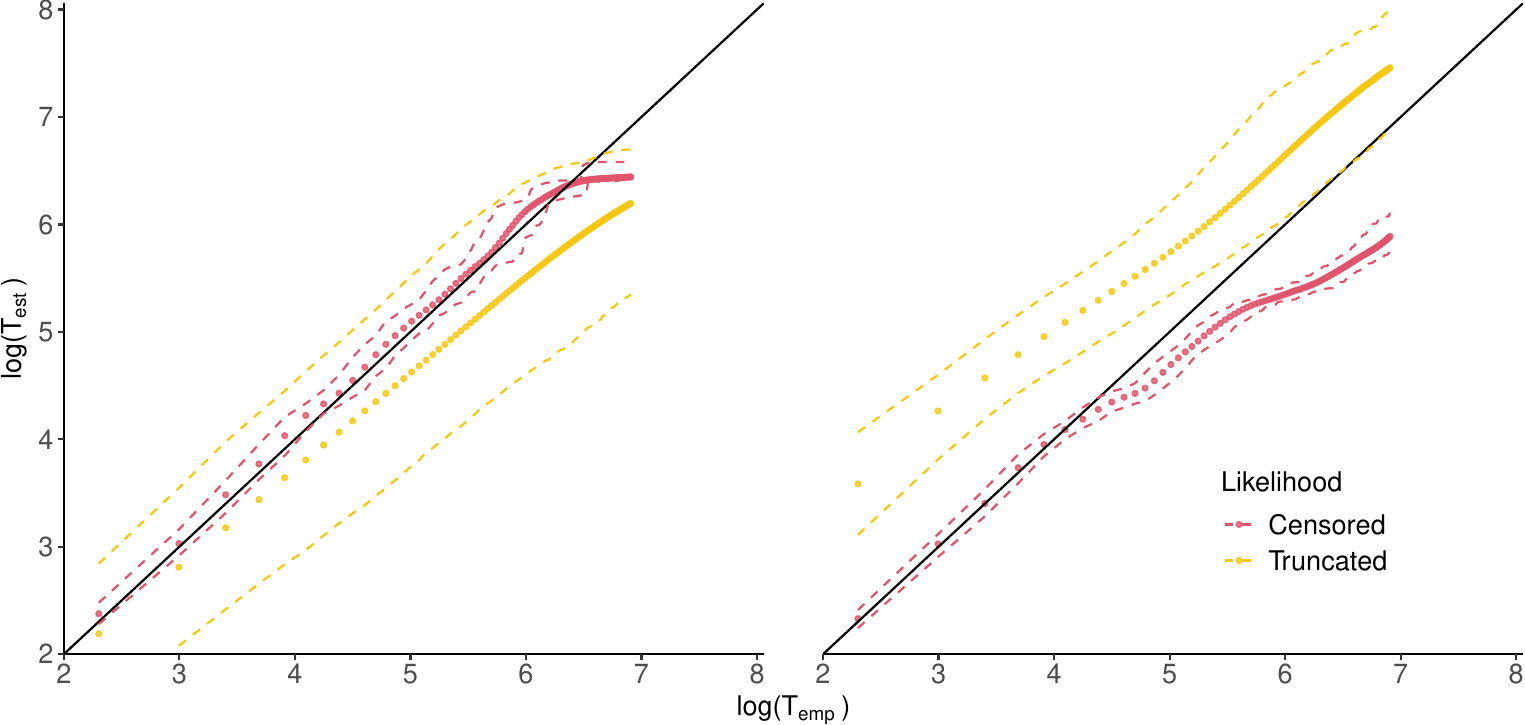}
         \caption{True estimated return curve, constant angular density}
         \label{fig:mix_return_curves}
     \end{subfigure}
     \begin{subfigure}{0.85\textwidth}
         \centering
         \includegraphics[width=\textwidth]{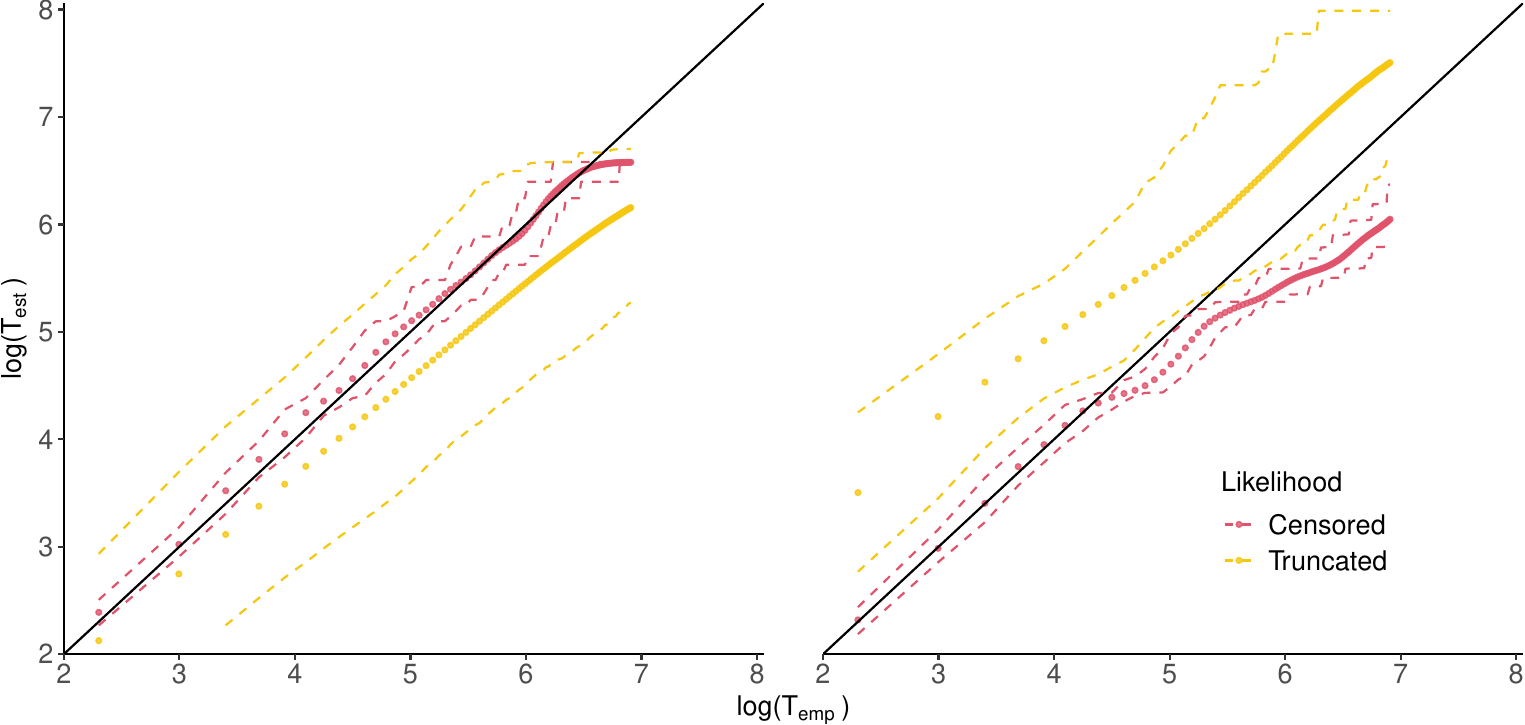}
         \caption{Diagnostic return curves, varying star-shaped density}
         \label{fig:star_return_curves}
     \end{subfigure}
        \caption{Estimated proportion of exceedances compared with the empirical for Friend Mountain (left) and Redstone (right). Dashed colored lines are the $95\%$ credible intervals, and the identity is in black. Axes are on the log scale.}
        \label{fig:return_curves}
\end{figure}

Figure~\ref{fig:return_curves} shows good agreement between Friend Mountain's estimated return-level sets and the empirical expected exceedances, with the censored likelihood outperforming the truncated. Redstone exhibits poorer performance, particularly under the truncated likelihood; the censored likelihood performs better, though not as well as Friend Mountain. However, the fit only begins to diverge after the 99th percentile ($\log(T) \approx 4.6$).

\subsubsection{Angular Density}\label{sec:ang_diag}
\begin{figure}[!htbp]
    \centering
    \includegraphics[width=0.95\linewidth]{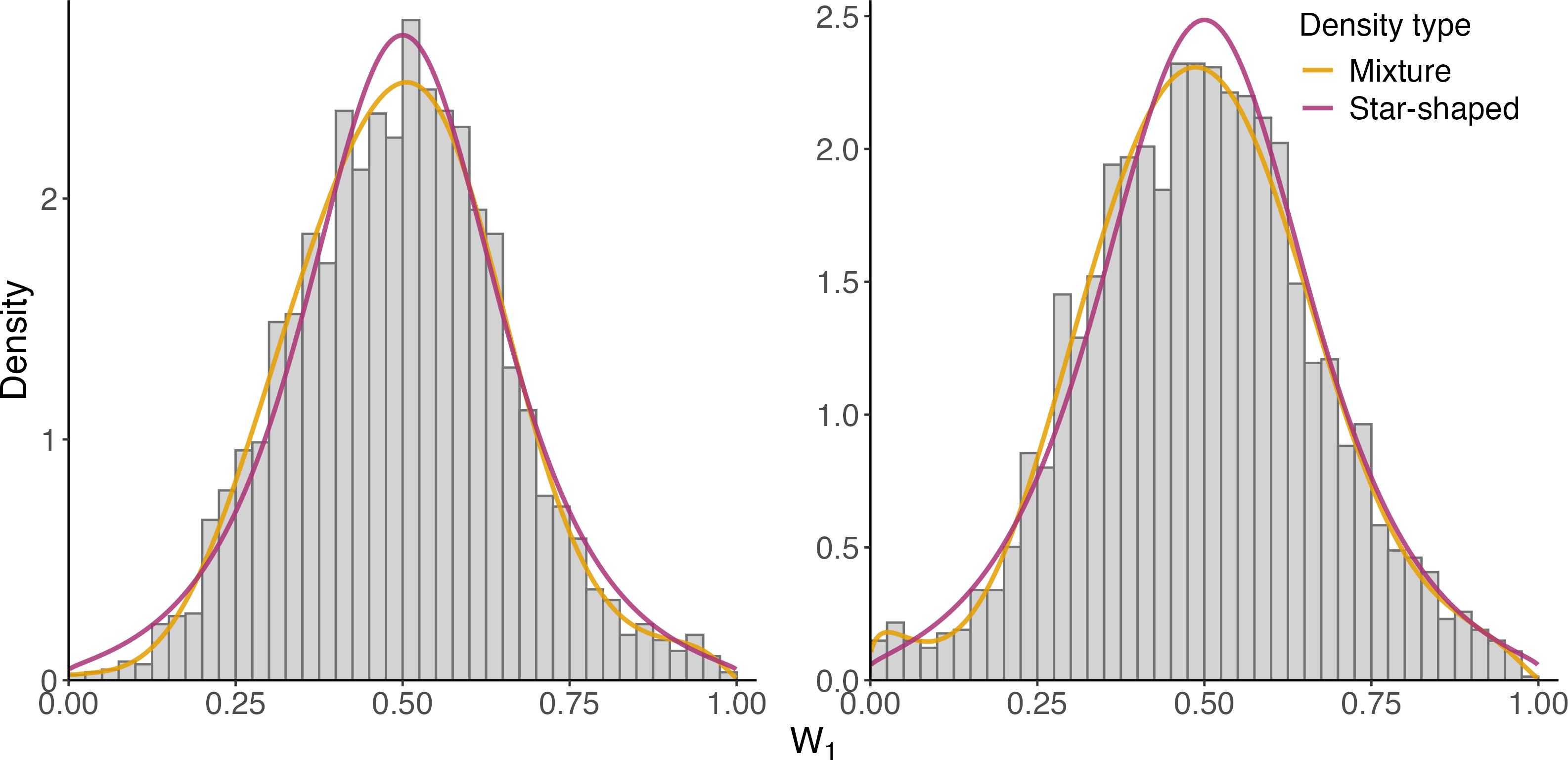}
    \caption{Empirical distribution of angles for Friend Mountain (left) and Redstone (right), with the marginal mixture density and the diagnostic model-weighted star-shaped density overlayed.}
    \label{fig:angular_diag_plot}
\end{figure}

The empirical distribution of angles can be compared to the fitted angular densities using density plots. As with the PP, QQ, and return-level plots, we construct Bayesian model-averaged diagnostics for the star-shaped density, with analogous caveats and interpretation. The form of the true marginal mixture density requires no additional manipulation. Figure~\ref{fig:angular_diag_plot} shows good agreement between the empirical and fitted angular distributions for both weather stations.

\begin{figure}[!htbp]
    \centering
    \includegraphics[width=0.95\linewidth]{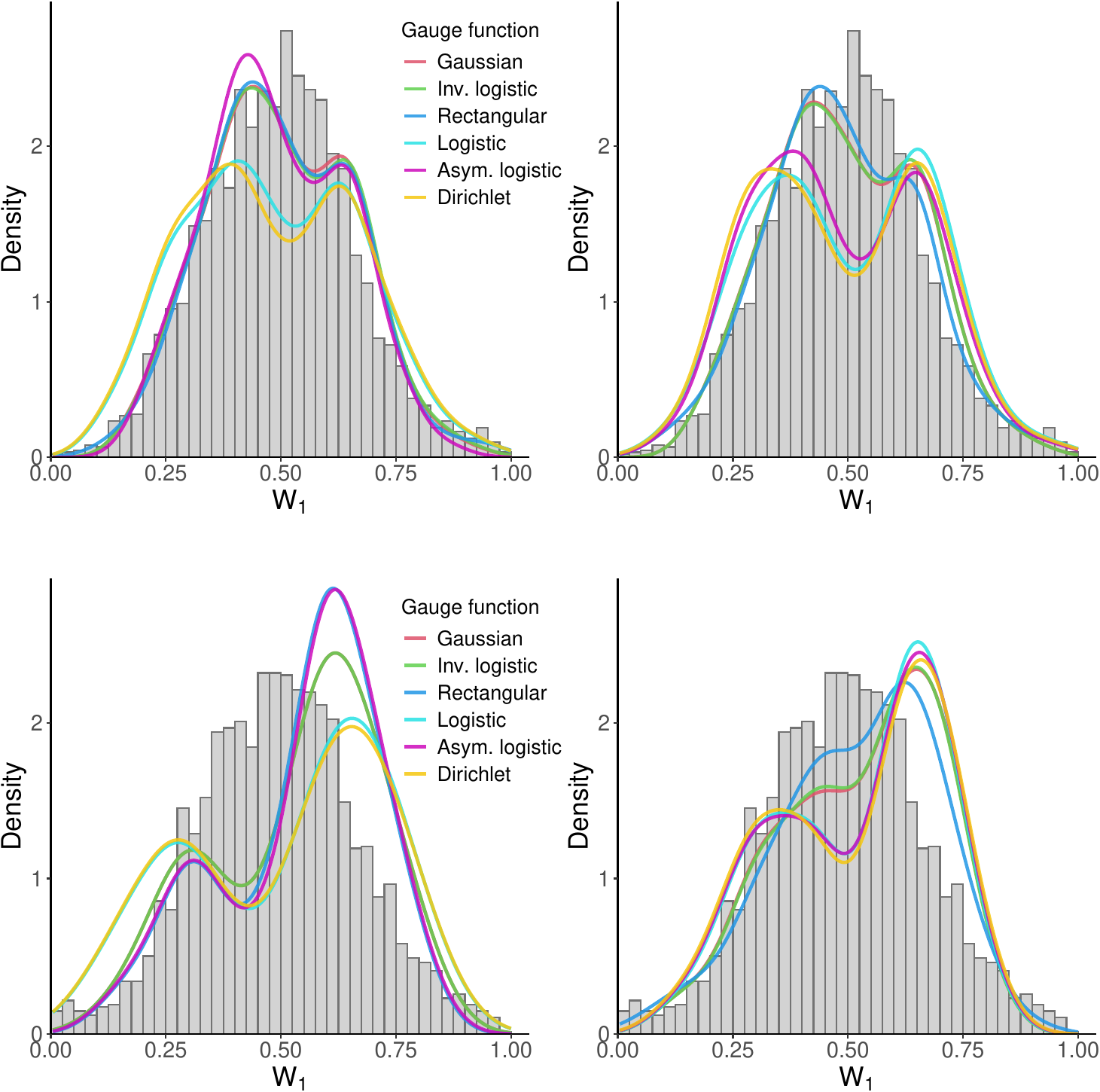}
    \caption{Histograms of the empirical distribution of angles for Friend Mountain (top) and Redstone (bottom). Within each location, the histogram remains the same across columns, as all angles were used in model fitting. Each column visualizes the density of $\bW$ above a high threshold determined via gamma quantile estimation. Since there are six radial fits (one per gauge function), we obtain six thresholds and thus six resulting densities for $\bW$. The left column corresponds to the truncated likelihood, and the right to the censored.}
    \label{fig:check_quant_assump}
\end{figure}

Finally, we assess Assumption~\ref{assump:w_above_quant_thres}, as we did in Section~\ref{sec:IS scheme}. While Figure~\ref{fig:angles_above_threshold} checks this assumption using the data-generating process, we now do so using the fitted radial models. For both weather stations and both likelihoods, we compute a gamma-based quantile threshold for each of the six gauge functions. We then plot a kernel density estimate of the angles that exceed each threshold (Figure~\ref{fig:check_quant_assump}). Some discrepancies are observed between the fitted densities and the empirical angular distribution, particularly for Redstone. The deviations appear less pronounced for Friend Mountain, possibly due to the slightly higher number of observations exceeding the 95th percentile threshold. This diagnostic suggests we may be introducing some additional bias in our prediction procedures by relying on Assumption~\ref{assump:w_above_quant_thres}.

\subsection{Predictions}
\begin{figure}[!htbp]
    \centering
    \includegraphics[width=0.95\linewidth]{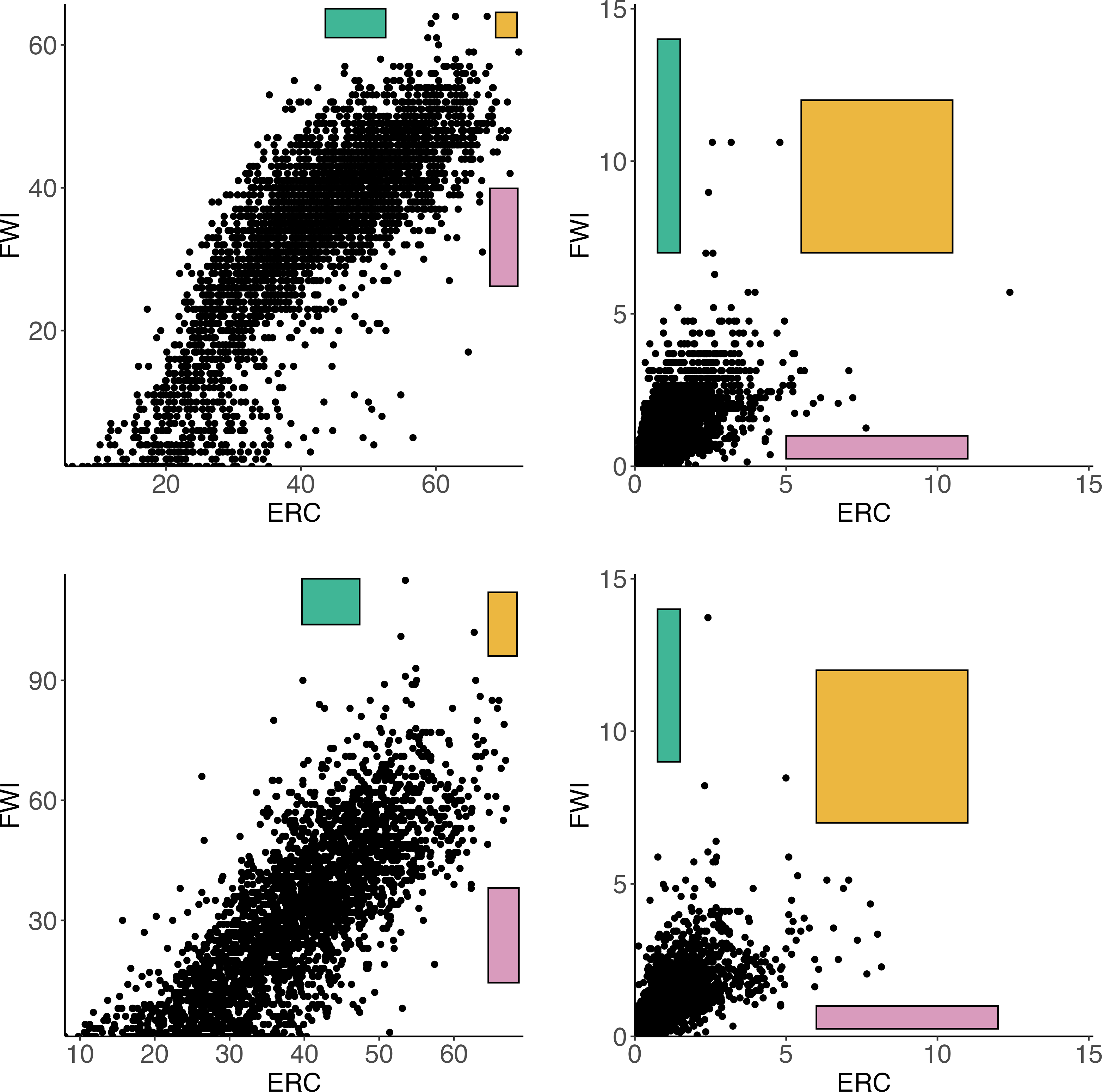}
    \caption{Illustration of the three sets we predict the probability of being in for Friend Mountain (top) and Redstone (bottom). The left column is in original margins, the right in exponential margins.}
    \label{fig:pred_task_boxes_fire}
\end{figure}

We evaluate predictive performance in three distinct regions for both RAWS locations, corresponding to scenarios where one or both indices are extreme (Figure~\ref{fig:pred_task_boxes_fire}). Specifically:
\begin{itemize}
    \item $B_1$ (teal): captures cases with extreme fire weather but less extreme burn potential
    \item $B_2$ (gold): corresponds to joint extremes in both fire weather and burn potential
    \item $B_3$ (pink): captures high burn potential with less extreme fire weather
\end{itemize}
Notably, no observed point falls in any of these boxes, so using a classical approach (instead of an extreme value approach) would yield an empirical estimate of 0 for the probability of lying in the region.

We generate full posterior distributions of predicted probabilities for each region, under all combinations of likelihood, angular density, and BMA weighting method. Tables~\ref{tab:friendmtn_preds} and \ref{tab:redstone_preds} report the median and $95\%$ credible intervals. Given the simulation study results and consistently unreliable performance of the truncated likelihood across both weather stations, we recommend disregarding those predictions entirely. The remainder of our discussion focuses on results from the censored likelihood.

There is general agreement between the star-shaped density and the mixture when predicting in $B_2$, which centers around $w_1 \approx 0.5$ --- where the empirical angular distribution is most concentrated. In contrast, predictions under the star-shaped density are consistently larger than those from the mixture in regions $B_1$ and $B_3$, which correspond to $w_1 \text{ near } 1 \text{ or } 0$, respectively, i.e., close to the axes on exponential margins. Specifically, star-shaped predictions are $5.6$ and $2.6$ times larger than mixture predictions for Redstone, and $4.5$ and $2.6$ for Friend Mountain, in $B_1$ and $B_3$, respectively. However, these differences are actually on the scale $10^{-5}$, so in terms of the original magnitude, the predictions are still similar.



\begin{table}[!htbp]
    \centering
\begin{tabular}[t]{llcccc}
\toprule
\multicolumn{3}{c}{ } & \multicolumn{3}{c}{Median ($95\%$ credible interval) $\times \,10^5$} \\
\cmidrule(lr){4-6}
Box & Likelihood & \shortstack{Ang.\\Dens.} & Pseudo-BMA & Pseudo-BMA+ & Stacking\\
\midrule
 &  & mix & 1.36 (0.702, 3.22) & 1.56 (0.928, 3.29) & 1.33 (0.676, 3.20)\\

 & \multirow{-2}{*}{\raggedright\arraybackslash Censored} & star & 6.08 (3.59, 8.84) & 6.08 (3.59, 8.84) & 6.02 (3.58, 8.68)\\
\cmidrule(lr){2-6}
 &  & mix & 4.26 (1.39, 13.5) & 4.65 (1.67, 13.4) & 3.01 (0.259, 17.7)\\

\multirow{-4}{*}[0.5\dimexpr\aboverulesep+\belowrulesep+\cmidrulewidth]{\raggedright\arraybackslash $B_1$} & \multirow{-2}{*}{\raggedright\arraybackslash Truncated}  & star & 6.49 (1.14, 21.8) & 6.53 (1.20, 22.0) & 6.64 (1.30, 21.8)\\
\specialrule{0.5pt}{2pt}{2pt}
 &  & mix & 7.59 (5.34, 10.5) & 7.69 (5.62, 10.4) & 7.57 (5.30, 10.5)\\

 & \multirow{-2}{*}{\raggedright\arraybackslash Censored} & star & 9.38 (4.52, 16.5) & 9.38 (4.52, 16.5) & 9.29 (4.60, 16.2)\\
\cmidrule(lr){2-6}
 &  & mix & 6.84 (2.63, 16.6) & 8.24 (3.09, 20.5) & 4.22 (0.583, 19.5)\\

\multirow{-4}{*}[0.5\dimexpr\aboverulesep+\belowrulesep+\cmidrulewidth]{\raggedright\arraybackslash $B_2$} & \multirow{-2}{*}{\raggedright\arraybackslash Truncated}& star & 5.71 (1.46, 17.6) & 5.67 (1.49, 17.5) & 5.71 (1.49, 17.6)\\
\specialrule{0.5pt}{2pt}{2pt}
 &  & mix & 15.1 (8.91, 29.6) & 17.2 (11.1, 30.8) & 14.9 (8.65, 29.4)\\

 & \multirow{-2}{*}{\raggedright\arraybackslash Censored} & star & 39.6 (24.0, 59.2) & 39.6 (24.0, 59.2) & 39.6 (24.4, 58.6)\\
\cmidrule(lr){2-6}
 &  & mix & 41.7 (15.3, 120.) & 44.9 (18.7, 119.) & 28.2 (3.36, 157.)\\

\multirow{-4}{*}[0.5\dimexpr\aboverulesep+\belowrulesep+\cmidrulewidth]{\raggedright\arraybackslash $B_3$} & \multirow{-2}{*}{\raggedright\arraybackslash Truncated}  & star & 45.8 (9.84, 132.) & 46.4 (10.5, 134.) & 47.1 (11.0, 134.)\\
\bottomrule
\end{tabular}
    \caption{Probability predictions for Friend Mountain RAWS in the three regions illustrated in the top row of Figure~\ref{fig:pred_task_boxes_fire}.}
    \label{tab:friendmtn_preds}
\end{table}

\begin{table}[!htbp]
    \centering
\begin{tabular}[t]{llcccc}
\toprule
\multicolumn{3}{c}{ } & \multicolumn{3}{c}{Median ($95\%$ credible interval) $\times \,10^5$} \\
\cmidrule(lr){4-6}
  Box & Likelihood & \shortstack{Ang.\\Dens.} & Pseudo-BMA & Pseudo-BMA+ & Stacking \\
\midrule
 &  & mix & 0.0543 (0.0258, 0.123) & 0.0876 (0.0438, 0.165) & 0.136 (0.0541, 0.274)\\

 & \multirow{-2}{*}{\raggedright\arraybackslash Censored} & star & 0.306 (0.100, 0.613) & 0.304 (0.100, 0.606) & 0.254 (0.0907, 0.494)\\
\cmidrule(lr){2-6}
 &  & mix & 6.01 (2.45, 13.3) & 6.14 (2.61, 13.0) & 4.84 (0.397, 19.2)\\

\multirow{-4}{*}[0.5\dimexpr\aboverulesep+\belowrulesep+\cmidrulewidth]{\raggedright\arraybackslash $B_1$} & \multirow{-2}{*}{\raggedright\arraybackslash Truncated}& star & 6.17 (1.38, 18.4) & 6.60 (1.56, 18.2) & 5.49 (0.631, 19.6)\\
\specialrule{0.5pt}{2pt}{2pt}
 &  & mix & 6.91 (5.08, 8.88) & 7.92 (5.93, 10.2) & 9.81 (6.92, 13.9)\\

 & \multirow{-2}{*}{\raggedright\arraybackslash Censored} & star & 13.3 (8.06, 20.7) & 13.2 (8.03, 20.6) & 11.7 (7.60, 17.6)\\
\cmidrule(lr){2-6}
 &  & mix & 64.6 (25.3, 159.) & 74.4 (29.3, 193.) & 27.4 (7.72, 77.7)\\

\multirow{-4}{*}[0.5\dimexpr\aboverulesep+\belowrulesep+\cmidrulewidth]{\raggedright\arraybackslash $B_2$} & \multirow{-2}{*}{\raggedright\arraybackslash Truncated} & star & 33.5 (11.8, 93.7) & 33.2 (12.6, 88.1) & 34.2 (9.07, 107.)\\
\specialrule{0.5pt}{2pt}{2pt}
 &  & mix & 2.67 (1.50, 5.25) & 3.50 (2.05, 5.98) & 4.78 (2.38, 8.35)\\

 & \multirow{-2}{*}{\raggedright\arraybackslash Censored} & star & 7.01 (2.97, 12.4) & 6.96 (2.96, 12.3) & 6.10 (2.88, 10.3)\\
\cmidrule(lr){2-6}
 &  & mix & 101. (50.2, 185.) & 102. (53.7, 182.) & 83.4 (14.2, 234.)\\

\multirow{-4}{*}[0.5\dimexpr\aboverulesep+\belowrulesep+\cmidrulewidth]{\raggedright\arraybackslash $B_3$} & \multirow{-2}{*}{\raggedright\arraybackslash Truncated} & star & 89.6 (26.3, 214.) & 93.1 (29.4, 210.) & 84.2 (14.4, 228.)\\
\bottomrule
\end{tabular}
    \caption{Probability predictions for Redstone RAWS in the three regions illustrated in the bottom row of Figure~\ref{fig:pred_task_boxes_fire}.}
    \label{tab:redstone_preds}
\end{table}

\section{Discussion}\label{sec:discussion}
Using BMA, we have introduced flexibility into a parametric modeling framework, providing an alternative to more complex semi-parametric approaches. This also allows us to bypass the uncertainty and potential bias associated with model selection. Incorporating the censored likelihood led to superior predictive performance in nearly all simulation settings---and consistently better results with real data---highlighting the value of the censored approach in small sample size scenarios. Yet the increase in bias that potentially results from the censored approach should be considered in the context of the application. While we were initially concerned that the star-shaped angular density may be overly rigid, it performed quite well, likely due to BMA.

One might conjecture that averaging across the factorial design of angular models and gauge functions, rather than reporting results from each angular model separately as we have done here, might lead to improved prediction performance.  This seemed promising, but in practice the difference was marginal (see Supplementary Material ~\ref{append:sim_study_BMA_angular}).

We were somewhat surprised by the poorer performance of stacking weights, especially under the H\"usler-Reiss dependence structure setting, since stacking should work well when the data-generating process is not included in the suite of models being averaged over. While stacking is asymptotically optimal when the data-generating model is not under consideration, this advantage may not hold in finite sample scenarios such as ours. 

Extensions to higher dimensions present some challenges within our framework. Each increase in dimension introduces more parameters for the gauge functions; as an extreme example, the asymmetric logistic model has 128 possible forms in three dimensions. The mixture angular model would generalize to a Dirichlet mixture, but the effectiveness of the stick-breaking process in higher dimensions is unclear. Despite these concerns, we plan to explore this framework in three dimensions.

Assumption~\ref{assump:w_above_quant_thres} may break down in small sample sizes. One way to address this would be to fit angular densities only to angles exceeding a high radial threshold. If a marginal threshold is used initially, this approach would require fitting the radial model to then identify the quantile-based radial threshold $r_q(\bw)$, above which the associated angles are used to fit the angular model.

Currently, the marginal transformation, radial likelihoods, and angular likelihoods are modeled independently, even though the latter two depend on the transformation. Future work could focus on a hierarchical model that jointly incorporates all three components, allowing for information sharing across model components and accounting for uncertainty propagated forward by the marginal transformation.

Finally, although we attempted to account for the varying exceedance ratios (Figure~\ref{fig:exceed_ratio}) with self-normalized importance sampling, the BMA framework itself suggests another possibility. \cite{yao2022} propose hierarchical stacking weights that vary with covariates. In our framework, this would entail allowing the stacking weights to depend on the angular values. Implementation requires discretization of the angular space, perhaps with groupings informed by the exceedance ratios. The hierarchical weights are then estimated in a Bayesian model using the discretized angles and the leave-one-out likelihoods as inputs. Though this approach is still new to us, it presents a promising direction for future research.

\bibliographystyle{apalike}
\bibliography{references}

\appendix
\section{Asymptotic approximation}\label{append:asym_approx_results}
\renewcommand{\thefigure}{A\arabic{figure}}
\renewcommand{\thetable}{A\arabic{table}}
\setcounter{figure}{0}
\setcounter{table}{0}
As described in Section~\ref{sec:radial}, we conducted an experiment to determine how well the gamma approximation performs in finite settings. Recall that data were generated according to Section~\ref{sec:sim_study} and fit using the data-generating gauge function, looking at the posterior estimates of the unknown dependence parameter compared with the truth. For ease of reference, we again provide the two figures we showed previously, in this supplement, along with the remaining figures. See Section~\ref{sec:radial} for interpretation.

\subsection{Gaussian joint density}
\begin{figure}[!htbp]
    \centering
    \includegraphics[width=0.5\linewidth]{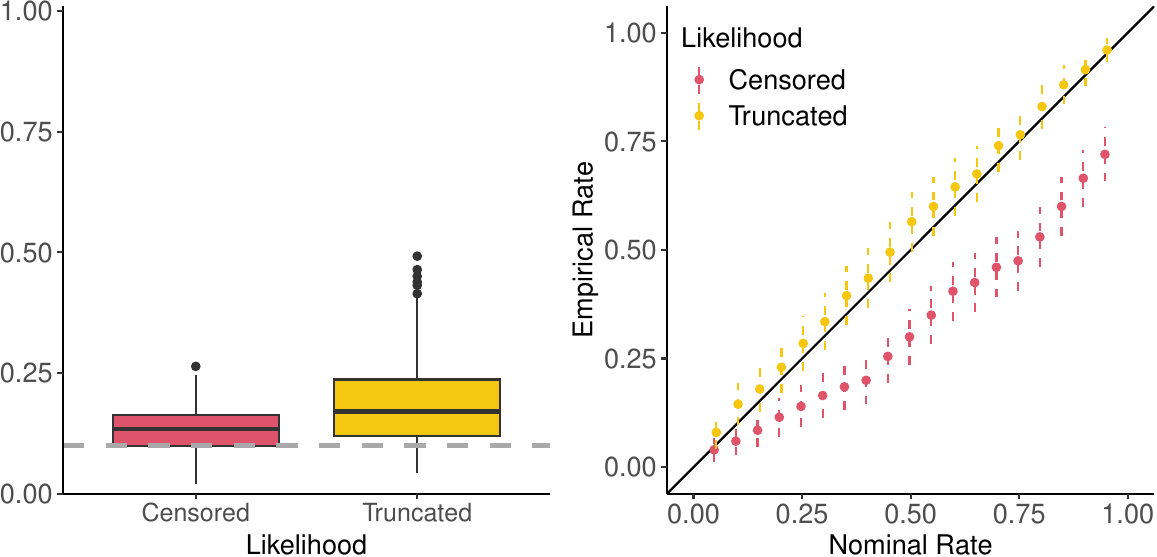}
    \caption{Gaussian dependence, $\rho = 0.1$. Grey dotted line is the true parameter value.}
    \label{fig:gauss_low_cov}
\end{figure}

\begin{figure}[!htbp]
    \centering
    \includegraphics[width=0.5\linewidth]{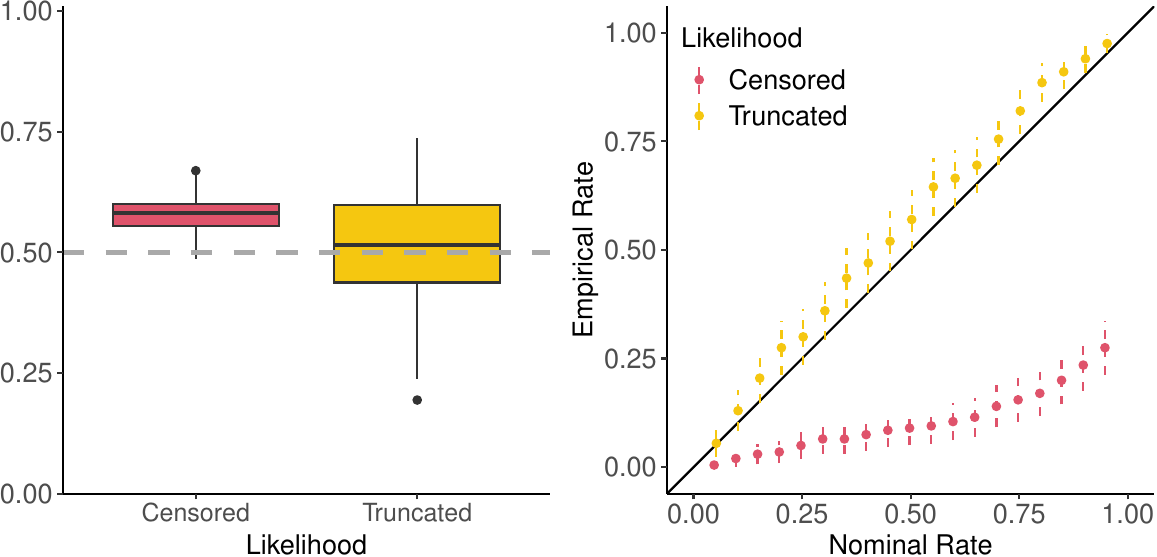}
    \caption{Gaussian dependence, $\rho = 0.5$. This plot is the same as Figure~\ref{fig:gauss_coverage}, but repeated here for convenience.}
    \label{fig:gauss_mid_cov}
\end{figure}

\begin{figure}[!htbp]
    \centering
    \includegraphics[width=0.5\linewidth]{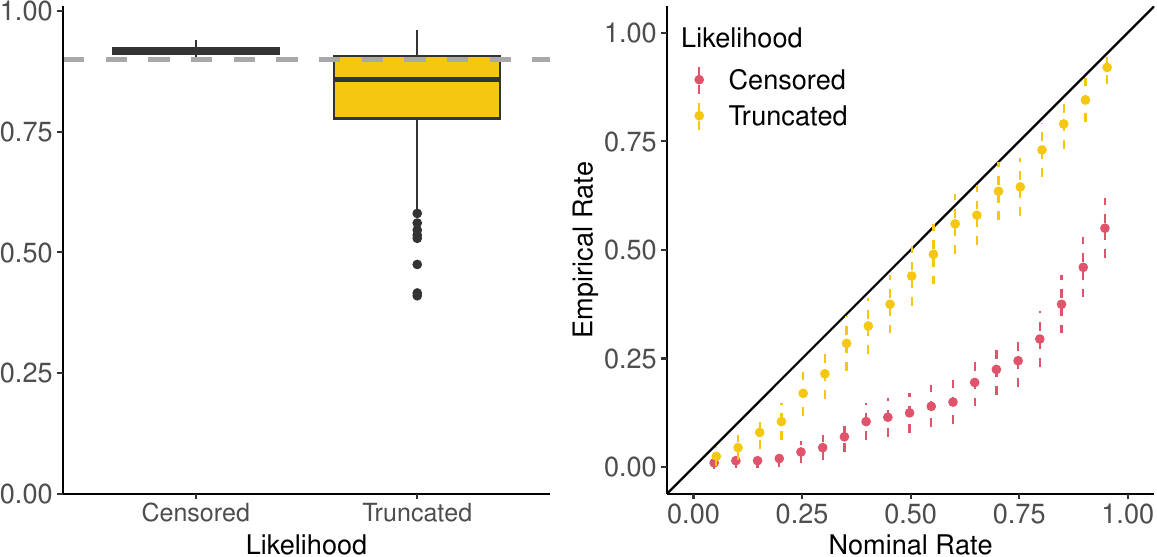}
    \caption{Gaussian dependence, $\rho = 0.9$.}
    \label{fig:gauss_high_cov}
\end{figure}
\clearpage
\subsection{Logistic joint density}
\begin{figure}[!htbp]
    \centering
    \includegraphics[width=0.5\linewidth]{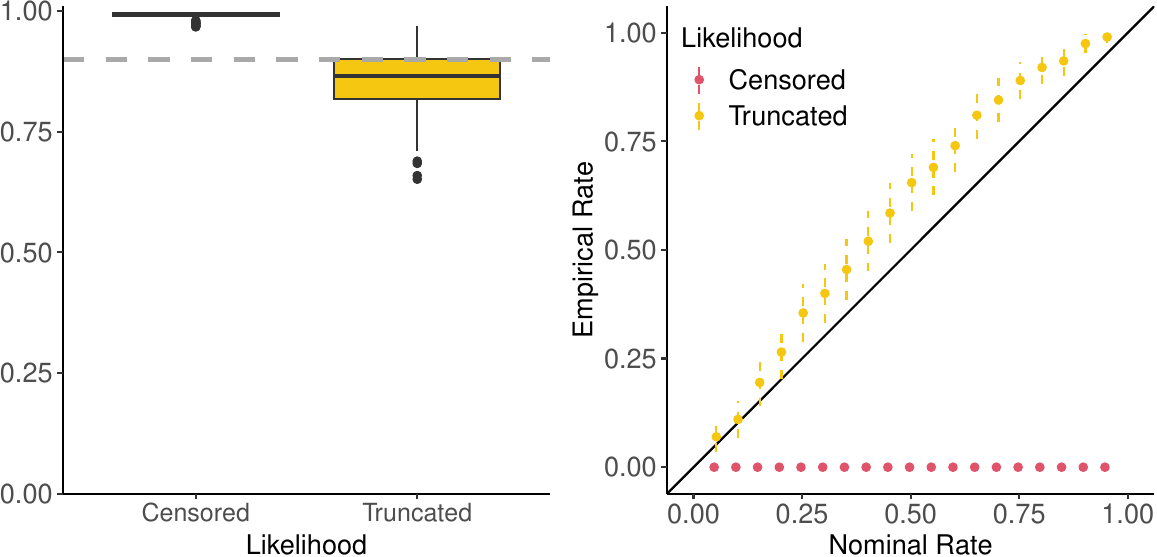}
    \caption{Logistic dependence, $\rho = 0.1$. Grey dotted line is the true parameter value.}
    \label{fig:logistic_low_cov}
\end{figure}

\begin{figure}[!htbp]
    \centering
    \includegraphics[width=0.5\linewidth]{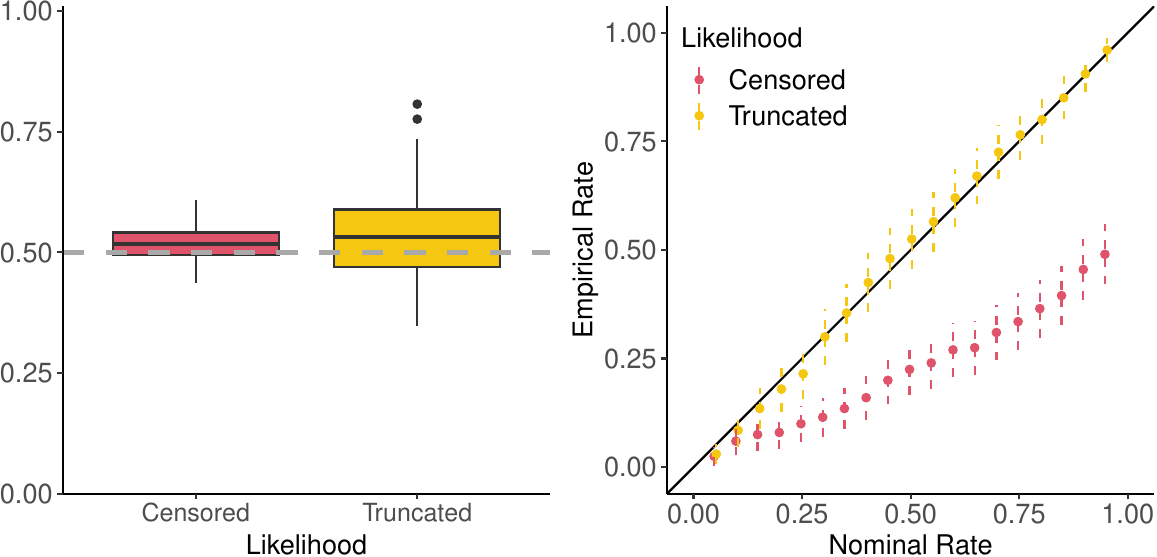}
    \caption{Logistic dependence, $\rho = 0.5$. This plot is the same as Figure~\ref{fig:logistic_coverage}, but repeated here for convenience.}
    \label{fig:logistic_mid_cov}
\end{figure}

\begin{figure}[!htbp]
    \centering
    \includegraphics[width=0.5\linewidth]{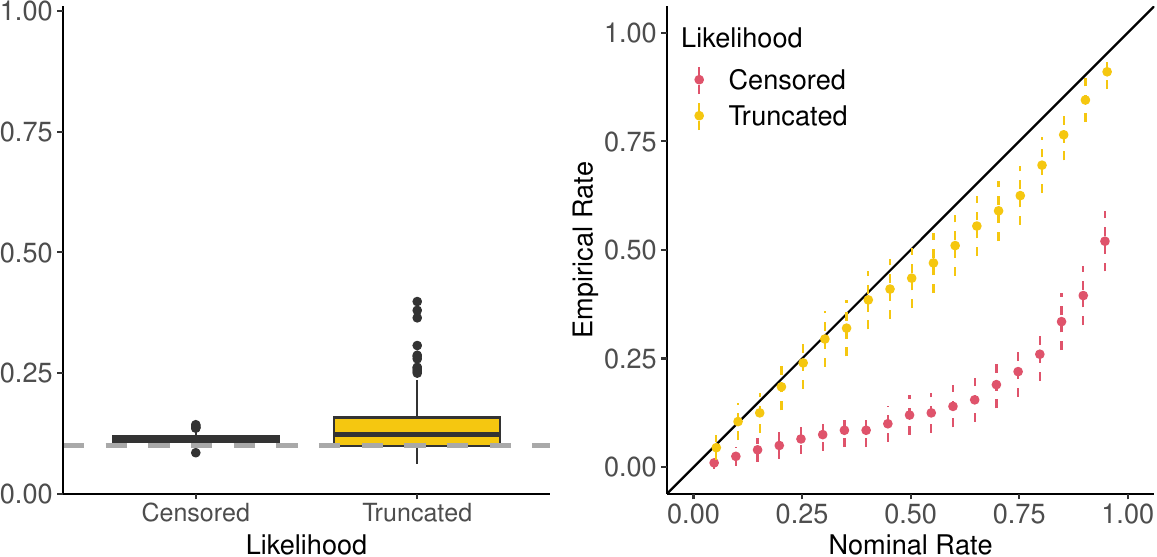}
    \caption{Logistic dependence, $\rho = 0.9$.}
    \label{fig:logistic_high_cov}
\end{figure}

\clearpage
\section{Simulation study results}\label{append:sim_study_results}
\renewcommand{\thefigure}{B\arabic{figure}}
\renewcommand{\thetable}{B\arabic{table}}
\setcounter{figure}{0}
\setcounter{table}{0}

\begin{figure}[!htbp]
    \centering
    \includegraphics[width=0.95\linewidth]{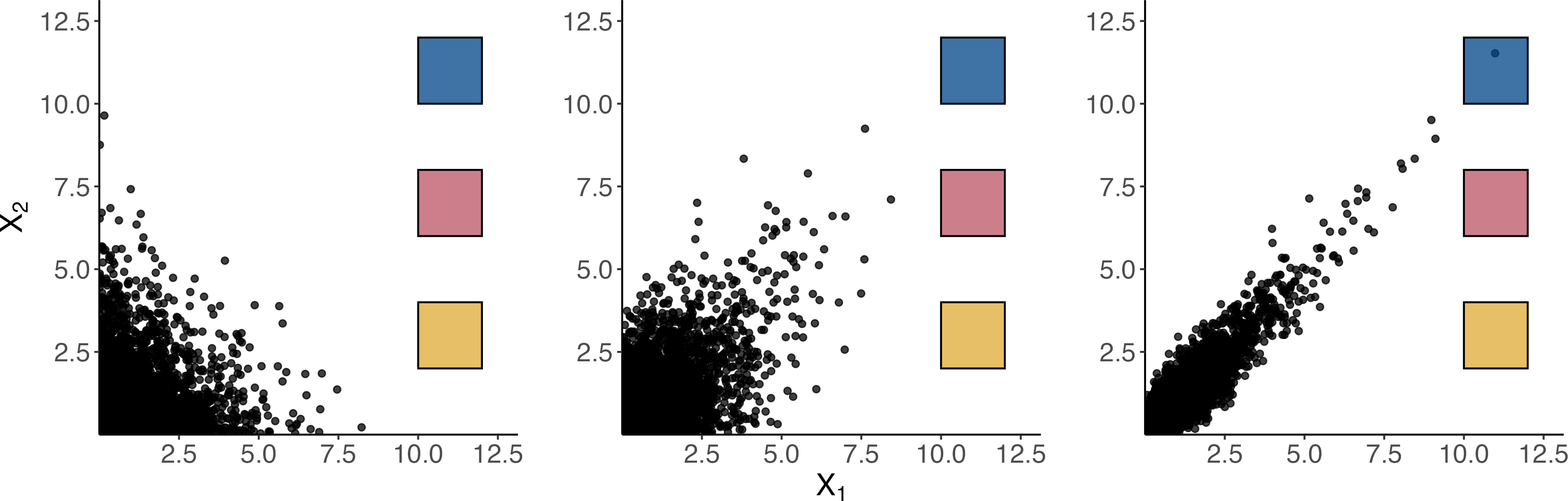}
    \caption{Prediction regions under the simulation study: $B_1=(10,12)\times(10,12)$, $B_2=(10,12)\times(6,8)$, and $B_3=(10,12)\times(2,4)$, represented by the blue, red, and yellow boxes, respectively. Levels of dependence $\lambda = 0.1, 1, \text{and }3$, corresponding to low, mid, and high, are shown from left to right. Data are generated according to \ref{sec:data_gen}.}
    \label{fig:hr_pred_task}
\end{figure}

In Section~\ref{sec:sim_study}, we outline the dependence structures and levels we examine and give a visualization of the Gaussian and logistic datasets (Figure~\ref{fig:sim_pred_task}). For reference throughout this section, Figure~\ref{fig:hr_pred_task} visualizes the three dependence levels we consider for the H\"usler-Reiss structure.

\subsection{Distributions of predictions (boxplots)}\label{append:pred_boxplots}
\subsubsection{Gaussian dependence structures}
\begin{figure}[!htbp]
    \centering
    \includegraphics[width=0.65\linewidth]{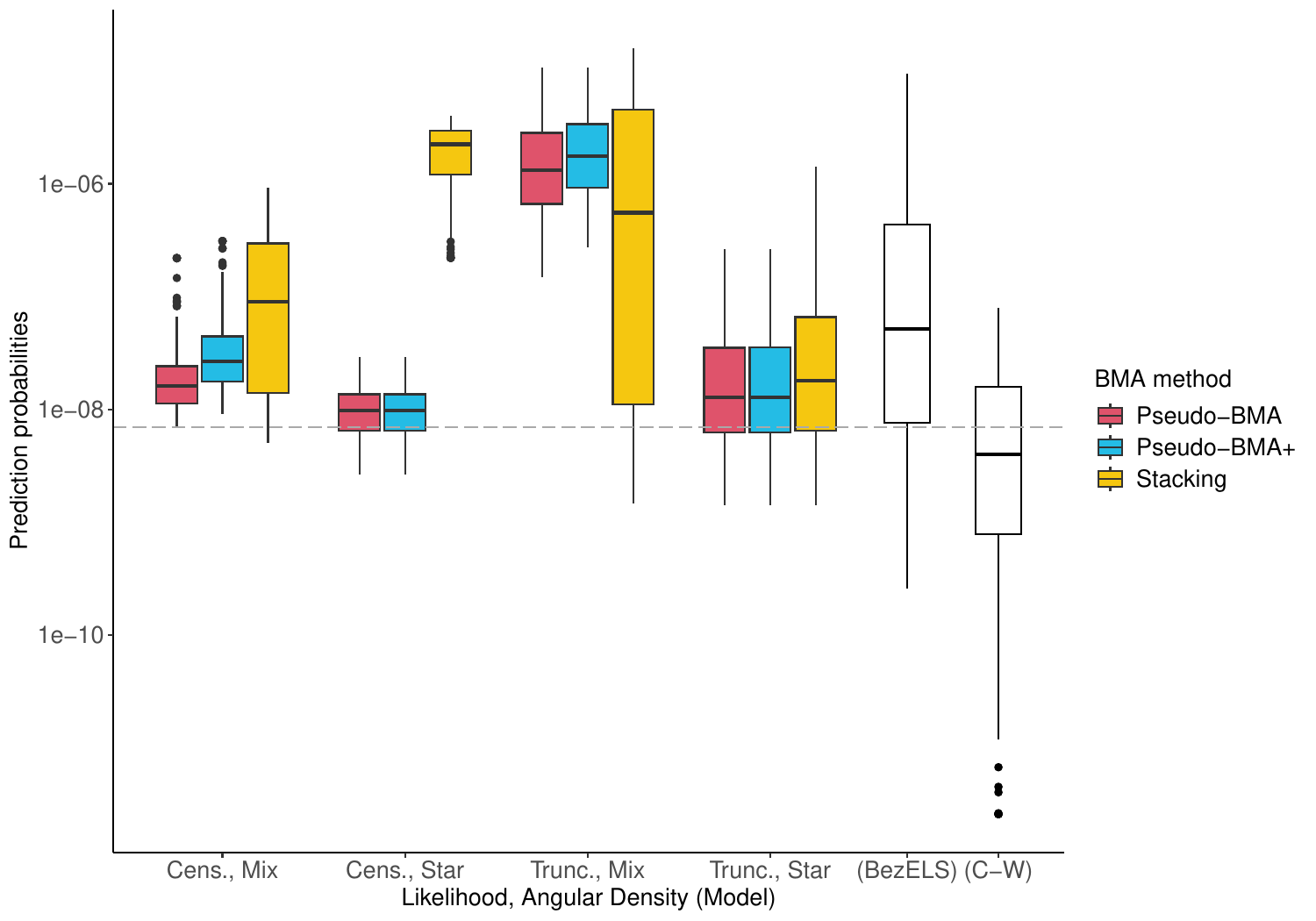}
    \caption{Gaussian dependence, with $\rho = 0.1$. Predictions in $B_1=(10,12)\times(10,12)$}
    \label{fig:gauss_low_b1}
\end{figure}

\begin{figure}[!htbp]
    \centering
    \includegraphics[width=0.65\linewidth]{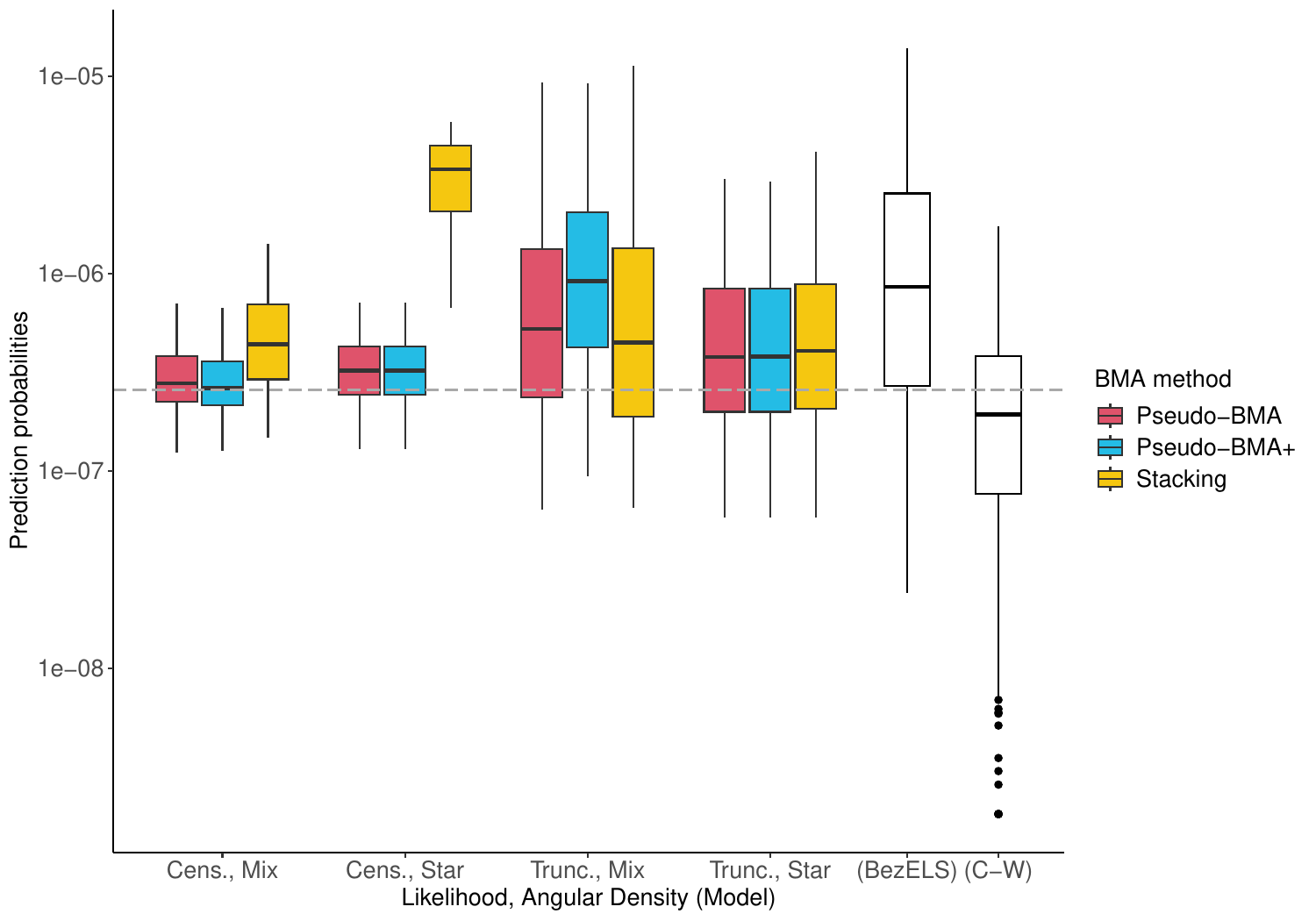}
    \caption{Gaussian dependence, with $\rho = 0.1$. Predictions in $B_2=(10,12)\times(6,8)$}
    \label{fig:gauss_low_b2}
\end{figure}

\begin{figure}[!htbp]
    \centering
    \includegraphics[width=0.65\linewidth]{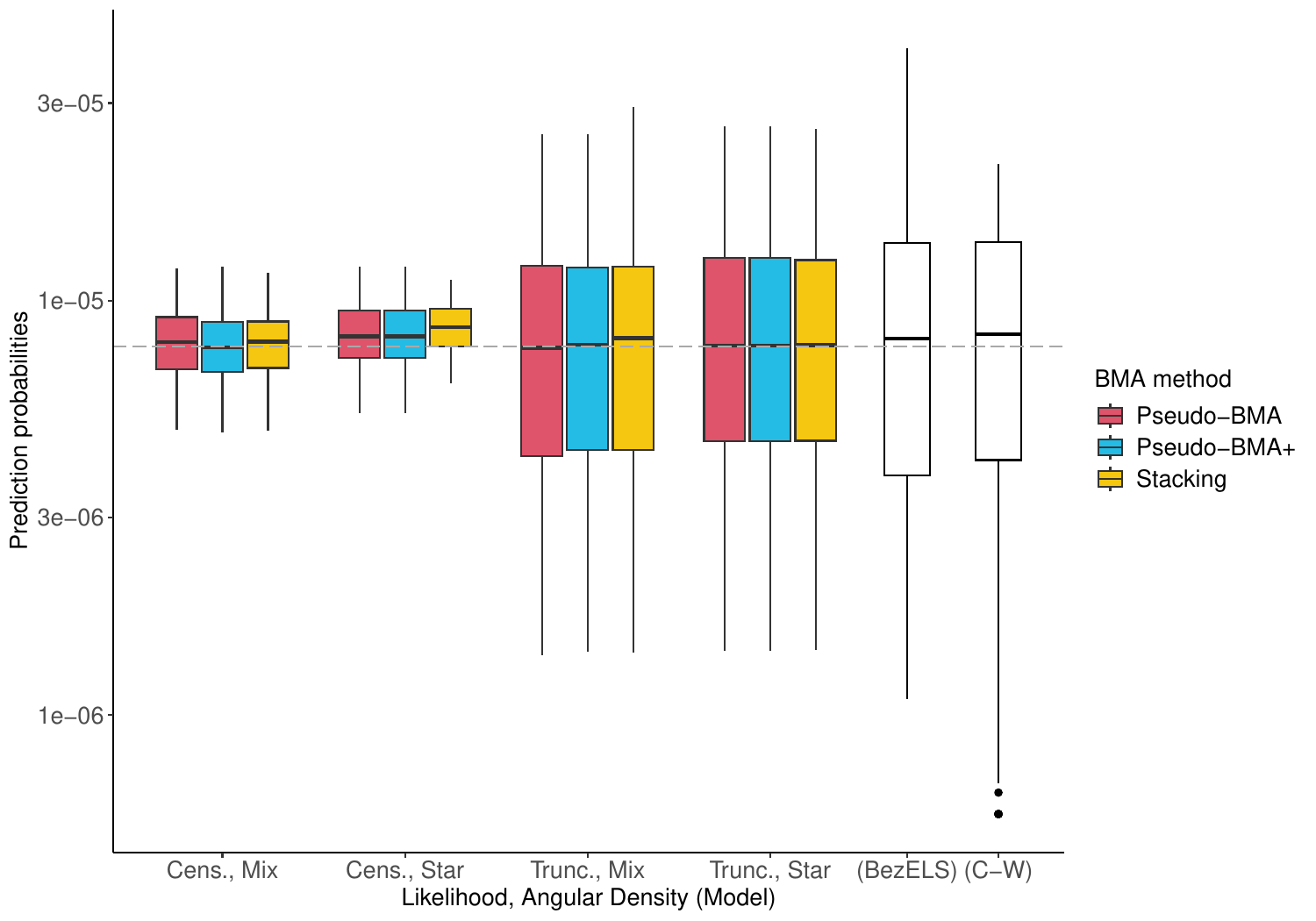}
    \caption{Gaussian dependence, with $\rho = 0.1$. Predictions in $B_3=(10,12)\times(2,4)$}
    \label{fig:gauss_low_b3}
\end{figure}

\begin{figure}[!htbp]
    \centering
    \includegraphics[width=0.65\linewidth]{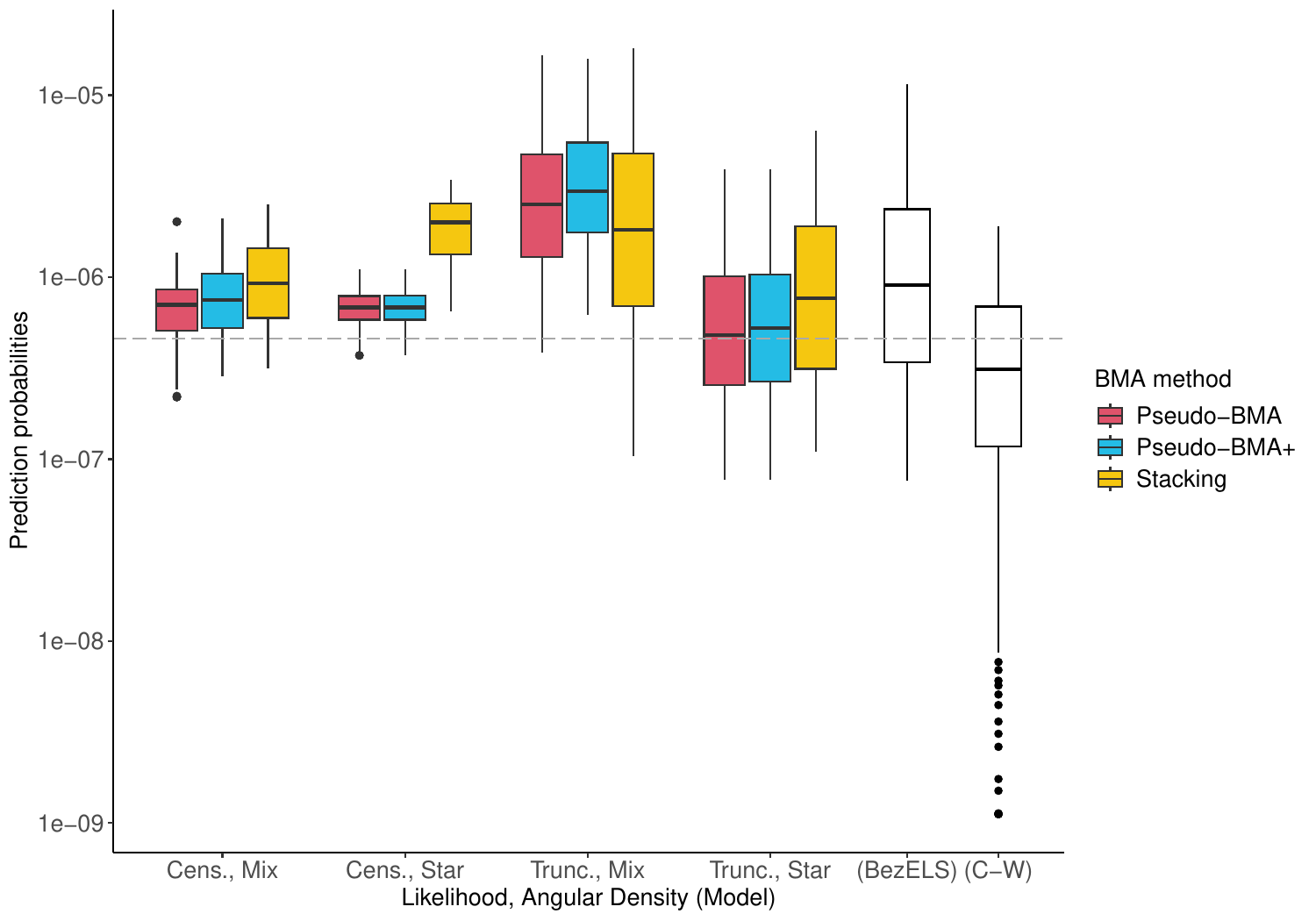}
    \caption{Gaussian dependence, with $\rho = 0.5$. Predictions in $B_1=(10,12)\times(10,12)$}
    \label{fig:gauss_mid_b1}
\end{figure}

\begin{figure}[!htbp]
    \centering
    \includegraphics[width=0.65\linewidth]{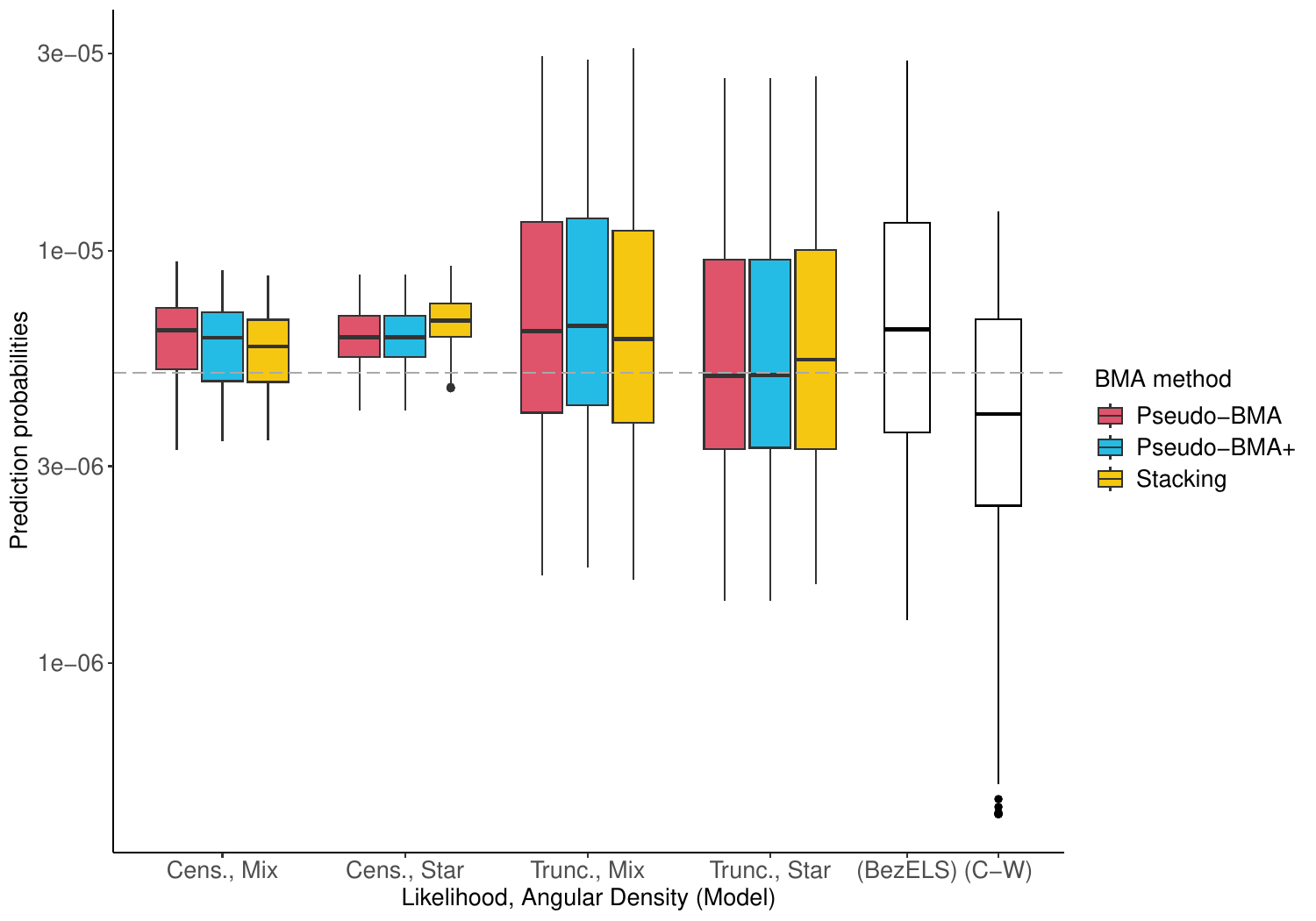}
    \caption{Gaussian dependence, with $\rho = 0.5$. Predictions in $B_2=(10,12)\times(6,8)$}
    \label{fig:gauss_mid_b2}
\end{figure}

\begin{figure}[!htbp]
    \centering
    \includegraphics[width=0.65\linewidth]{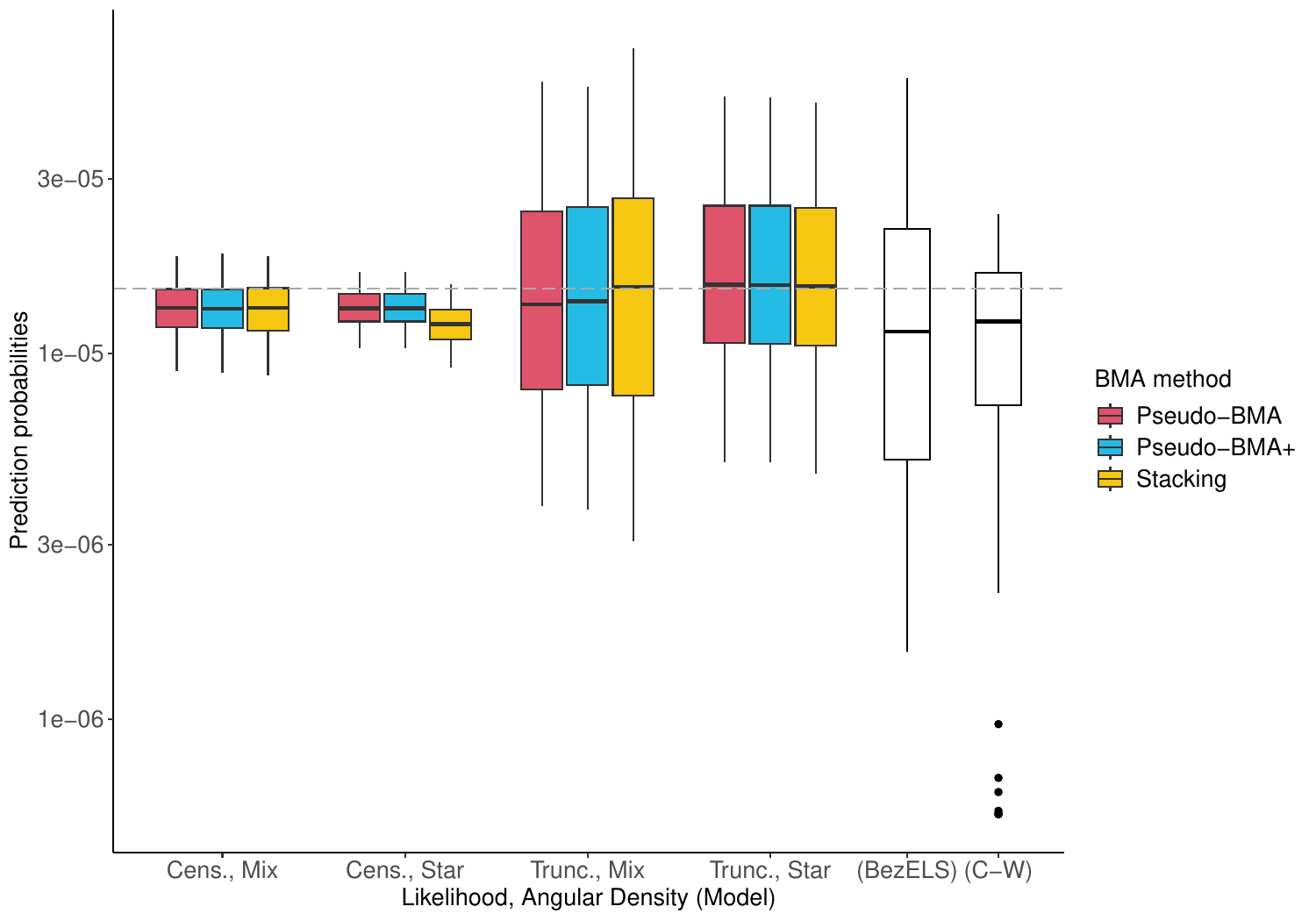}
    \caption{Gaussian dependence, with $\rho = 0.5$. Predictions in $B_3=(10,12)\times(2,4)$}
    \label{fig:gauss_mid_b3}
\end{figure}

\begin{figure}[!htbp]
    \centering
    \includegraphics[width=0.65\linewidth]{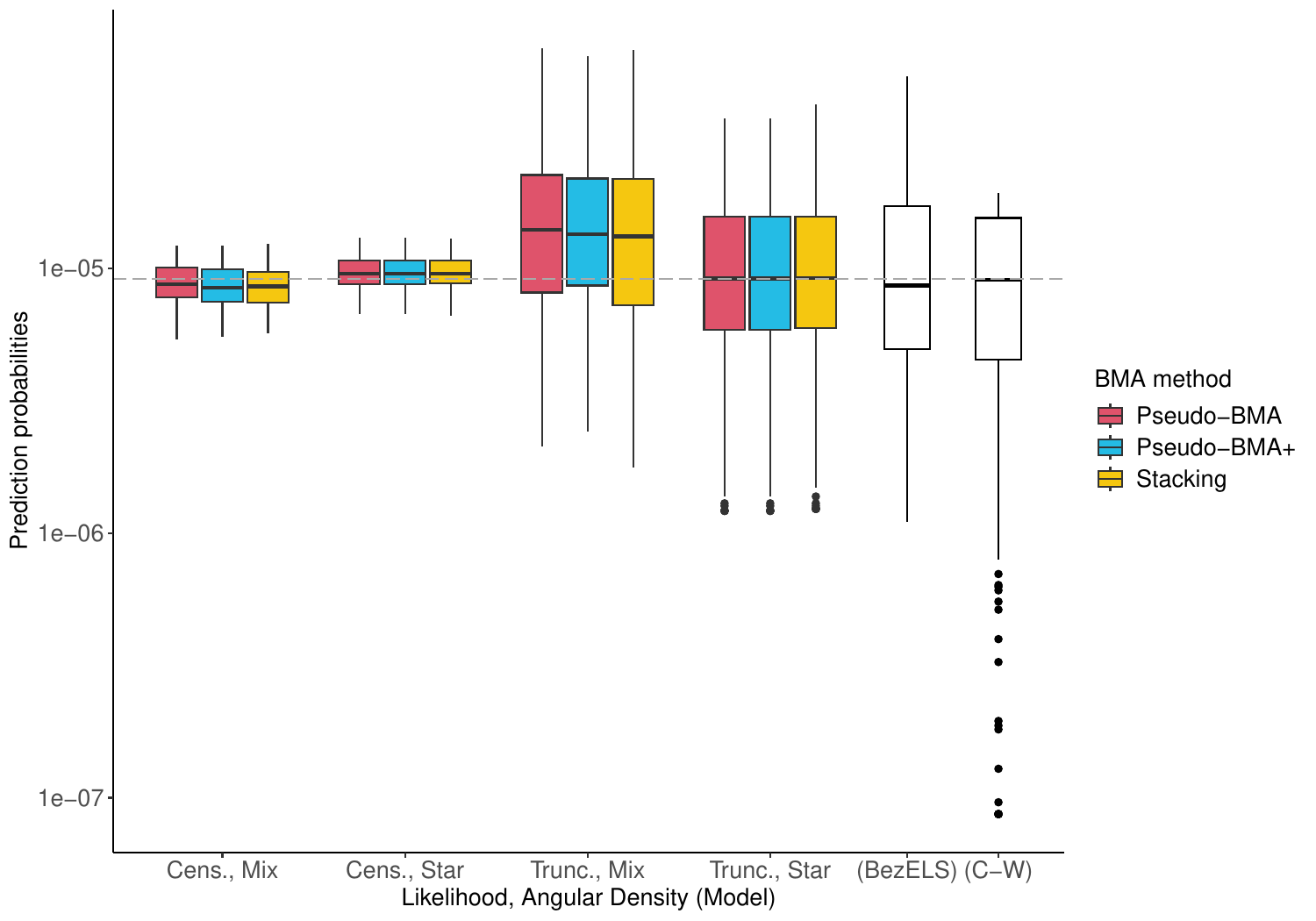}
    \caption{Gaussian dependence, with $\rho = 0.9$. Predictions in $B_1=(10,12)\times(10,12)$}
    \label{fig:gauss_high_b1}
\end{figure}

\begin{figure}[!htbp]
    \centering
    \includegraphics[width=0.65\linewidth]{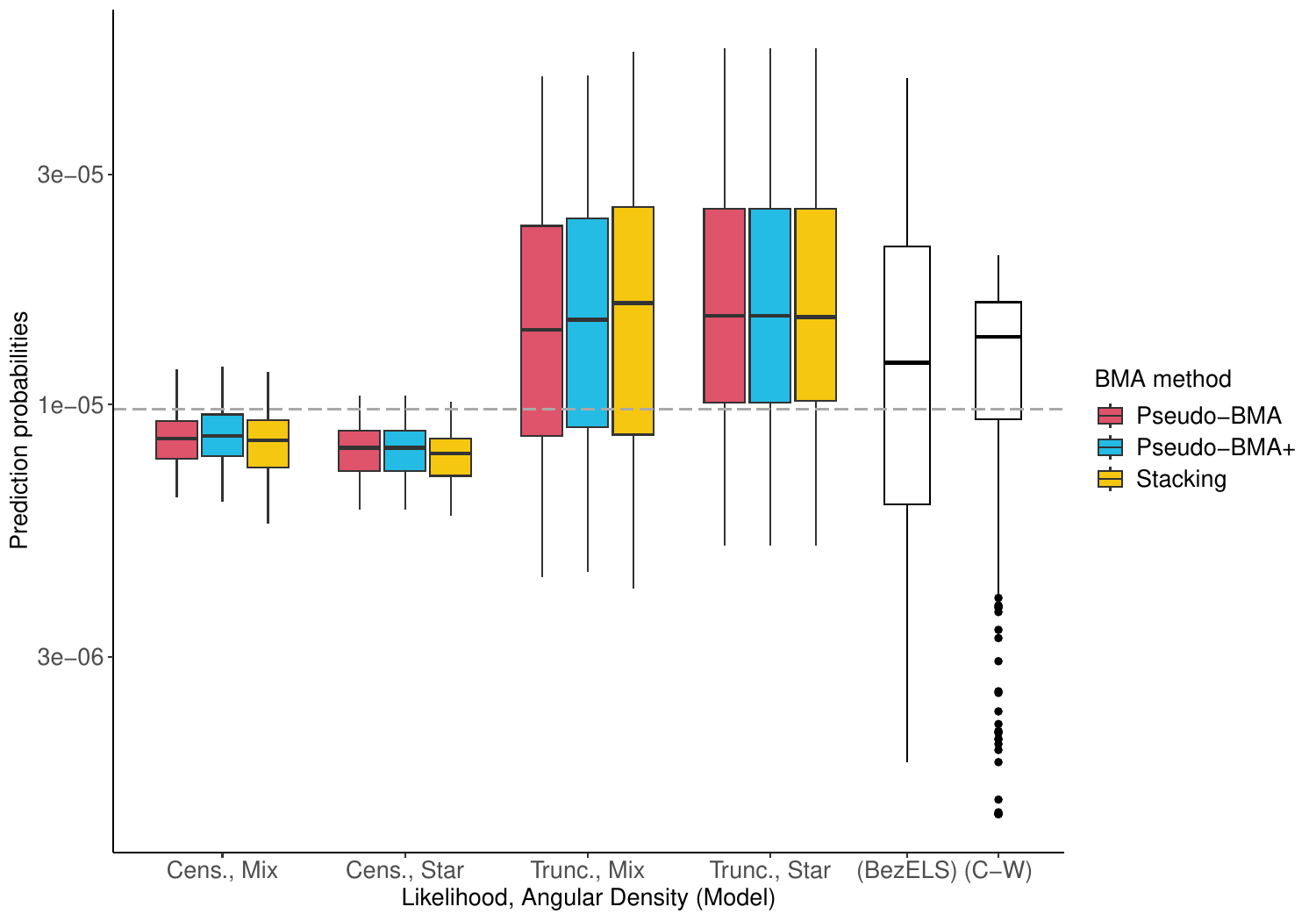}
    \caption{Gaussian dependence, with $\rho = 0.9$. Predictions in $B_2=(10,12)\times(6,8)$}
    \label{fig:gauss_high_b2}
\end{figure}

\begin{figure}[!htbp]
    \centering
    \includegraphics[width=0.65\linewidth]{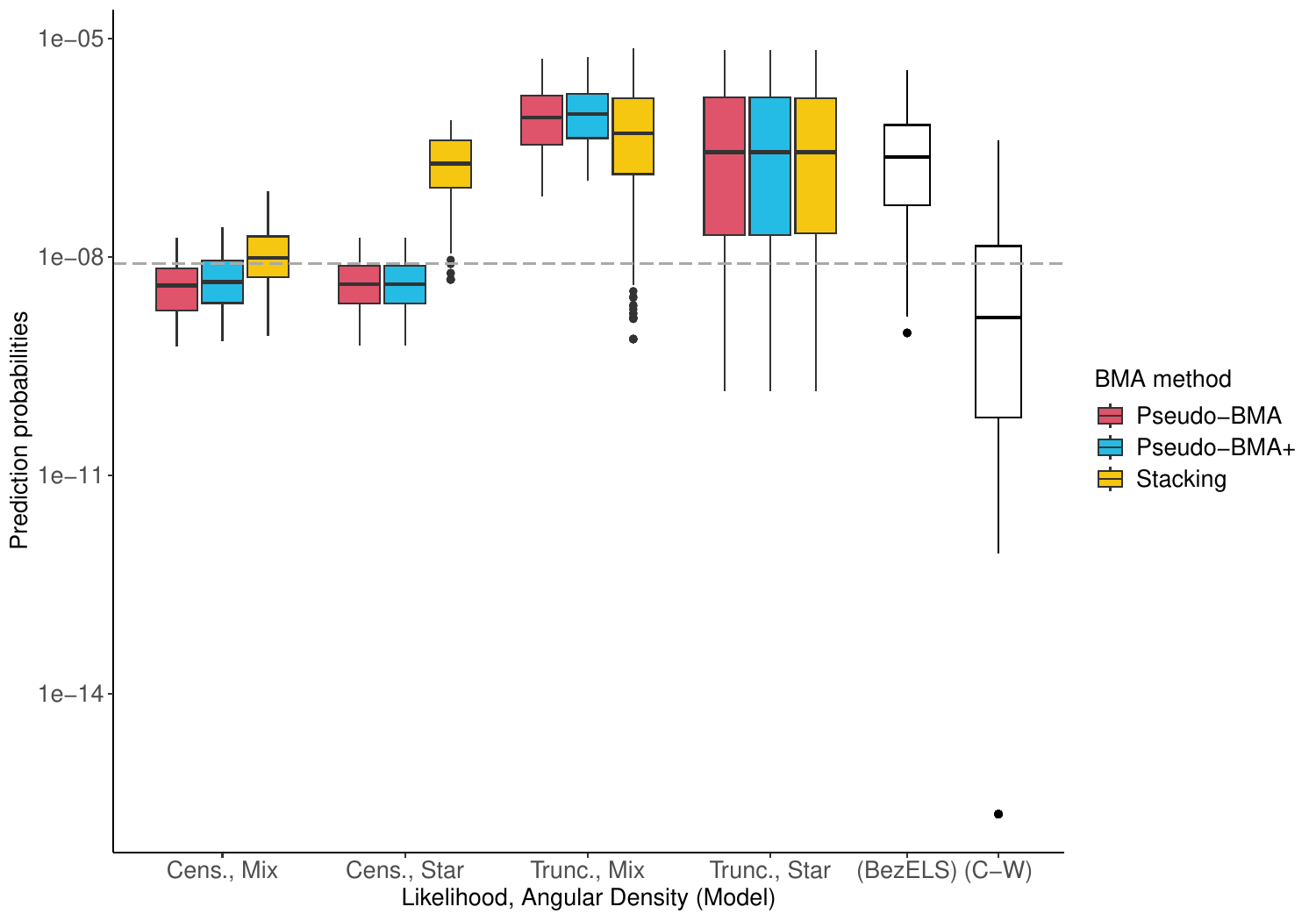}
    \caption{Gaussian dependence, with $\rho = 0.9$. Predictions in $B_3=(10,12)\times(2,4)$}
    \label{fig:gauss_high_b3}
\end{figure}
\clearpage
\subsubsection{Logistic dependence structures}
\begin{figure}[!htbp]
\centering
\includegraphics[width=0.65\linewidth]{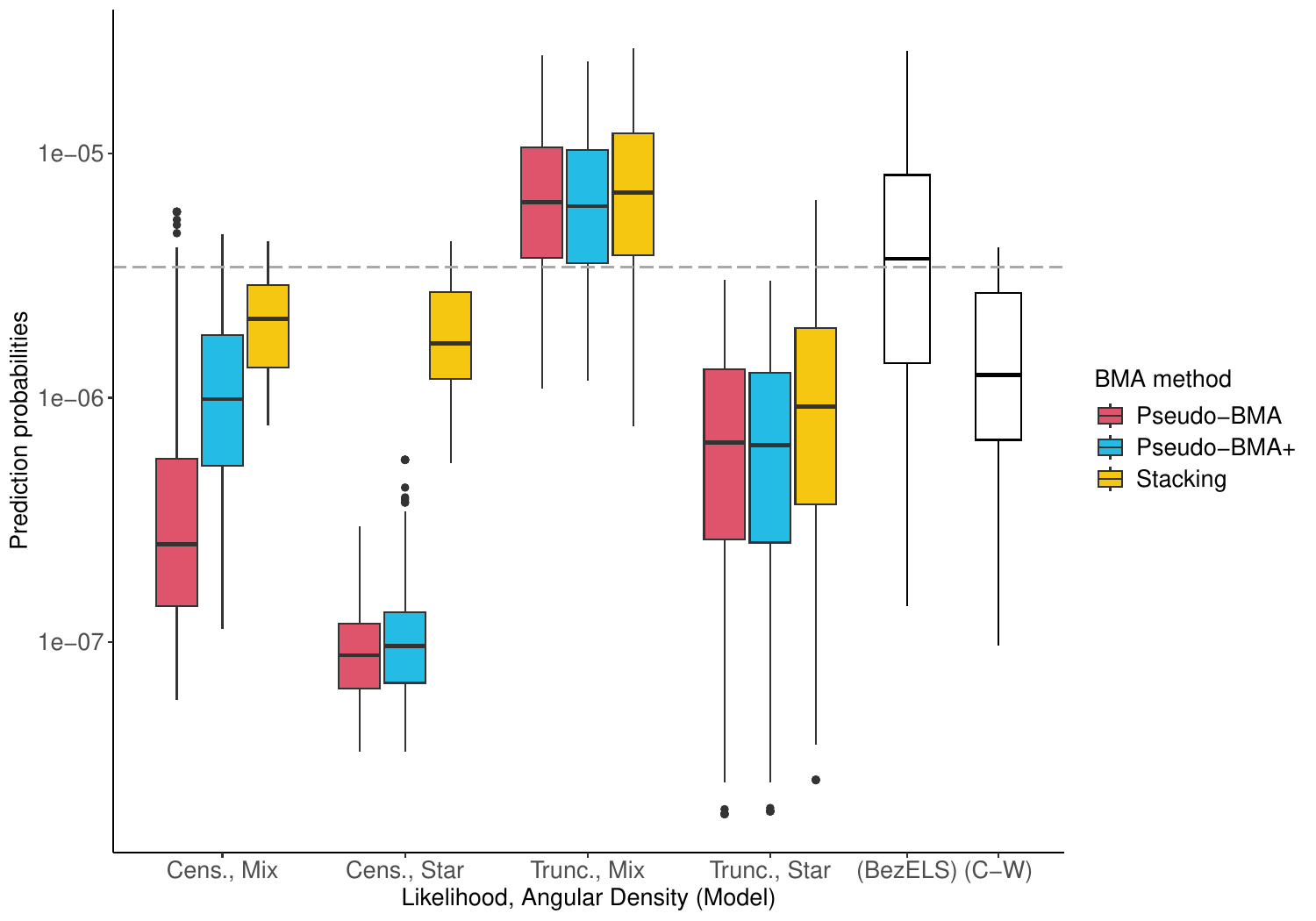}
\caption{Logistic dependence, with $\lambda = 0.9$. Predictions in $B_1=(10,12)\times(10,12)$}
\label{fig:logistic_low_b1}
\end{figure}

\begin{figure}[!htbp]
\centering
\includegraphics[width=0.65\linewidth]{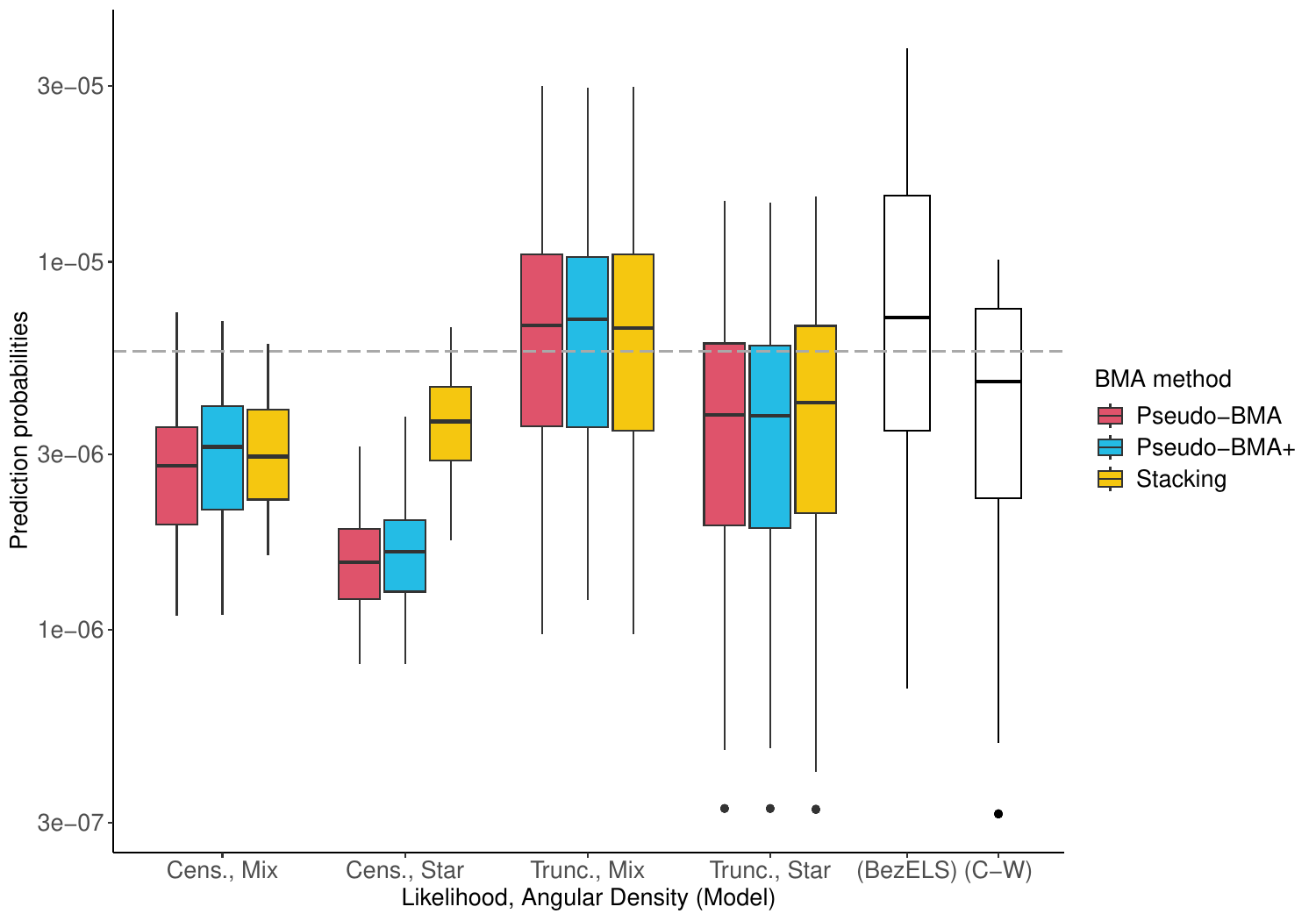}
\caption{Logistic dependence, with $\lambda = 0.9$. Predictions in $B_2=(10,12)\times(6,8)$}
\label{fig:logistic_low_b2}
\end{figure}

\begin{figure}[!htbp]
\centering
\includegraphics[width=0.65\linewidth]{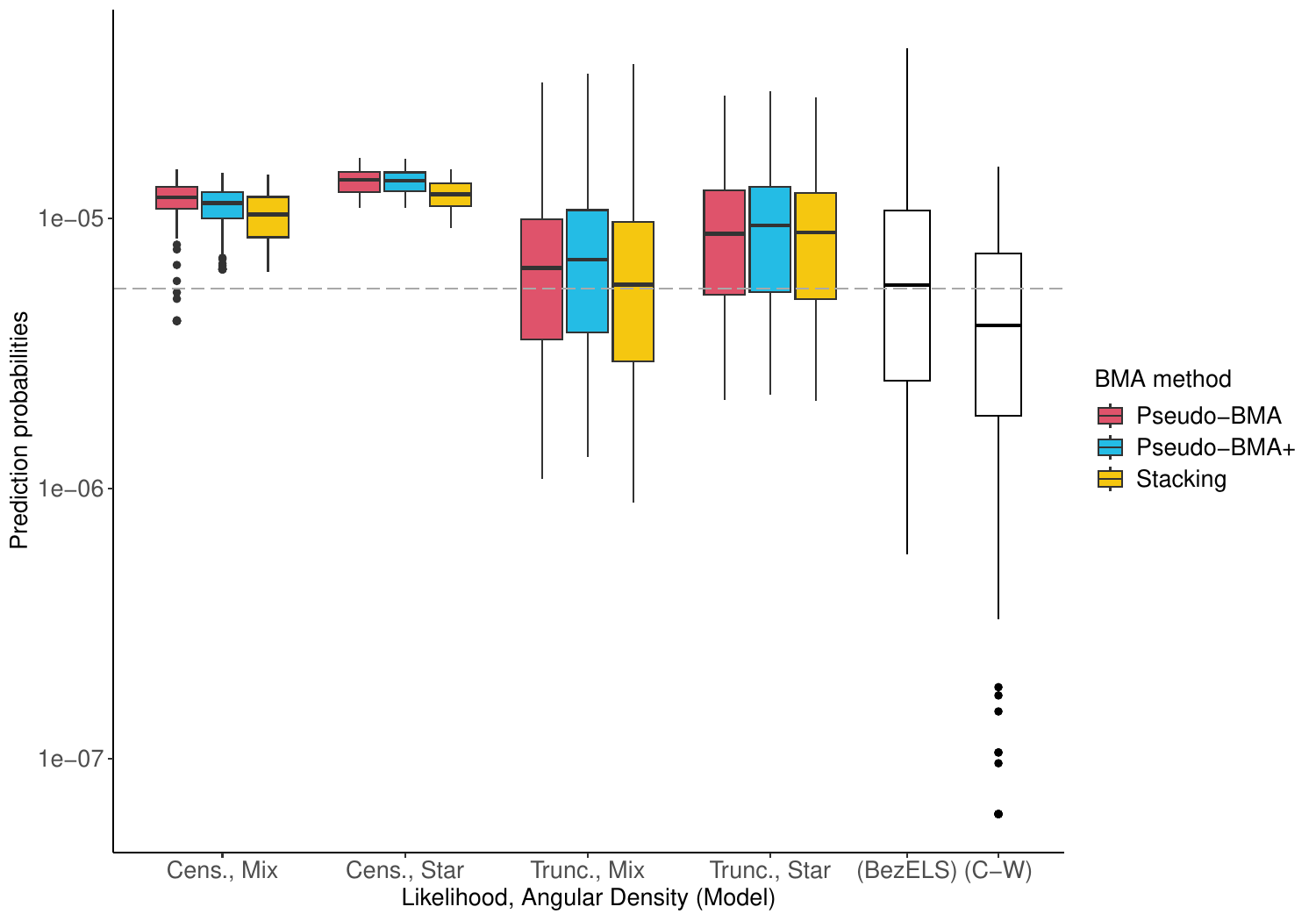}
\caption{Logistic dependence, with $\lambda = 0.9$. Predictions in $B_3=(10,12)\times(2,4)$}
\label{fig:logistic_low_b3}
\end{figure}

\begin{figure}[!htbp]
\centering
\includegraphics[width=0.65\linewidth]{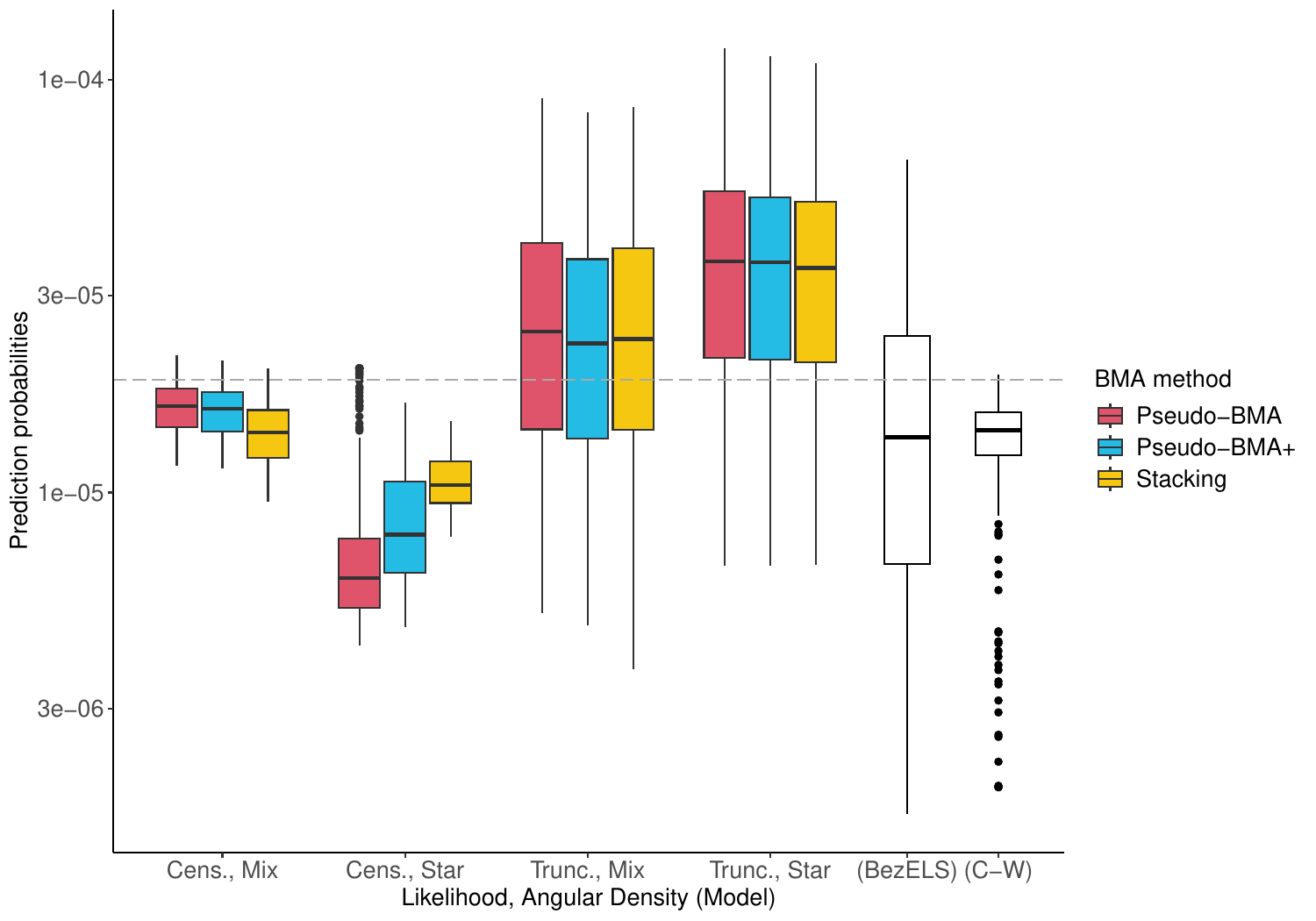}
\caption{Logistic dependence, with $\lambda = 0.5$. Predictions in $B_1=(10,12)\times(10,12)$}
\label{fig:logistic_mid_b1}
\end{figure}

\begin{figure}[!htbp]
\centering
\includegraphics[width=0.65\linewidth]{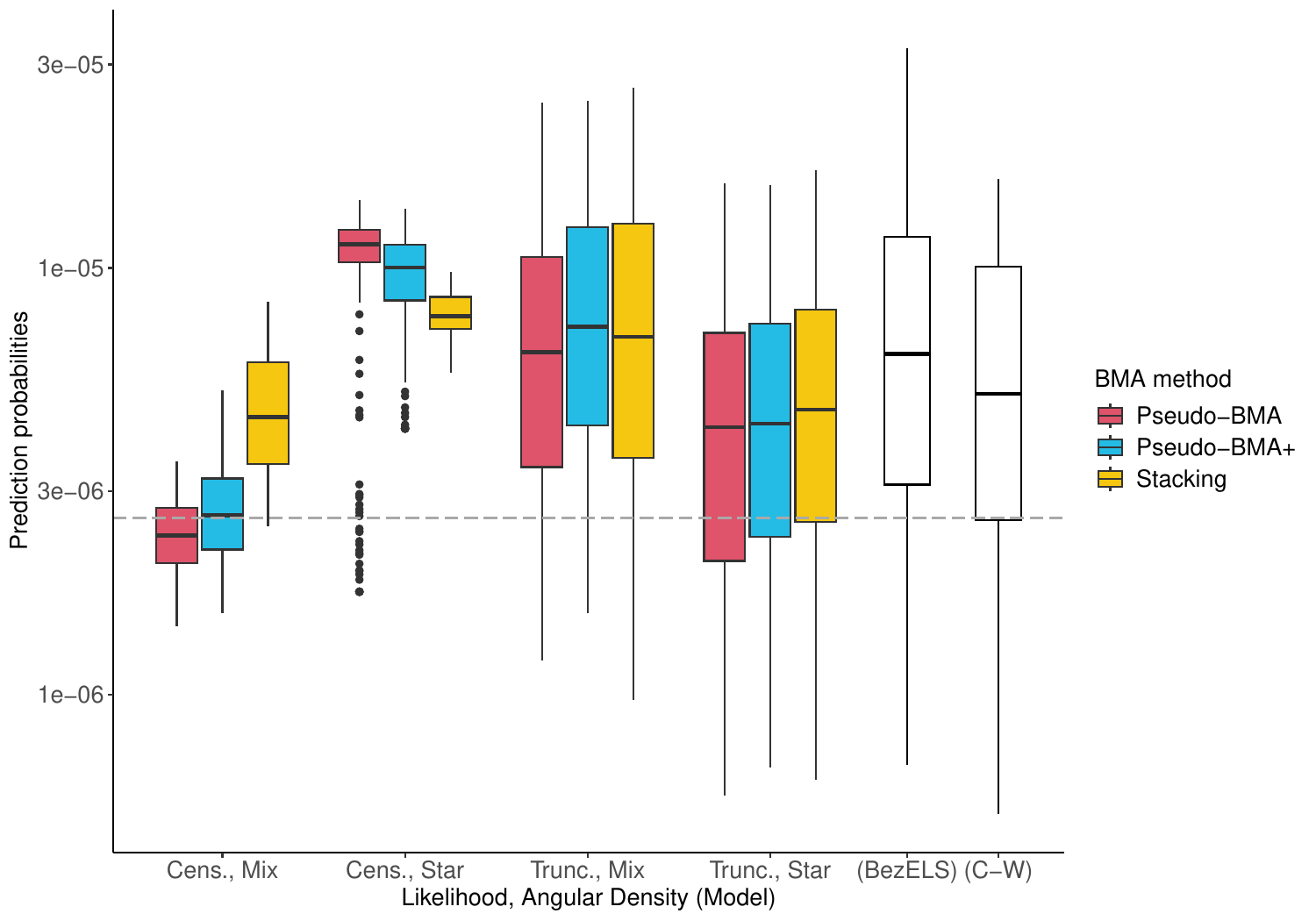}
\caption{Logistic dependence, with $\lambda = 0.5$. Predictions in $B_2=(10,12)\times(6,8)$}
\label{fig:logistic_mid_b2}
\end{figure}

\begin{figure}[!htbp]
\centering
\includegraphics[width=0.65\linewidth]{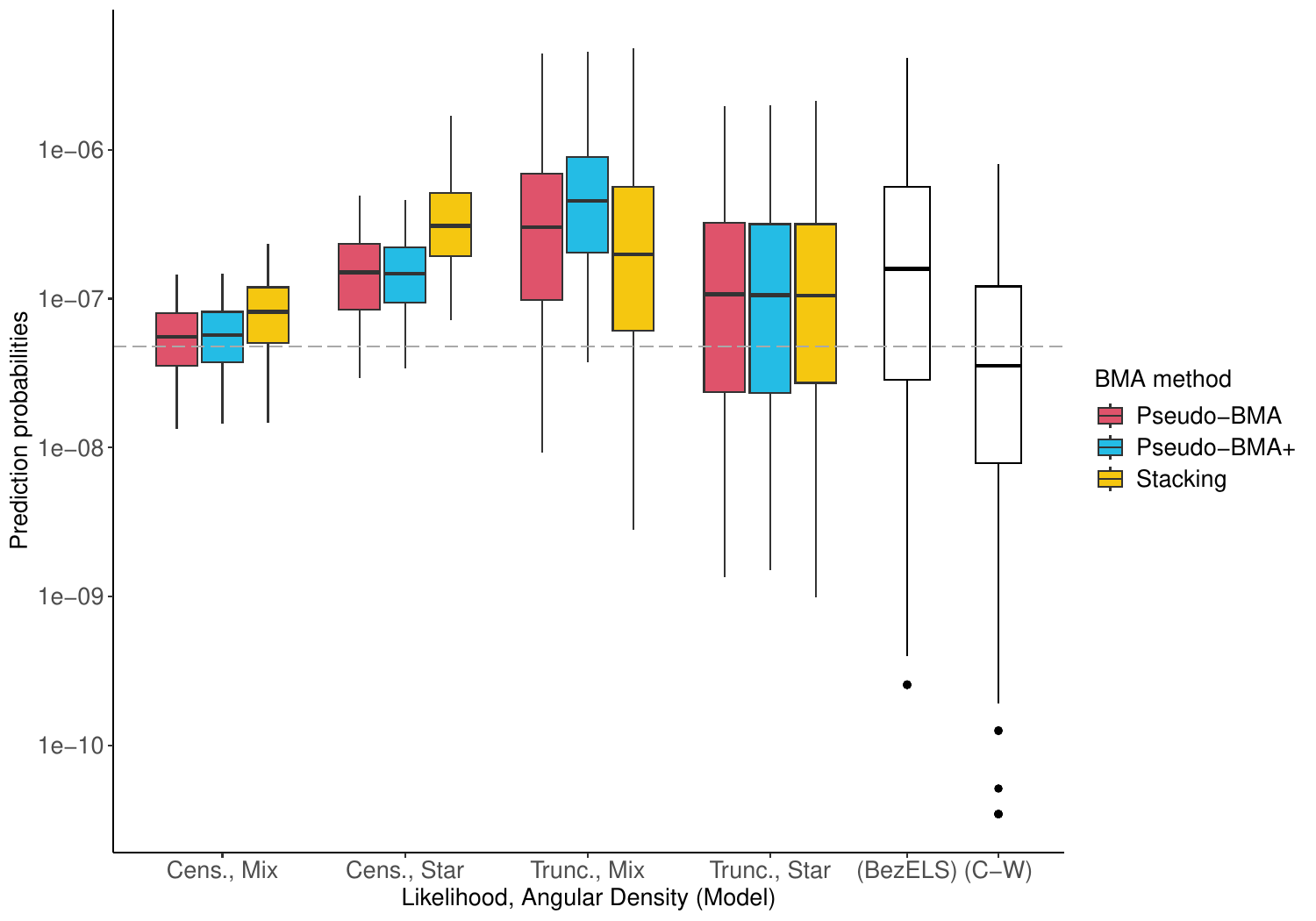}
\caption{Logistic dependence, with $\lambda = 0.5$. Predictions in $B_3=(10,12)\times(2,4)$}
\label{fig:logistic_mid_b3}
\end{figure}

\begin{figure}[!htbp]
\centering
\includegraphics[width=0.65\linewidth]{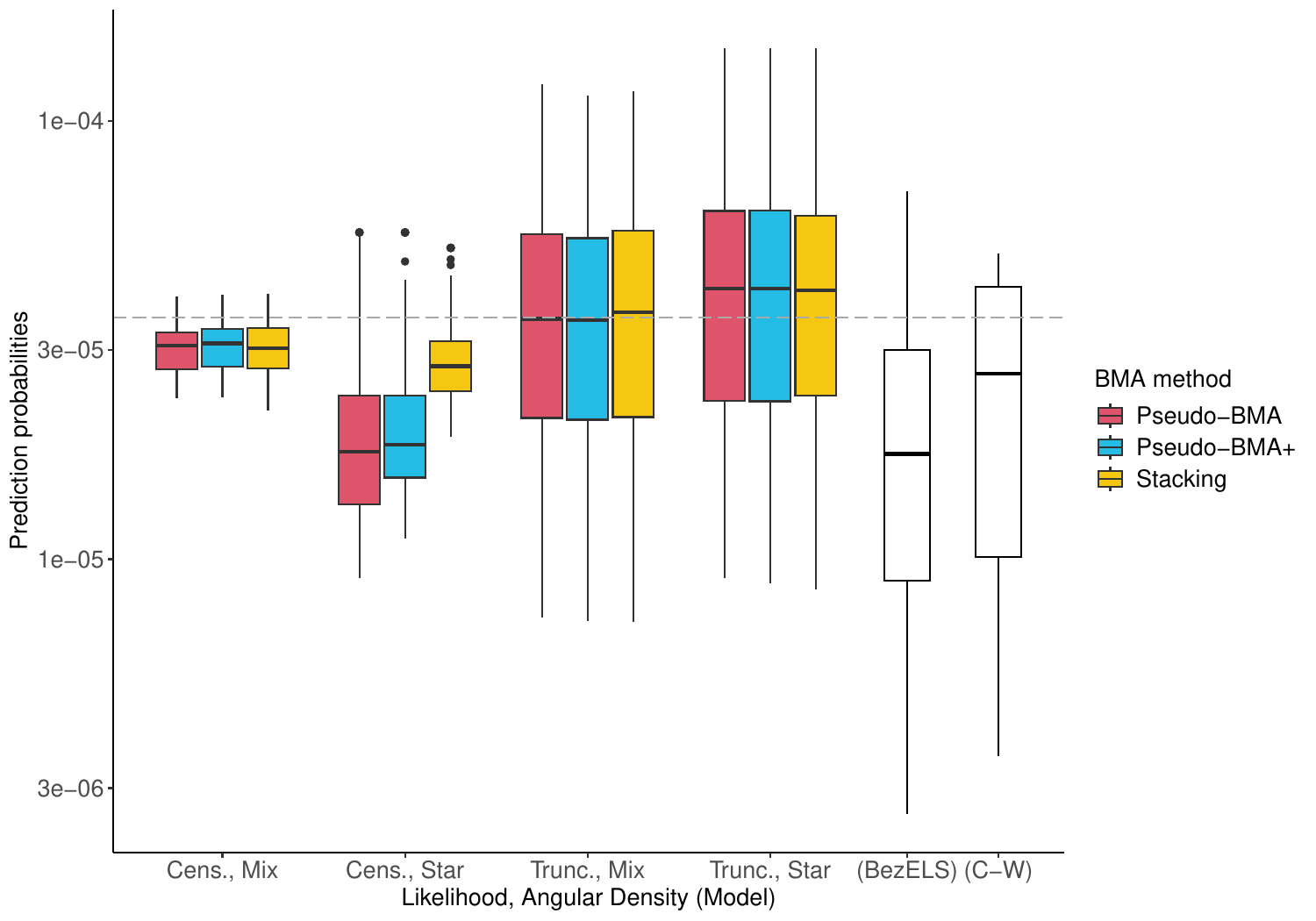}
\caption{Logistic dependence, with $\lambda = 0.1$. Predictions in $B_1=(10,12)\times(10,12)$}
\label{fig:logistic_high_b1}
\end{figure}

\begin{figure}[!htbp]
\centering
\includegraphics[width=0.65\linewidth]{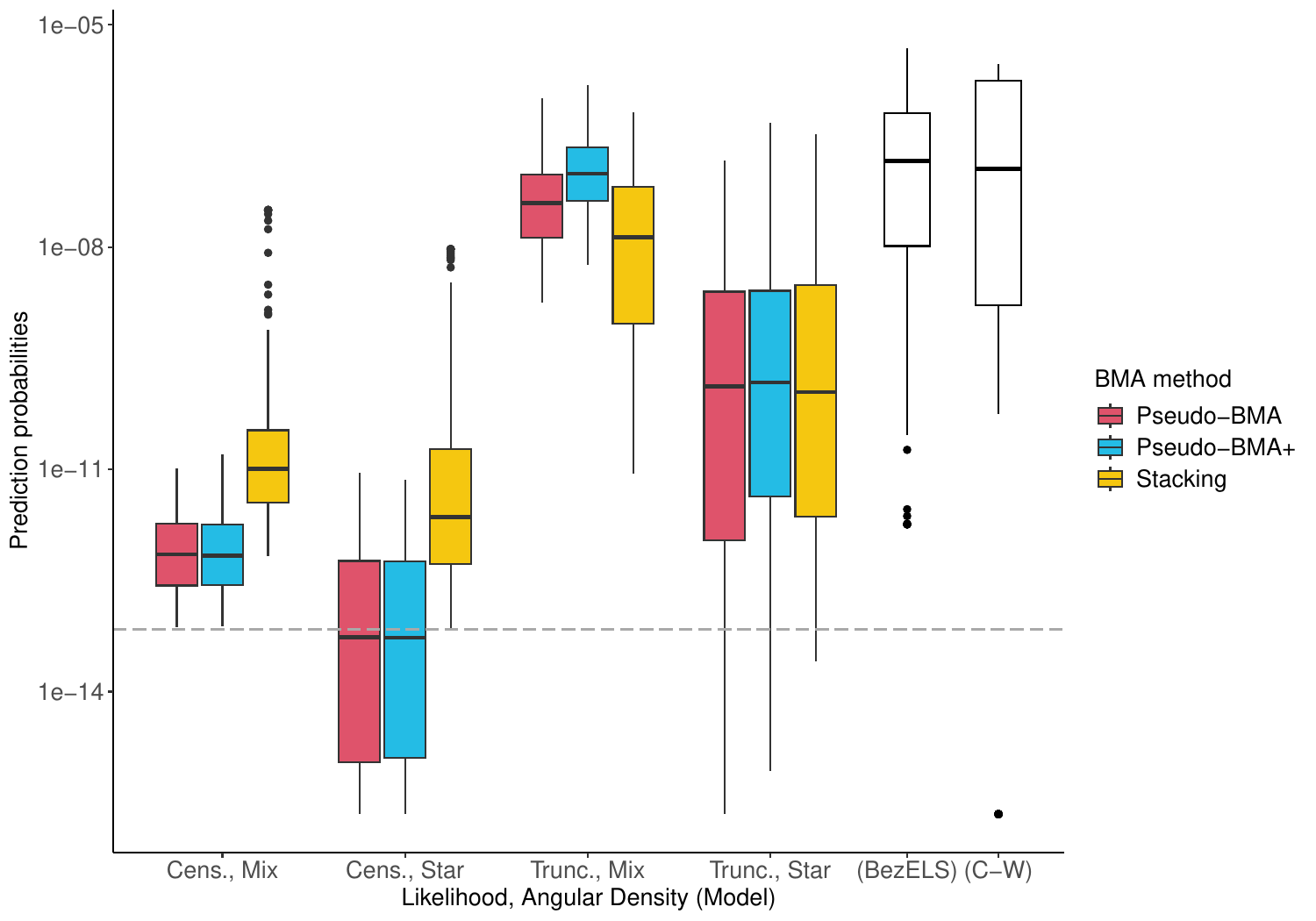}
\caption{Logistic dependence, with $\lambda = 0.1$. Predictions in $B_2=(10,12)\times(6,8)$}
\label{fig:logistic_high_b2}
\end{figure}

\clearpage
\subsubsection{H\"usler-Reiss dependence structures}
\begin{figure}[!htbp]
\centering
\includegraphics[width=0.65\linewidth]{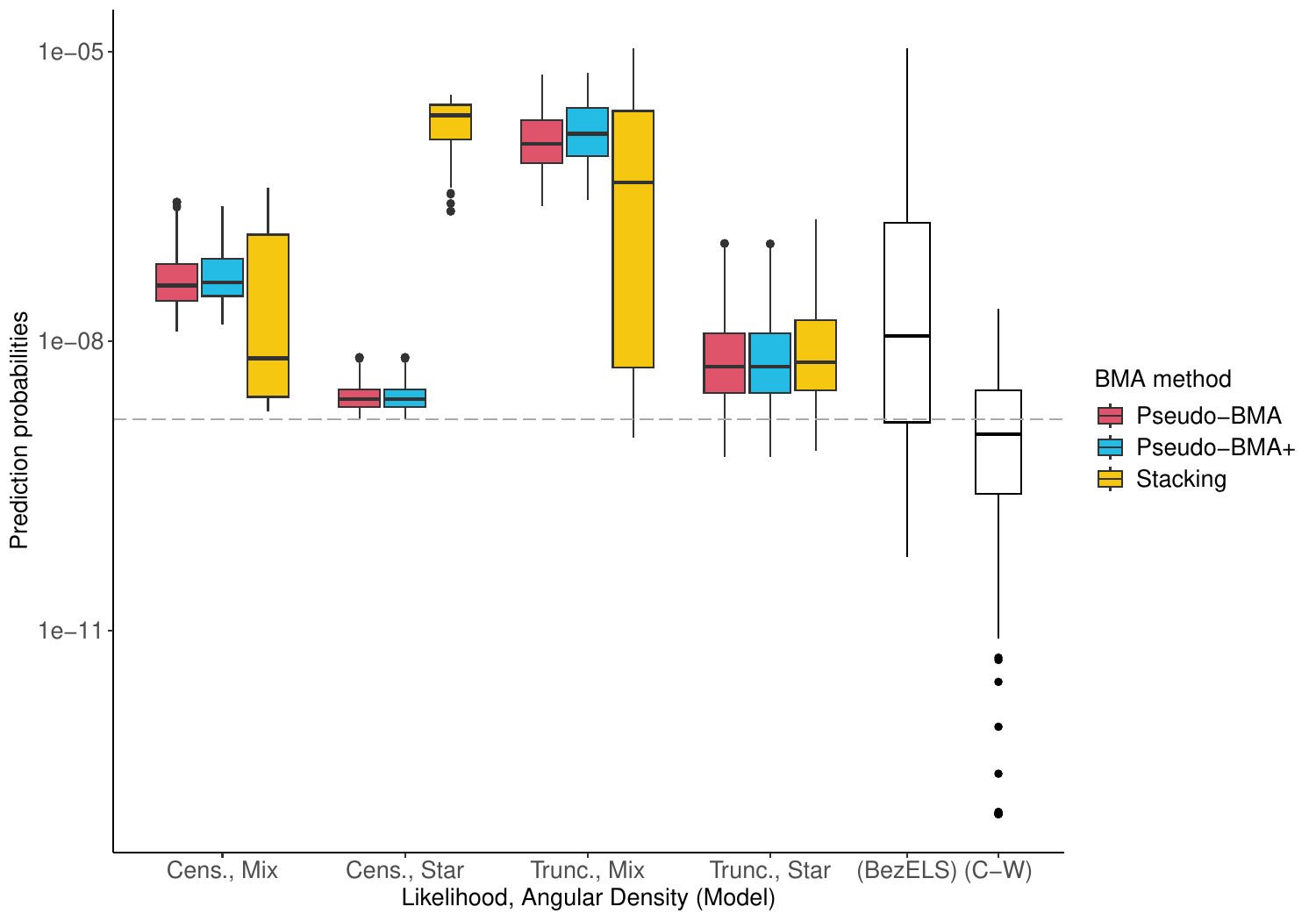}
\caption{H\"usler-Reiss dependence, with $\lambda = 0.1$. Predictions in $B_1=(10,12)\times(10,12)$}
\label{fig:husler_reiss_low_b1}
\end{figure}

\begin{figure}[!htbp]
\centering
\includegraphics[width=0.65\linewidth]{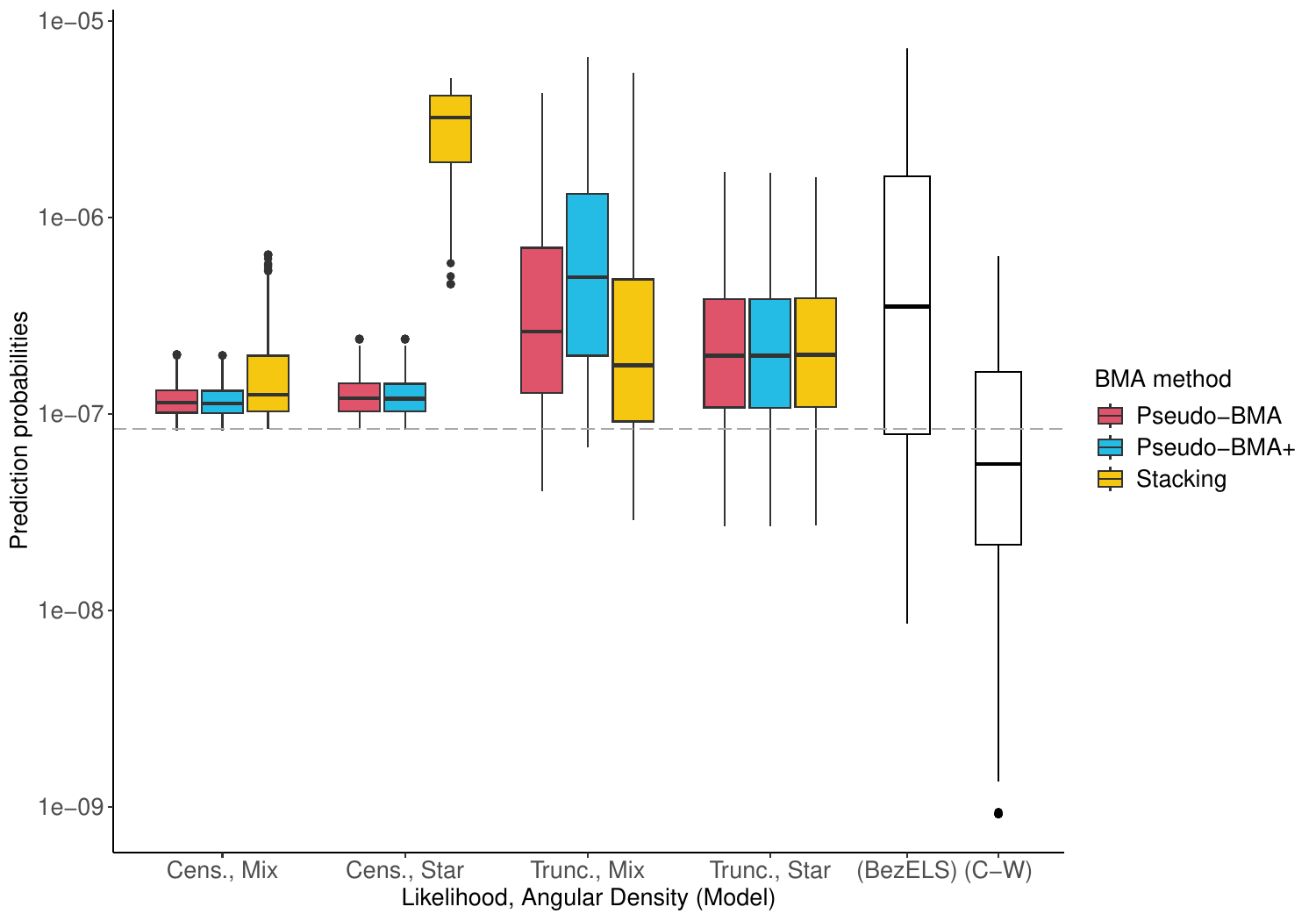}
\caption{H\"usler-Reiss dependence, with $\lambda = 0.1$. Predictions in $B_2=(10,12)\times(6,8)$}
\label{fig:husler_reiss_low_b2}
\end{figure}

\begin{figure}[!htbp]
\centering
\includegraphics[width=0.65\linewidth]{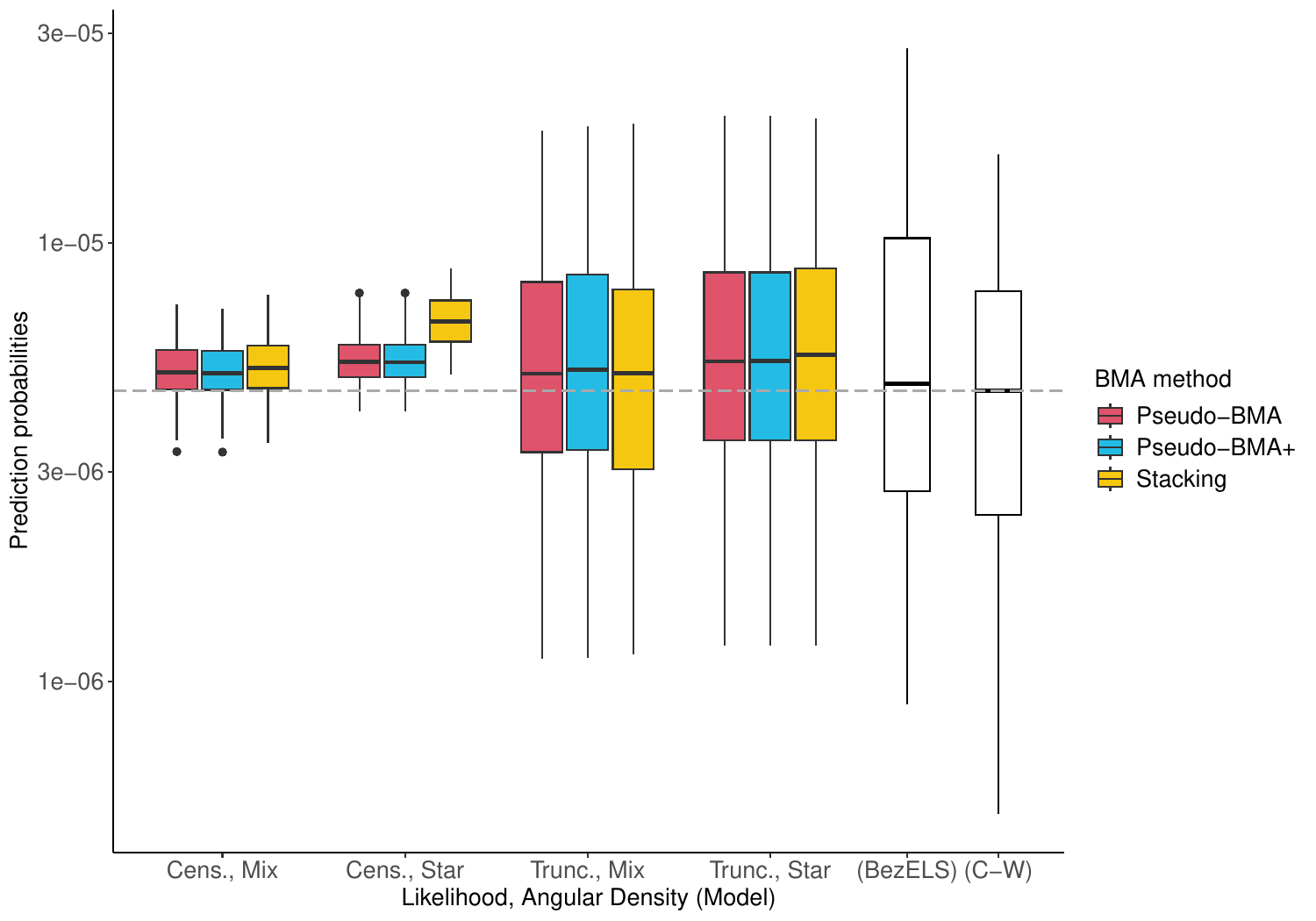}
\caption{H\"usler-Reiss dependence, with $\lambda = 0.1$. Predictions in $B_3=(10,12)\times(2,4)$}
\label{fig:husler_reiss_low_b3}
\end{figure}

\begin{figure}[!htbp]
\centering
\includegraphics[width=0.65\linewidth]{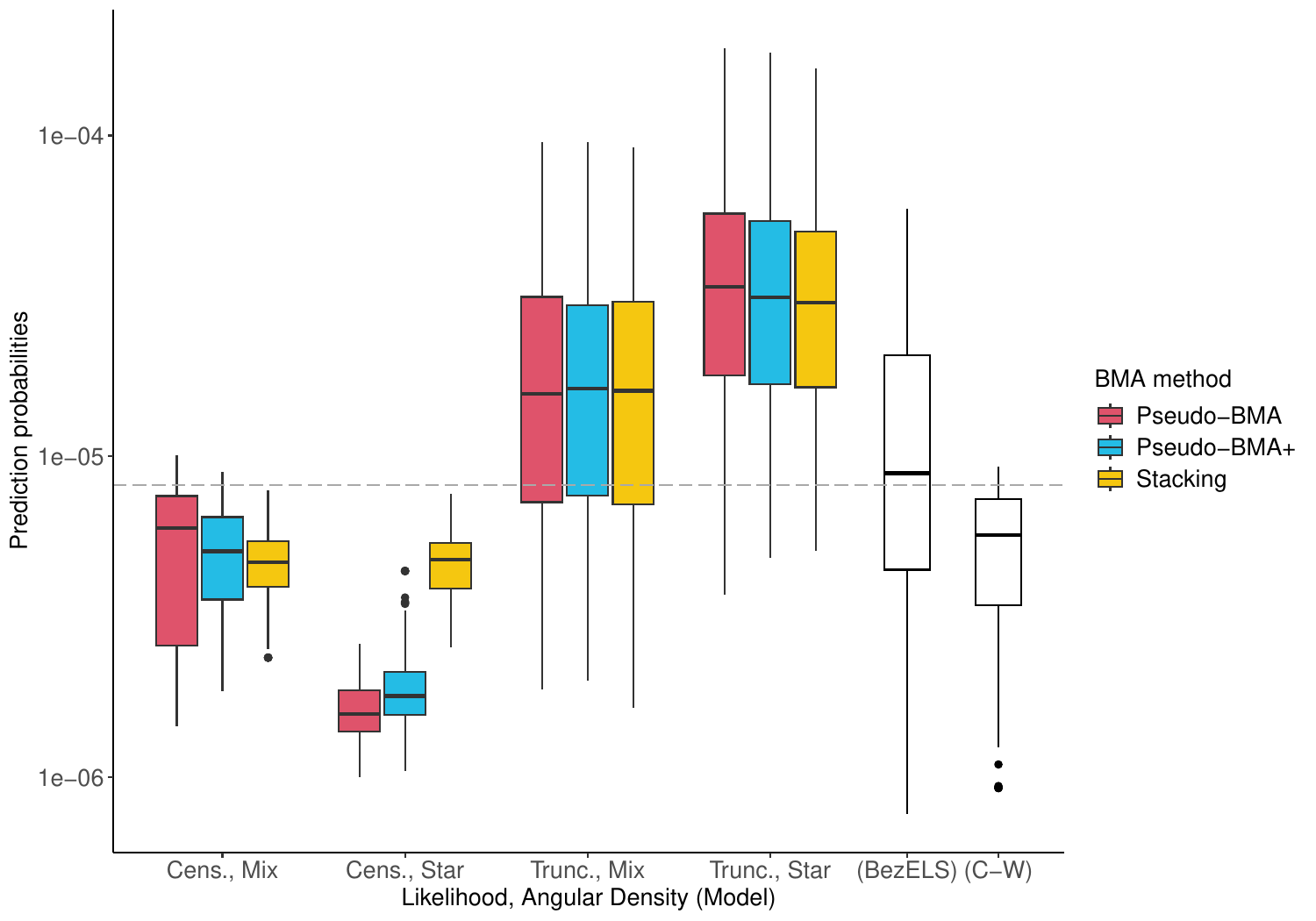}
\caption{H\"usler-Reiss dependence, with $\lambda = 1$. Predictions in $B_1=(10,12)\times(10,12)$}
\label{fig:husler_reiss_mid_b1}
\end{figure}

\begin{figure}[!htbp]
\centering
\includegraphics[width=0.65\linewidth]{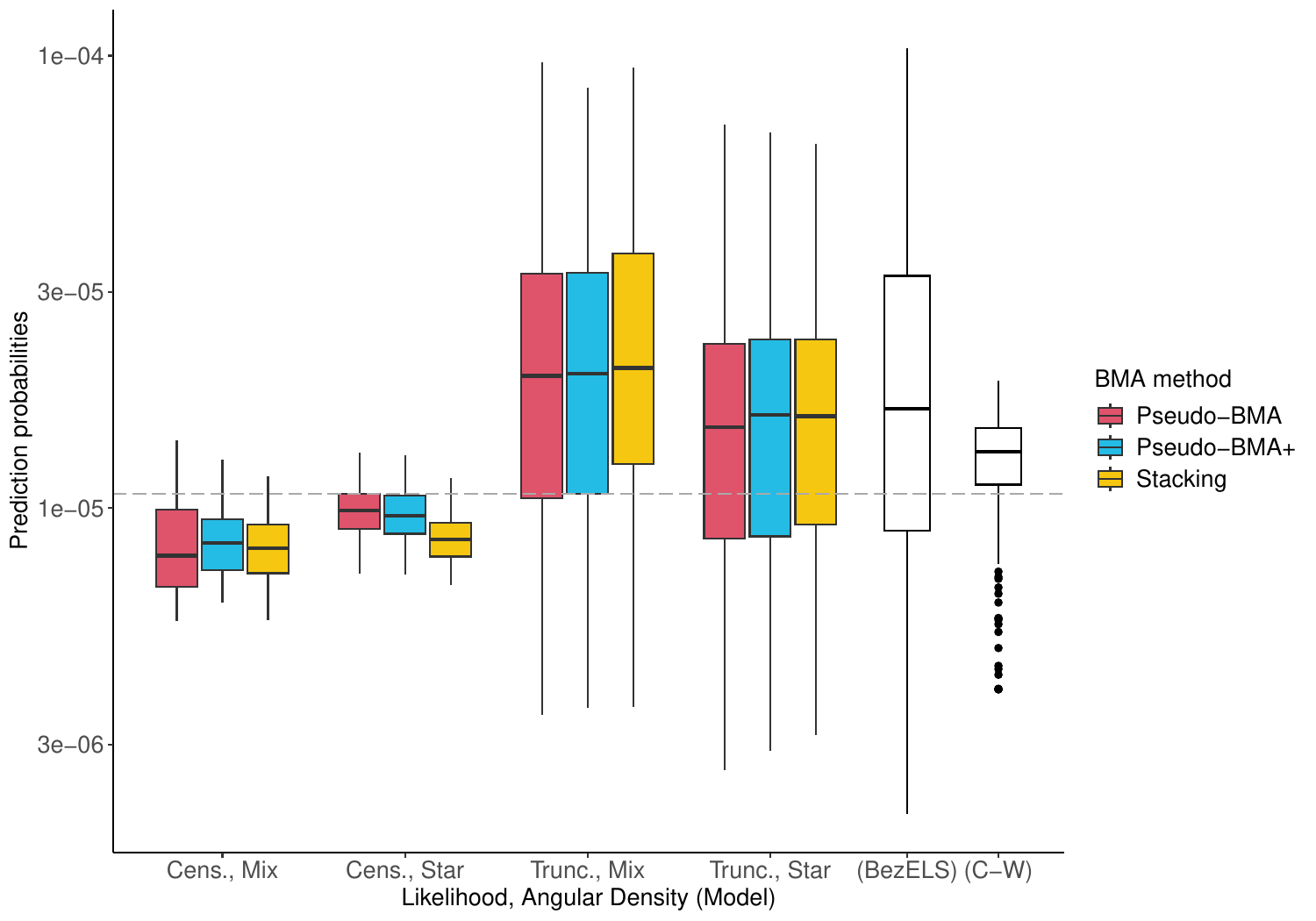}
\caption{H\"usler-Reiss dependence, with $\lambda = 1$. Predictions in $B_2=(10,12)\times(6,8)$}
\label{fig:husler_reiss_mid_b2}
\end{figure}

\begin{figure}[!htbp]
\centering
\includegraphics[width=0.65\linewidth]{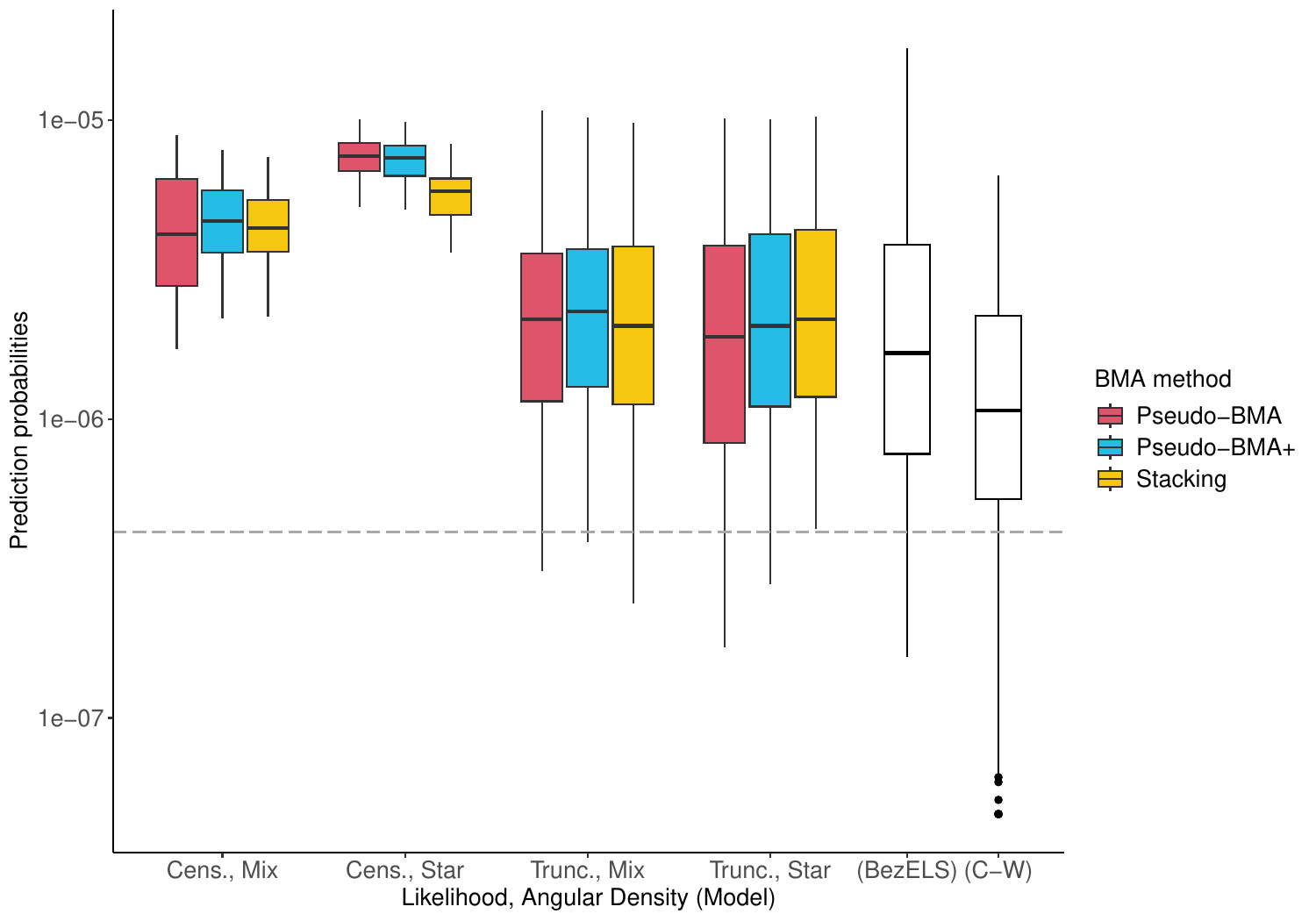}
\caption{H\"usler-Reiss dependence, with $\lambda = 1$. Predictions in $B_3=(10,12)\times(2,4)$}
\label{fig:husler_reiss_mid_b3}
\end{figure}

\begin{figure}[!htbp]
\centering
\includegraphics[width=0.65\linewidth]{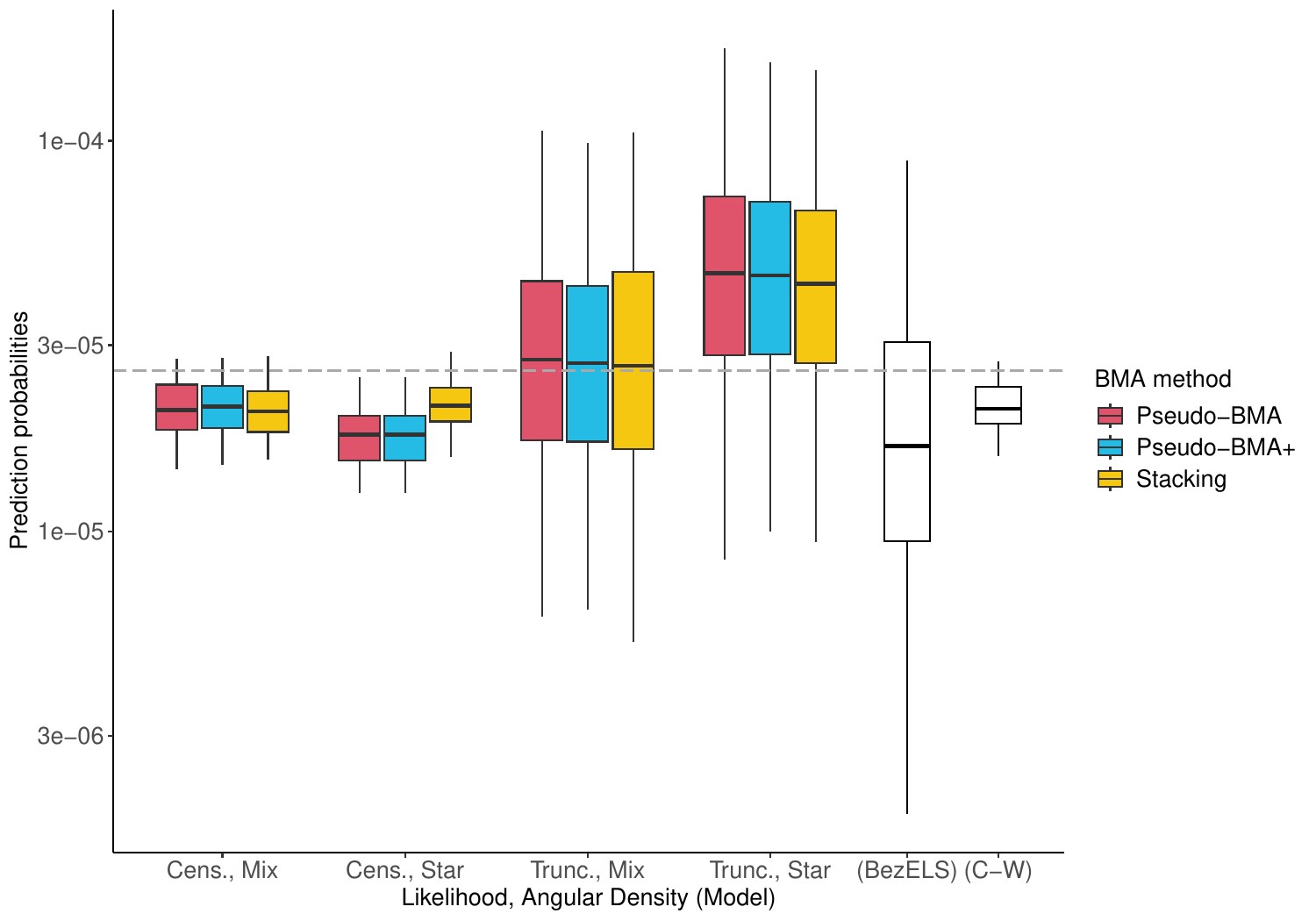}
\caption{H\"usler-Reiss dependence, with $\lambda = 3$. Predictions in $B_1=(10,12)\times(10,12)$}
\label{fig:husler_reiss_high_b1}
\end{figure}

\begin{figure}[!htbp]
\centering
\includegraphics[width=0.65\linewidth]{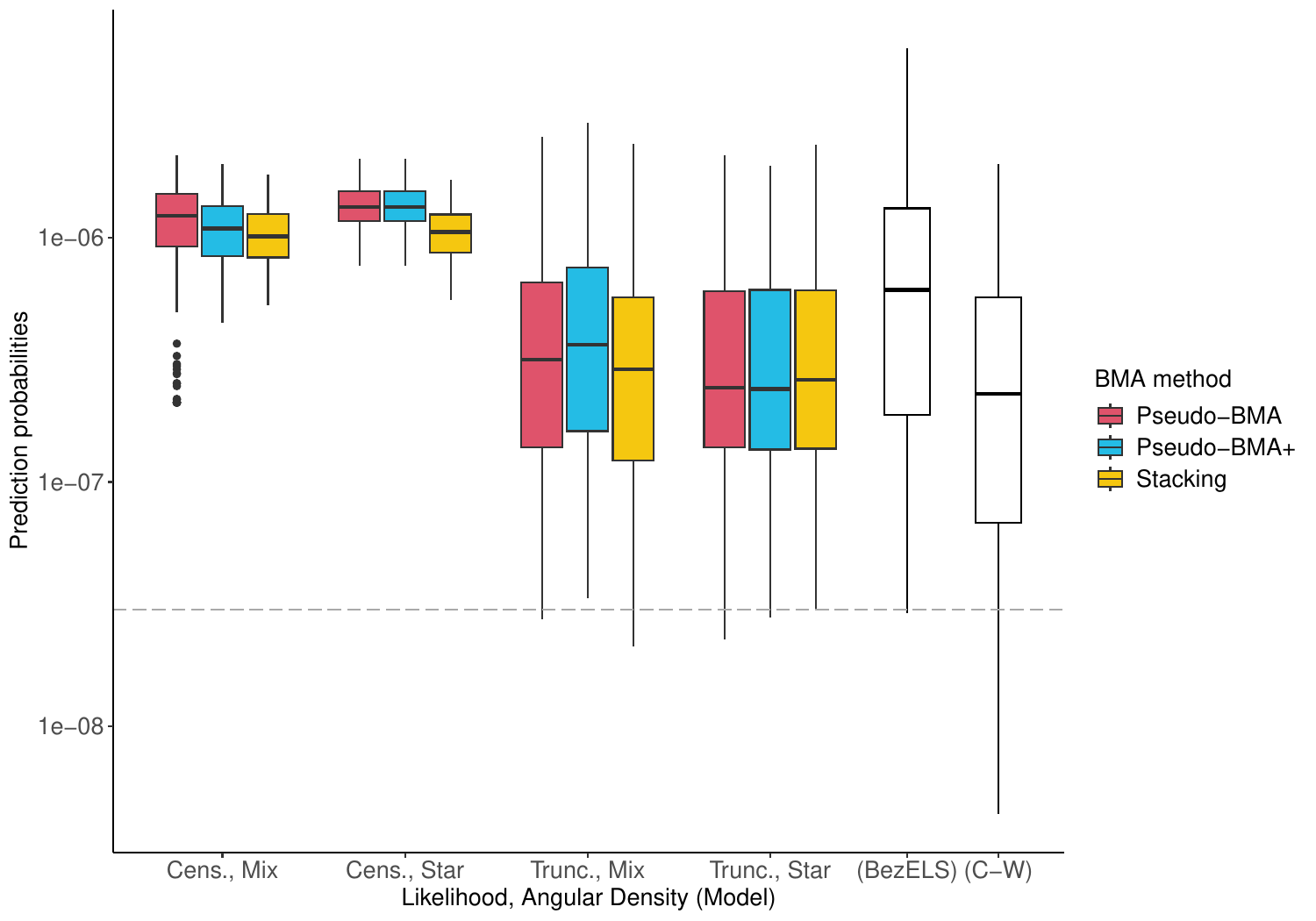}
\caption{Hüsler-Reiss dependence, with $\lambda = 3$. Predictions in $B_2=(10,12)\times(6,8)$}
\label{fig:husler_reiss_high_b2}
\end{figure}


\subsection{Averaging Across Angular Models} \label{append:sim_study_BMA_angular}
\begin{table}[!htbp]
\centering
\begin{tabular}[t]{c|c|c||c|c|c||c|c|c}
\toprule
\multicolumn{3}{c}{\textbf{Low}} & \multicolumn{3}{c}{\textbf{Mid}} & \multicolumn{3}{c}{\textbf{High}} \\
\cmidrule(l{3pt}r{3pt}){1-3} \cmidrule(l{3pt}r{3pt}){4-6} \cmidrule(l{3pt}r{3pt}){7-9}
B1 & B2 & B3 & B1 & B2 & B3 & B1 & B2 & B3\\
\midrule
\cellcolor[HTML]{2297E6}$\star\,\textit{1.110}$ & \cellcolor[HTML]{2297E6}$\circledast\,\textit{0.558}$ & \cellcolor[HTML]{F5C710}$\star\,\textit{0.220}$ & \cellcolor[HTML]{DF536B}$\star\,\textit{0.646}$ & \cellcolor[HTML]{F5C710}$\bigcirc\,\textit{0.315}$ & \cellcolor[HTML]{2297E6}$\star\,\textit{0.151}$ & \cellcolor[HTML]{F5C710}$\star\,\textit{0.187}$ & \cellcolor[HTML]{DF536B}$\circledast\,\textit{0.185}$ & \cellcolor[HTML]{DF536B}$\star\,\textit{0.668}$\\
\cellcolor[HTML]{DF536B}$\star\,\textit{1.112}$ & \cellcolor[HTML]{2297E6}$\bigcirc\,\textit{0.559}$ & \cellcolor[HTML]{F5C710}$\circledast\,\textit{0.225}$ & \cellcolor[HTML]{2297E6}$\star\,\textit{0.647}$ & \cellcolor[HTML]{DF536B}$\star\,\textit{0.337}$ & \cellcolor[HTML]{DF536B}$\star\,\textit{0.151}$ & \cellcolor[HTML]{DF536B}$\star\,\textit{0.190}$ & \cellcolor[HTML]{2297E6}$\circledast\,\textit{0.185}$ & \cellcolor[HTML]{2297E6}$\star\,\textit{0.668}$\\
\cellcolor[HTML]{FFFFFF}$4.830$ & \cellcolor[HTML]{DF536B}$\circledast\,\textit{0.607}$ & \cellcolor[HTML]{2297E6}$\circledast\,\textit{0.229}$ & \cellcolor[HTML]{DF536B}$\bigcirc\,\textit{1.079}$ & \cellcolor[HTML]{2297E6}$\star\,\textit{0.337}$ & \cellcolor[HTML]{DF536B}$\bigcirc\,\textit{0.189}$ & \cellcolor[HTML]{2297E6}$\star\,\textit{0.190}$ & \cellcolor[HTML]{2297E6}$\bigcirc\,\textit{0.188}$ & \cellcolor[HTML]{DF536B}$\circledast\,\textit{0.679}$\\
\cellcolor[HTML]{2297E6}$\star\,\textbf{10.924}$ & \cellcolor[HTML]{DF536B}$\bigcirc\,\textit{0.611}$ & \cellcolor[HTML]{F5C710}$\bigcirc\,\textit{0.233}$ & \cellcolor[HTML]{DF536B}$\circledast\,\textit{1.080}$ & \cellcolor[HTML]{2297E6}$\circledast\,\textit{0.354}$ & \cellcolor[HTML]{2297E6}$\bigcirc\,\textit{0.191}$ & \cellcolor[HTML]{DF536B}$\circledast\,\textit{0.194}$ & \cellcolor[HTML]{DF536B}$\bigcirc\,\textit{0.188}$ & \cellcolor[HTML]{DF536B}$\bigcirc\,\textit{0.688}$\\
\cellcolor[HTML]{DF536B}$\star\,\textbf{11.056}$ & \cellcolor[HTML]{2297E6}$\star\,\textit{0.674}$ & \cellcolor[HTML]{2297E6}$\bigcirc\,\textit{0.238}$ & \cellcolor[HTML]{FFFFFF}$1.201$ & \cellcolor[HTML]{2297E6}$\bigcirc\,\textit{0.354}$ & \cellcolor[HTML]{F5C710}$\bigcirc\,\textit{0.191}$ & \cellcolor[HTML]{DF536B}$\bigcirc\,\textit{0.199}$ & \cellcolor[HTML]{DF536B}$\star\,\textit{0.203}$ & \cellcolor[HTML]{2297E6}$\circledast\,\textit{0.800}$\\
\cellcolor[HTML]{DF536B}$\circledast\,\textit{12.888}$ & \cellcolor[HTML]{DF536B}$\star\,\textit{0.675}$ & \cellcolor[HTML]{2297E6}$\star\,\textit{0.239}$ & \cellcolor[HTML]{2297E6}$\bigcirc\,\textit{1.330}$ & \cellcolor[HTML]{F5C710}$\circledast\,\textit{0.363}$ & \cellcolor[HTML]{DF536B}$\circledast\,\textit{0.194}$ & \cellcolor[HTML]{F5C710}$\bigcirc\,\textit{0.203}$ & \cellcolor[HTML]{2297E6}$\star\,\textit{0.203}$ & \cellcolor[HTML]{2297E6}$\bigcirc\,\textit{0.840}$\\
\cellcolor[HTML]{DF536B}$\bigcirc\,\textit{13.253}$ & \cellcolor[HTML]{F5C710}$\bigcirc\,\textit{1.729}$ & \cellcolor[HTML]{DF536B}$\circledast\,\textit{0.239}$ & \cellcolor[HTML]{2297E6}$\circledast\,\textit{1.330}$ & \cellcolor[HTML]{DF536B}$\circledast\,\textit{0.393}$ & \cellcolor[HTML]{2297E6}$\circledast\,\textit{0.196}$ & \cellcolor[HTML]{2297E6}$\circledast\,\textit{0.204}$ & \cellcolor[HTML]{F5C710}$\bigcirc\,\textit{0.209}$ & \cellcolor[HTML]{F5C710}$\bigcirc\,\textit{3.333}$\\
\cellcolor[HTML]{2297E6}$\bigcirc\,\textit{16.663}$ & \cellcolor[HTML]{FFFFFF}$2.536$ & \cellcolor[HTML]{DF536B}$\star\,\textit{0.239}$ & \cellcolor[HTML]{F5C710}$\bigcirc\,\textit{1.885}$ & \cellcolor[HTML]{DF536B}$\bigcirc\,\textit{0.393}$ & \cellcolor[HTML]{F5C710}$\star\,\textit{0.213}$ & \cellcolor[HTML]{F5C710}$\circledast\,\textit{0.206}$ & \cellcolor[HTML]{F5C710}$\star\,\textit{0.224}$ & \cellcolor[HTML]{FFFFFF}$12.863$\\
\cellcolor[HTML]{2297E6}$\circledast\,\textit{17.294}$ & \cellcolor[HTML]{2297E6}$\star\,\textbf{3.448}$ & \cellcolor[HTML]{DF536B}$\bigcirc\,\textit{0.249}$ & \cellcolor[HTML]{DF536B}$\star\,\textbf{2.306}$ & \cellcolor[HTML]{F5C710}$\star\,\textit{0.413}$ & \cellcolor[HTML]{F5C710}$\circledast\,\textit{0.241}$ & \cellcolor[HTML]{2297E6}$\bigcirc\,\textit{0.207}$ & \cellcolor[HTML]{F5C710}$\circledast\,\textit{0.267}$ & \cellcolor[HTML]{F5C710}$\star\,\textit{41.923}$\\
\cellcolor[HTML]{F5C710}$\bigcirc\,\textit{46.497}$ & \cellcolor[HTML]{DF536B}$\star\,\textbf{3.458}$ & \cellcolor[HTML]{FFFFFF}$0.827$ & \cellcolor[HTML]{2297E6}$\star\,\textbf{2.327}$ & \cellcolor[HTML]{FFFFFF}$0.658$ & \cellcolor[HTML]{FFFFFF}$0.466$ & \cellcolor[HTML]{FFFFFF}$0.684$ & \cellcolor[HTML]{FFFFFF}$0.624$ & \cellcolor[HTML]{F5C710}$\circledast\,\textit{78.495}$\\
\cellcolor[HTML]{F5C710}$\star\,\textbf{77.498}$ & \cellcolor[HTML]{F5C710}$\star\,\textbf{4.459}$ & \cellcolor[HTML]{2297E6}$\star\,\textbf{0.867}$ & \cellcolor[HTML]{F5C710}$\star\,\textit{3.665}$ & \cellcolor[HTML]{2297E6}$\star\,\textbf{1.347}$ & \cellcolor[HTML]{F5C710}$\star\,\textbf{0.866}$ & \cellcolor[HTML]{FFFFFF}$\bigcirc\,1.454$ & \cellcolor[HTML]{DF536B}$\bigcirc\,\textbf{1.434}$ & \cellcolor[HTML]{FFFFFF}$\bigcirc\,141.723$\\
\cellcolor[HTML]{F5C710}$\circledast\,\textit{138.477}$ & \cellcolor[HTML]{F5C710}$\circledast\,\textit{6.893}$ & \cellcolor[HTML]{DF536B}$\star\,\textbf{0.867}$ & \cellcolor[HTML]{F5C710}$\circledast\,\textit{4.424}$ & \cellcolor[HTML]{DF536B}$\star\,\textbf{1.347}$ & \cellcolor[HTML]{2297E6}$\star\,\textbf{0.904}$ & \cellcolor[HTML]{2297E6}$\bigcirc\,\textbf{1.897}$ & \cellcolor[HTML]{2297E6}$\bigcirc\,\textbf{1.488}$ & \cellcolor[HTML]{2297E6}$\bigcirc\,\textbf{249.855}$\\
\cellcolor[HTML]{F5C710}$\star\,\textit{336.152}$ & \cellcolor[HTML]{DF536B}$\bigcirc\,\textbf{11.292}$ & \cellcolor[HTML]{F5C710}$\star\,\textbf{0.871}$ & \cellcolor[HTML]{F5C710}$\star\,\textbf{4.795}$ & \cellcolor[HTML]{F5C710}$\star\,\textbf{1.394}$ & \cellcolor[HTML]{DF536B}$\star\,\textbf{0.906}$ & \cellcolor[HTML]{DF536B}$\bigcirc\,\textbf{1.973}$ & \cellcolor[HTML]{DF536B}$\star\,\textbf{1.597}$ & \cellcolor[HTML]{DF536B}$\bigcirc\,\textbf{252.395}$\\
\cellcolor[HTML]{FFFFFF}$\bigcirc\,414.870$ & \cellcolor[HTML]{2297E6}$\bigcirc\,\textbf{12.725}$ & \cellcolor[HTML]{DF536B}$\bigcirc\,\textbf{0.906}$ & \cellcolor[HTML]{FFFFFF}$\bigcirc\,8.841$ & \cellcolor[HTML]{F5C710}$\bigcirc\,\textbf{1.561}$ & \cellcolor[HTML]{2297E6}$\bigcirc\,\textbf{0.978}$ & \cellcolor[HTML]{F5C710}$\bigcirc\,\textbf{2.078}$ & \cellcolor[HTML]{2297E6}$\star\,\textbf{1.597}$ & \cellcolor[HTML]{F5C710}$\bigcirc\,\textbf{296.203}$\\
\cellcolor[HTML]{DF536B}$\bigcirc\,\textbf{511.159}$ & \cellcolor[HTML]{F5C710}$\star\,\textit{12.981}$ & \cellcolor[HTML]{2297E6}$\bigcirc\,\textbf{0.912}$ & \cellcolor[HTML]{2297E6}$\bigcirc\,\textbf{12.164}$ & \cellcolor[HTML]{2297E6}$\bigcirc\,\textbf{1.577}$ & \cellcolor[HTML]{DF536B}$\bigcirc\,\textbf{1.015}$ & \cellcolor[HTML]{2297E6}$\star\,\textbf{2.534}$ & \cellcolor[HTML]{F5C710}$\star\,\textbf{1.598}$ & \cellcolor[HTML]{F5C710}$\star\,\textbf{325.473}$\\
\cellcolor[HTML]{2297E6}$\bigcirc\,\textbf{538.582}$ & \cellcolor[HTML]{F5C710}$\bigcirc\,\textbf{14.384}$ & \cellcolor[HTML]{F5C710}$\bigcirc\,\textbf{1.054}$ & \cellcolor[HTML]{DF536B}$\bigcirc\,\textbf{12.513}$ & \cellcolor[HTML]{DF536B}$\bigcirc\,\textbf{1.583}$ & \cellcolor[HTML]{FFFFFF}$\bigcirc\,1.106$ & \cellcolor[HTML]{DF536B}$\star\,\textbf{2.535}$ & \cellcolor[HTML]{FFFFFF}$\bigcirc\,1.617$ & \cellcolor[HTML]{2297E6}$\star\,\textbf{326.152}$\\
\cellcolor[HTML]{F5C710}$\bigcirc\,\textbf{777.740}$ & \cellcolor[HTML]{FFFFFF}$\bigcirc\,18.873$ & \cellcolor[HTML]{FFFFFF}$\bigcirc\,1.579$ & \cellcolor[HTML]{F5C710}$\bigcirc\,\textbf{12.655}$ & \cellcolor[HTML]{FFFFFF}$\bigcirc\,1.721$ & \cellcolor[HTML]{F5C710}$\bigcirc\,\textbf{1.221}$ & \cellcolor[HTML]{F5C710}$\star\,\textbf{2.542}$ & \cellcolor[HTML]{F5C710}$\bigcirc\,\textbf{1.661}$ & \cellcolor[HTML]{DF536B}$\star\,\textbf{326.195}$\\
\bottomrule
\end{tabular}
    \caption{Comparisons across low, mid, and high levels of Gaussian dependence structures, in three boxes of interest (Figure~\ref{fig:sim_pred_task}). This table builds upon Table~\ref{tab:gauss_mse} with the addition of the censored likelihood averaged across both angular densities, indicated by $\circledast$.}
    \label{tab:gauss_mse_angle_avg}
\end{table}

\begin{table}[!htbp]
    \centering
\centering
\begin{tabular}[t]{c|c|c||c|c|c||c|c|c}
\toprule
\multicolumn{3}{c}{\textbf{Low}} & \multicolumn{3}{c}{\textbf{Mid}} & \multicolumn{3}{c}{\textbf{High}} \\
\cmidrule(l{3pt}r{3pt}){1-3} \cmidrule(l{3pt}r{3pt}){4-6} \cmidrule(l{3pt}r{3pt}){7-9}
B1 & B2 & B3 & B1 & B2 & B3 & B1 & B2 & B3\\
\midrule
\cellcolor[HTML]{F5C710}$\bigcirc\,\textit{0.467}$ & \cellcolor[HTML]{F5C710}$\star\,\textit{0.399}$ & \cellcolor[HTML]{FFFFFF}$0.810$ & \cellcolor[HTML]{DF536B}$\bigcirc\,\textit{0.193}$ & \cellcolor[HTML]{DF536B}$\bigcirc\,\textit{0.339}$ & \cellcolor[HTML]{DF536B}$\circledast\,\textit{0.810}$ & \cellcolor[HTML]{2297E6}$\circledast\,\textit{0.179}$ & \cellcolor[HTML]{2297E6}$\star\,\textit{38.069}$ & --\\
\cellcolor[HTML]{F5C710}$\circledast\,\textit{0.488}$ & \cellcolor[HTML]{F5C710}$\circledast\,\textit{0.451}$ & \cellcolor[HTML]{F5C710}$\bigcirc\,\textit{1.010}$ & \cellcolor[HTML]{DF536B}$\circledast\,\textit{0.196}$ & \cellcolor[HTML]{DF536B}$\circledast\,\textit{0.345}$ & \cellcolor[HTML]{DF536B}$\bigcirc\,\textit{0.824}$ & \cellcolor[HTML]{2297E6}$\bigcirc\,\textit{0.180}$ & \cellcolor[HTML]{DF536B}$\star\,\textit{41.279}$ & --\\
\cellcolor[HTML]{F5C710}$\star\,\textit{0.524}$ & \cellcolor[HTML]{F5C710}$\bigcirc\,\textit{0.485}$ & \cellcolor[HTML]{2297E6}$\circledast\,\textit{1.098}$ & \cellcolor[HTML]{2297E6}$\bigcirc\,\textit{0.206}$ & \cellcolor[HTML]{2297E6}$\bigcirc\,\textit{0.420}$ & \cellcolor[HTML]{2297E6}$\circledast\,\textit{0.831}$ & \cellcolor[HTML]{DF536B}$\circledast\,\textit{0.187}$ & \cellcolor[HTML]{DF536B}$\circledast\,\textit{51.554}$ & --\\
\cellcolor[HTML]{FFFFFF}$0.635$ & \cellcolor[HTML]{2297E6}$\bigcirc\,\textit{0.503}$ & \cellcolor[HTML]{2297E6}$\bigcirc\,\textit{1.127}$ & \cellcolor[HTML]{2297E6}$\circledast\,\textit{0.208}$ & \cellcolor[HTML]{2297E6}$\circledast\,\textit{0.425}$ & \cellcolor[HTML]{2297E6}$\bigcirc\,\textit{0.839}$ & \cellcolor[HTML]{DF536B}$\bigcirc\,\textit{0.189}$ & \cellcolor[HTML]{DF536B}$\bigcirc\,\textit{56.271}$ & --\\
\cellcolor[HTML]{2297E6}$\bigcirc\,\textit{0.695}$ & \cellcolor[HTML]{2297E6}$\circledast\,\textit{0.504}$ & \cellcolor[HTML]{DF536B}$\circledast\,\textit{1.213}$ & \cellcolor[HTML]{F5C710}$\bigcirc\,\textit{0.285}$ & \cellcolor[HTML]{F5C710}$\bigcirc\,\textit{1.082}$ & \cellcolor[HTML]{F5C710}$\bigcirc\,\textit{1.564}$ & \cellcolor[HTML]{F5C710}$\bigcirc\,\textit{0.197}$ & \cellcolor[HTML]{2297E6}$\circledast\,\textit{69.413}$ & --\\
\cellcolor[HTML]{2297E6}$\circledast\,\textit{0.695}$ & \cellcolor[HTML]{DF536B}$\bigcirc\,\textit{0.541}$ & \cellcolor[HTML]{F5C710}$\circledast\,\textit{1.241}$ & \cellcolor[HTML]{FFFFFF}$0.376$ & \cellcolor[HTML]{DF536B}$\star\,\textbf{1.919}$ & \cellcolor[HTML]{2297E6}$\star\,\textit{3.537}$ & \cellcolor[HTML]{F5C710}$\star\,\textit{0.310}$ & \cellcolor[HTML]{2297E6}$\bigcirc\,\textit{87.881}$ & --\\
\cellcolor[HTML]{F5C710}$\star\,\textbf{0.771}$ & \cellcolor[HTML]{DF536B}$\circledast\,\textit{0.542}$ & \cellcolor[HTML]{DF536B}$\bigcirc\,\textit{1.241}$ & \cellcolor[HTML]{F5C710}$\star\,\textit{0.437}$ & \cellcolor[HTML]{2297E6}$\star\,\textbf{1.981}$ & \cellcolor[HTML]{DF536B}$\star\,\textit{3.711}$ & \cellcolor[HTML]{F5C710}$\circledast\,\textit{0.423}$ & \cellcolor[HTML]{F5C710}$\star\,\textit{4.303e+04}$ & --\\
\cellcolor[HTML]{DF536B}$\star\,\textbf{0.777}$ & \cellcolor[HTML]{FFFFFF}$0.550$ & \cellcolor[HTML]{F5C710}$\star\,\textit{1.306}$ & \cellcolor[HTML]{2297E6}$\star\,\textit{0.560}$ & \cellcolor[HTML]{F5C710}$\star\,\textit{2.041}$ & \cellcolor[HTML]{FFFFFF}$5.682$ & \cellcolor[HTML]{2297E6}$\star\,\textit{0.492}$ & \cellcolor[HTML]{F5C710}$\bigcirc\,\textit{1.207e+05}$ & --\\
\cellcolor[HTML]{2297E6}$\star\,\textbf{0.781}$ & \cellcolor[HTML]{2297E6}$\star\,\textit{0.708}$ & \cellcolor[HTML]{DF536B}$\bigcirc\,\textbf{1.484}$ & \cellcolor[HTML]{F5C710}$\circledast\,\textit{0.595}$ & \cellcolor[HTML]{F5C710}$\star\,\textbf{2.209}$ & \cellcolor[HTML]{F5C710}$\star\,\textit{11.044}$ & \cellcolor[HTML]{DF536B}$\star\,\textit{0.522}$ & \cellcolor[HTML]{DF536B}$\star\,\textbf{1.076e+06}$ & --\\
\cellcolor[HTML]{DF536B}$\bigcirc\,\textit{0.875}$ & \cellcolor[HTML]{DF536B}$\star\,\textit{0.723}$ & \cellcolor[HTML]{2297E6}$\star\,\textit{1.561}$ & \cellcolor[HTML]{DF536B}$\star\,\textit{0.623}$ & \cellcolor[HTML]{FFFFFF}$2.302$ & \cellcolor[HTML]{2297E6}$\star\,\textbf{16.293}$ & \cellcolor[HTML]{FFFFFF}$0.530$ & \cellcolor[HTML]{2297E6}$\star\,\textbf{1.627e+06}$ & --\\
\cellcolor[HTML]{DF536B}$\circledast\,\textit{0.877}$ & \cellcolor[HTML]{2297E6}$\star\,\textbf{0.724}$ & \cellcolor[HTML]{DF536B}$\star\,\textit{1.577}$ & \cellcolor[HTML]{FFFFFF}$\bigcirc\,1.006$ & \cellcolor[HTML]{2297E6}$\star\,\textit{2.877}$ & \cellcolor[HTML]{DF536B}$\star\,\textbf{16.579}$ & \cellcolor[HTML]{FFFFFF}$\bigcirc\,0.660$ & \cellcolor[HTML]{F5C710}$\star\,\textbf{2.356e+06}$ & --\\
\cellcolor[HTML]{2297E6}$\star\,\textit{0.961}$ & \cellcolor[HTML]{DF536B}$\star\,\textbf{0.725}$ & \cellcolor[HTML]{2297E6}$\bigcirc\,\textbf{1.617}$ & \cellcolor[HTML]{2297E6}$\bigcirc\,\textbf{1.432}$ & \cellcolor[HTML]{F5C710}$\circledast\,\textit{2.966}$ & \cellcolor[HTML]{F5C710}$\star\,\textbf{16.831}$ & \cellcolor[HTML]{2297E6}$\bigcirc\,\textbf{0.750}$ & \cellcolor[HTML]{F5C710}$\bigcirc\,\textbf{3.127e+06}$ & --\\
\cellcolor[HTML]{DF536B}$\star\,\textit{0.966}$ & \cellcolor[HTML]{F5C710}$\star\,\textbf{0.738}$ & \cellcolor[HTML]{F5C710}$\bigcirc\,\textbf{1.659}$ & \cellcolor[HTML]{F5C710}$\bigcirc\,\textbf{1.485}$ & \cellcolor[HTML]{DF536B}$\bigcirc\,\textbf{3.189}$ & \cellcolor[HTML]{F5C710}$\circledast\,\textit{17.068}$ & \cellcolor[HTML]{DF536B}$\bigcirc\,\textbf{0.779}$ & \cellcolor[HTML]{F5C710}$\circledast\,\textit{4.968e+06}$ & --\\
\cellcolor[HTML]{FFFFFF}$\bigcirc\,2.092$ & \cellcolor[HTML]{2297E6}$\bigcirc\,\textbf{1.249}$ & \cellcolor[HTML]{F5C710}$\star\,\textbf{1.783}$ & \cellcolor[HTML]{DF536B}$\bigcirc\,\textbf{1.639}$ & \cellcolor[HTML]{DF536B}$\star\,\textit{3.250}$ & \cellcolor[HTML]{DF536B}$\bigcirc\,\textbf{27.024}$ & \cellcolor[HTML]{F5C710}$\bigcirc\,\textbf{0.809}$ & \cellcolor[HTML]{DF536B}$\bigcirc\,\textbf{5.292e+06}$ & --\\
\cellcolor[HTML]{2297E6}$\bigcirc\,\textbf{2.237}$ & \cellcolor[HTML]{DF536B}$\bigcirc\,\textbf{1.315}$ & \cellcolor[HTML]{DF536B}$\star\,\textbf{1.849}$ & \cellcolor[HTML]{F5C710}$\star\,\textbf{2.113}$ & \cellcolor[HTML]{2297E6}$\bigcirc\,\textbf{3.539}$ & \cellcolor[HTML]{2297E6}$\bigcirc\,\textbf{28.379}$ & \cellcolor[HTML]{F5C710}$\star\,\textbf{0.973}$ & \cellcolor[HTML]{2297E6}$\bigcirc\,\textbf{7.768e+06}$ & --\\
\cellcolor[HTML]{DF536B}$\bigcirc\,\textbf{2.439}$ & \cellcolor[HTML]{F5C710}$\bigcirc\,\textbf{1.328}$ & \cellcolor[HTML]{2297E6}$\star\,\textbf{1.905}$ & \cellcolor[HTML]{2297E6}$\star\,\textbf{2.173}$ & \cellcolor[HTML]{F5C710}$\bigcirc\,\textbf{3.755}$ & \cellcolor[HTML]{F5C710}$\bigcirc\,\textbf{28.885}$ & \cellcolor[HTML]{2297E6}$\star\,\textbf{1.017}$ & \cellcolor[HTML]{FFFFFF}$1.922e+07$ & --\\
\cellcolor[HTML]{F5C710}$\bigcirc\,\textbf{2.652}$ & \cellcolor[HTML]{FFFFFF}$\bigcirc\,1.996$ & \cellcolor[HTML]{FFFFFF}$\bigcirc\,1.955$ & \cellcolor[HTML]{DF536B}$\star\,\textbf{2.221}$ & \cellcolor[HTML]{FFFFFF}$\bigcirc\,4.034$ & \cellcolor[HTML]{FFFFFF}$\bigcirc\,30.066$ & \cellcolor[HTML]{DF536B}$\star\,\textbf{1.033}$ & \cellcolor[HTML]{FFFFFF}$\bigcirc\,1.975e+07$ & --\\
\bottomrule
\end{tabular}
    \caption{Comparisons across low, mid, and high levels of logistic dependence structures, in three boxes of interest (Figure~\ref{fig:sim_pred_task}). The probability for $B_3$ under the high scenario is smaller than \texttt{R}'s machine precision, yielding zero as the true probability; as a result, the RMSE values are not informative and are excluded from this table. This table is identical to Table~\ref{tab:logistic_mse} but includes the results from averaging across the angular densities, indicated by $\circledast$.}
    \label{tab:logistic_mse_angle_avg}
\end{table}
\end{document}